\documentclass[12pt,a4paper]{article}

\usepackage[dvipsnames]{xcolor}

\usepackage[english]{babel}
\usepackage[authoryear]{natbib}
\usepackage[utf8]{inputenc}
\usepackage{algorithmic}
\usepackage{algorithm}
\usepackage[official]{eurosym}     
\usepackage{multirow}
\usepackage{amsmath, amsthm, amssymb, amsfonts, graphicx,tabularx,rotate}
\usepackage{longtable, lscape, adjustbox,empheq}
\usepackage[figuresright]{rotating}
\usepackage{multicol}
\usepackage{rotfloat,subcaption}
\usepackage[justification=justified, singlelinecheck=false]{caption}

\usepackage{afterpage,url}
\usepackage[tmargin=2.5cm,bmargin=2.25cm,lmargin=2.25cm,rmargin=2.25cm]{geometry}
\usepackage{setspace}
\usepackage{booktabs}

\usepackage[hidelinks]{hyperref}

\usepackage{placeins}

\makeatletter
\AtBeginDocument{%
  \expandafter\renewcommand\expandafter\subsection\expandafter
    {\expandafter\@fb@secFB\subsection}%
  \newcommand\@fb@secFB{\FloatBarrier
    \gdef\@fb@afterHHook{\@fb@topbarrier \gdef\@fb@afterHHook{}}}%
  \g@addto@macro\@afterheading{\@fb@afterHHook}%
  \gdef\@fb@afterHHook{}%
}
\makeatother
\title{\vspace{-60pt} \textbf{Forecasting the Price of Carbon with Macroeconomic and Financial variables}\thanks{\footnotesize
\noindent The authors thank Christiane Baumeister, Hilde Bjornland, Chiara Casoli, Jennifer Castle, Mattia Guerini, Diego Kanzig, Massimiliano Marcellino, Luca Pedini, Ivan Petrella, Francesco Ravazzolo, Francesco Vona, Sylvia Kaufmann and the participants at the 4th IWEEE in Bolzano, the Energy Finance 9 in Bari, the 12th IAERE in Pescara, the 4th DMM in Castelrotto, the ISF 2024 in Dijon, the ``Econometrics of Energy Transition'' workshop in FEEM, the 11th ICEEE in Palermo, the 2026 RCEA in Madrid and “Transition in a Changing World” workshop at BI Oslo for their suggestions. Luca Rossini acknowledges financial support from MUR under the PRIN-PNRR project Mapping and Pricing of Methane Emissions from the European Electricity Sector (grant P2022H483A). This research used the Computational resources provided by INDACO at the University of Milan. The paper previously circulated under the title ``\textit{What drives the European carbon market? Macroeconomic factors and forecasts}.''
}
}

\author{
Andrea Bastianin\thanks{University of Milan, Italy and Fondazione Eni Enrico Mattei (FEEM). \color{blue}\texttt{andrea.bastianin@unimi.it}}
\and
Elisabetta Mirto\thanks{Study Center Gerzensee, Foundation of the Swiss National Bank. Switzerland. {\color{blue}\texttt{elisabetta.mirto@szgerzensee.ch}} } 
\and 
Yan Qin\thanks{ClearBlue Markets, Norway. \color{blue}\texttt{qinmaria@gmail.com}}
\and 
Luca Rossini\thanks{University of Milan, Italy and Fondazione Eni Enrico Mattei (FEEM). \color{blue}\texttt{luca.rossini@unimi.it}}
}

\date{\vspace{-0.2cm} September 2026}

\makeatletter
\let\arxivSavedMakeTitle\maketitle
\let\arxivSavedInternalMakeTitle\@maketitle
\let\arxivSavedThanks\thanks
\let\arxivSavedTitleCmd\title
\let\arxivSavedAuthorCmd\author
\let\arxivSavedDateCmd\date
\let\arxivSavedAndCmd\and
\newcommand{\arxivRestoreMakeTitle}{%
  \let\maketitle\arxivSavedMakeTitle
  \let\@maketitle\arxivSavedInternalMakeTitle
  \let\thanks\arxivSavedThanks
  \let\title\arxivSavedTitleCmd
  \let\author\arxivSavedAuthorCmd
  \let\date\arxivSavedDateCmd
  \let\and\arxivSavedAndCmd
}
\makeatother
\begin{document}

\maketitle

\doublespacing
\vspace{-1.1cm}

\noindent\textbf{Abstract:} 
We tackle the issue of producing point, sign, and density forecasts for the monthly real price of carbon in the European Union Emissions Trading System (EU ETS). We show that Bayesian Vector Autoregressive (BVAR) models augmented with factors based on macroeconomic and financial variables yield improvements in point and density forecasting performance, with the largest point-forecast gains emerging at intermediate and longer horizons. 
In particular, the one-factor BVAR model provides lowest values at six and nine months ahead, while the baseline BVAR performs best at the one-year horizon.
By contrast, a large BVAR including all predictors individually does not improve point forecast accuracy, supporting the use of parsimonious model specifications. Simple forecast pooling delivers modest gains from medium horizons onward, but does not mitigate the time variation in relative forecast performance. We also provide a qualitative comparison of model-based forecasts with survey expectations and forecasts released by data providers, which further highlights that relative forecast performance varies over time. Moreover, we consider verified emissions and show that stochastic volatility improves point forecasts, with reductions of about 6.5--7.1\% at the one-year horizon. Lastly, we use model-based forecasts to construct market monitoring tools that track demand and price pressure in the EU ETS.



\pagenumbering{arabic}
\doublespacing
\normalsize
\clearpage

\clearpage

\defcitealias{ECB2021}{ECB, 2021}
\section{Introduction}\label{sec:introduction}
\noindent Climate change is one of the greatest challenges currently addressed by governments, central banks, and other national and supranational regulators. The size and complexity of the call to fight climate change are evident, even when -- setting aside its environmental consequences -- we focus solely on its direct socio-economic impacts \citep{carleton2016social}.
The literature has highlighted a multitude of effects and transmission channels that connect the environmental influences of climate change to various economic aggregates \citep{dell2014we, ciccarelli2021demand, liu2025same}.
Moreover, when designing mitigation and adaptation policies, a trade-off emerges between the stringency of measures and the undesired outcomes that these might induce \citep{fullerton2019bears, Hsiang_et_al2019}.

Among the policy measures implemented to achieve the target of net-zero emissions by 2050 -- as laid down in the 2015 Paris Agreement -- carbon taxes and carbon pricing have received a lot of attention from academics and policymakers interested in quantifying the desired and unintended macroeconomic effects of putting a price on carbon \citep[see e.g.][]{ECB2021, IMF2021, CrossReview}. Given this literature and the need to incorporate climate and carbon market modules into macroeconomic models used by central banks to produce scenarios, having access to reliable short- and medium-term forecasts of the real price of carbon is becoming increasingly important \citep{ECB2021press, NGFS}.

This paper produces point, directional, and density forecasts of the real price of carbon at monthly frequency in the world's largest carbon market, the EU Emission Trading Scheme (EU ETS), over horizons from one to twelve months ahead. Extending the horizon to one year aligns with the annual compliance cycle of the EU ETS, under which regulated entities must report and certify their greenhouse gas emissions each year. We focus on the \textit{real} price of carbon -- a natural choice in empirical and theoretical macroeconomic models \citep[e.g.][]{baumeister2022energy, baumeister2014central, bjornland2023, issler2014using, kilian2011responses, van2021optimal, wang2023commodity} -- which allows us to incorporate real monthly macroeconomic aggregates as predictors.
Since practitioners often refer to the nominal price \citep{icap, stateofeuets2023}, we also forecast the nominal price and compare our forecasts with expert and survey expectations. We further extend the forecasting exercise to verified emissions -- a topic that has received essentially no attention in the academic literature\footnote{A few papers address forecasting of total $CO_2$ emissions. \citet{auffhammer2012} and \citet{bennedsen2021modeling} analyse US $CO_2$ emissions; \citet{Ruiz} focuses on Spain.} -- and use model-based predictive distributions to construct real-time indices of demand pressure and price pressure for the EU ETS market.

The paper makes three contributions to the literature on carbon price forecasting
\citep[see][for a survey]{Chevallier2012, CrossReview}. First, we build a Bayesian forecasting framework for the \textit{real} monthly price
of carbon that incorporates macroeconomic and financial predictors via factor
augmentation and models time-varying volatility through Stochastic Volatility (SV)
dynamics. The existing forecasting literature relies predominantly on frequentist
methods applied to daily or weekly nominal data \citep[see
e.g.][]{chevallier2011favar, lei2022probability, tan2022IJF}: the only paper to use
Bayesian methods is \citet{koop2013JRSS}, which works at daily frequency. Working at
monthly frequency with real prices allows us to include real macroeconomic
aggregates as predictors. The most closely
related paper at monthly frequency is \citet{bjornland2023}, who estimate a Bayesian
structural VAR to identify the demand and supply shocks driving carbon prices and
emissions. Their goal is structural identification, not forecasting. We take their
baseline VAR as a starting point and show that augmenting it with factors extracted
from a broad panel of 20 predictors delivers out-of-sample accuracy gains, especially at intermediate horizons -- a question that their analysis does not address.

Second, we provide a broad and externally relevant evaluation of predictive accuracy.
We assess point forecasts (RMSFE), directional accuracy (Success Ratio), and density
forecasts (quantile-weighted CRPS, \citealt{gneiting2011comparing}), separately for the
center and both tails of the predictive distribution, over horizons from one month to
one year. We forecast both the real price -- the natural object of the macroeconomic
analysis -- and the nominal price, on which market participants and data providers
report. This dual focus lets us benchmark our forecasts not only against standard
time-series competitors, but also against the forecasts practitioners actually use:
the expert point forecasts and the market surveys produced by the Carbon Team at the
London Stock Exchange Group (LSEG). To our knowledge, such a comparison with expert and
survey expectations has not previously been undertaken in the carbon price forecasting
literature, and it is what makes the evaluation relevant beyond an internal model
comparison. This breadth is absent from \citet{koop2013JRSS}, which considers neither
macroeconomic predictors, directional accuracy, nor tail forecasting. Density forecasts
are directly relevant for policy institutions that embed carbon price trajectories into
macroeconomic scenarios \citep{ECB2021press, NGFS}.

Third, we extend the forecasting exercise beyond prices. We produce, to our
knowledge for the first time, out-of-sample forecasts of verified emissions at
monthly frequency -- directly relevant for compliance planning, since entities'
verified emissions net of free allowances represent their net demand for permits and
hence a key driver of prices. We further use model-based predictive distributions to
construct real-time indices of demand pressure and price pressure for the EU ETS,
providing a market monitoring tool with no precedent in this literature.

Our main findings are as follows. A BVAR model augmented with one factor -- capturing a combination of macroeconomic demand conditions and energy market pressures -- delivers accuracy gains over a broad set of benchmarks, including the random walk, at medium and long horizons, from six months ahead forecasts. At six and nine months ahead, the one-factor BFAVAR(1) delivers the lowest RMSFE, reducing forecast errors relative to the random walk by up to 3.9\%.
By contrast, a large BVAR including all predictors individually fails to improve upon the random walk at any horizon, supporting the benefits of parsimonious model specifications. Diebold-Mariano tests do not reject equal predictive accuracy, but fluctuations tests reveal substantial time-variation in relative forecast performance, with factor models performing particularly well during the COVID-19 recession and the 2021--2022 energy crisis.
Pooling parsimonious models yields only marginal gains: a simple equal-weight average beats the random walk from the six-month horizon onward, but does not improve on the best individual specifications nor offset the documented instability. The comparison with LSEG expert and survey forecasts confirms this time variation: during the 2021--2022 run-up, the factor model tracks the sharp increase in nominal prices more closely than survey and expert forecasts, and realized prices fall in the right tail of the model's predictive density. Incorporating SV does not improve forecasts of the real price of carbon, but it improves both point and density forecasts of verified emissions, reducing their RMSFE at the one-year horizon by 6.5--7.1\%. 
These findings are robust to alternative lag specifications, recursive factor selection, the use of an alternative shrinkage prior, and the use of nominal carbon prices.
Finally, demand and price pressure indices constructed from model forecasts track the major turning points in the EU ETS over the evaluation sample.

The rest of the paper is organized as follows. Section~\ref{sec:euets} offers an overview of the EU ETS; data and details of the forecasting exercise are illustrated in Section~\ref{sec:datamethods}. Main results and sensitivity checks are presented in Section~\ref{sec:results}, while extensions and further results are reported in Section~\ref{sec:robustness}. Section~\ref{sec:concl} concludes. A Supplement completes the paper.

\section{The EU Emissions Trading System}\label{sec:euets}
The EU ETS is a cap and trade system that started in 2005 intending to reduce carbon emissions. 
In this system, the maximum quantity of emissions (the cap) is set through unit permits (European Unit Allowances, EUA), which allow the owner of the permit to produce 1 ton of CO$_{2}$ or an equivalent quantity of other greenhouse gases. 
The European Commission sets a yearly cap on the total greenhouse gas emissions that can be produced by actors participating in the system. 
Since the aim is to decrease emissions over time, every year the cap is lower than the year before, and consequently, the maximum allocation of EUA is reduced. 

The EU ETS system is a financial market where actors can acquire EUA on the primary market through an auctioning system, and trade derivatives on the secondary market. 
A certain amount of permits is originally granted for free each year according to the needs of specific sector emissions, although the remaining amount of available allowances is allocated on the primary market through uniform price auctions with single rounds and sealed bids, conducted daily by the European Energy Exchange (EEX). 
Since EUA have been classified as financial instruments, %
the associated derivatives - such as spot, futures, options, and forward contracts - can be traded on secondary markets, both on exchange and over the counter. 
While auctions take place on the EEX, trading takes place also on the Intercontinental Exchange (ICE).

\begin{figure}[ht]
    \centering
    \caption{Allocated and verified emissions for all stationary installations in EU-27 countries (excluding the aviation sector).}
    \includegraphics[width=1\linewidth]{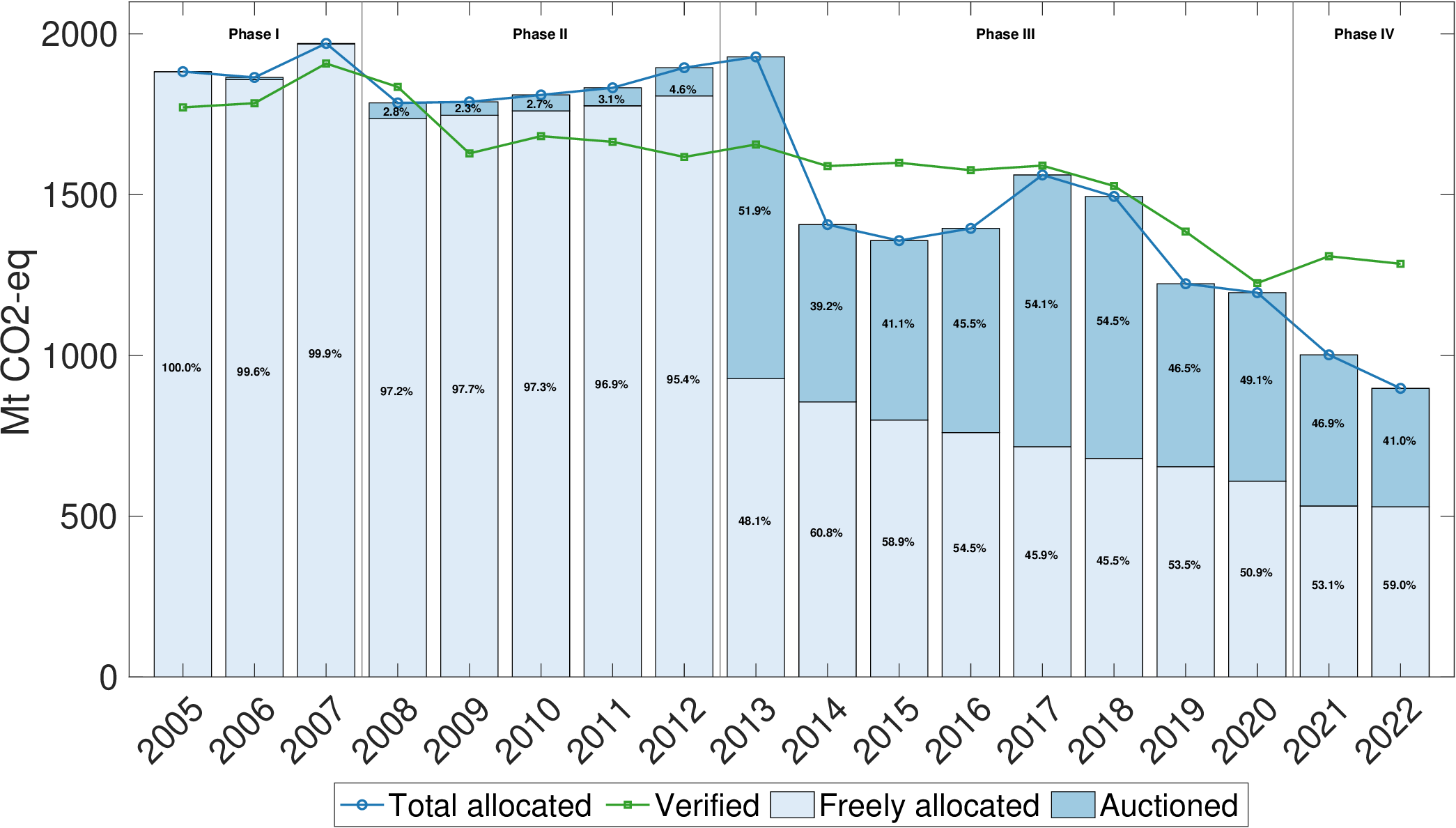}
    \caption*{\scriptsize{\textit{Notes:} authors' elaboration of data from \href{https://www.eea.europa.eu/data-and-maps/dashboards/emissions-trading-viewer-1}{https://www.eea.europa.eu}.}}
\label{fig:Fig02Allocated}
\end{figure}

EUA are distributed to the market through a system of benchmark-based allocation or auctions. 
If emissions at the end of the year prove to be lower than the cap set for each installation participating in the market, permits can be traded among actors for an economic value to be determined on the secondary market. 
In case emissions exceed the threshold, sanctions are applied to economic agents participating in the market. 

Actors participating in the EU ETS market entail industries belonging to high emissions sectors: electricity and heat generation, energy-intensive industry sectors (including oil refineries, steel works, and production of iron, aluminum, metals, paper, etc.), aviation within the European Economic Area (EEA) and, starting from 2024, maritime transport. 
Participation in the EU ETS is mandatory for companies in the covered sectors; however, for some sectors only production plants above a minimum size threshold are included.

Historically, the EU ETS evolution went through four phases. 
To meet the objectives set by the European Commission in terms of emissions reduction, each phase aims to reduce the number of EUA granted to each participating sector. 
The cap can be lowered by setting a decreasing number of allowances to be allocated each year or by establishing a yearly linear reduction factor, e.g. a linear reduction of 1.74\% and 2.2\% of the baseline 2008-2012 emissions have been set respectively from 2013 and 2021 onwards, with no end date, resulting in a year-on-year reduction of up to 43 million allowances \citep{icap}.

The pilot phase (2005-2007) aimed to verify rules, regulations, emission detection systems, as well as the regulatory framework. 
In this phase, the allocation system was \textit{grandfathering}: all EUA were allocated freely to industries, up to the cap set for each regulated sector. 
The second phase, which lasted from 2008 to 2012, was characterized by the introduction of the allocation of permits employing an auctioning system. 
In this phase, roughly 2 to 5\% of the total permits were allocated through auctions. 
This share increased to reach 54$\%$ in the third phase, which lasted from 2013 to 2020, and it includes more sectors and gases. 
Lastly, the fourth phase (2021-2030) has the aim of reducing net emissions by at least 55\% by 2030 compared to 1990, as set in the European Climate Law, by further lowering the cap and targeting the \textit{carbon leakage} phenomenon. 

In July 2021, the European Commission adopted a series of legislative proposals regarding EU ETS aimed at increasing the pace of emissions cuts. 
These include, among others, covering more sectors and gases, gradually lowering the number of emission allowances each year, and reinforcing the Market Stability Reserve (MSR), which aims at reducing the surplus of allowances in the market. 
An excessive allowances surplus would lead to lower carbon prices, thereby weakening 
the incentives of regulated entities to reduce their emissions. The MSR is automatically applied when the total number of allowances on the market exceeds a certain threshold.  
In Phase IV, the free EUA allocation system was granted a ten-year extension, and specific measures have been taken for sectors exposed to a higher risk of carbon leaking.%

The way allowance allocation has evolved throughout the years is described in Figure \ref{fig:Fig02Allocated}. 
In Phase II, auctions were introduced, but between 95\% and 97\% of allowances were still distributed for free. 
This changed completely during Phase III, where auctions became the predominant allocation method for most sectors, covering between 40\% and 55\% of the total allocated permits. 
Phase III also needed to deal with the aftermath of an excessive surplus of unused EUA in Phase II. 
The number of allocated EUA decreased in 2014, leading to an increase in the price of allowances and a reduction in the number of verified emissions and unused EUA \citep{bjornland2023}. 
In Phase IV, auctions remain the main allocation method on the primary market. 
Figure \ref{fig:Fig02Allocated} also shows verified emissions. 
These data, available on an annual basis, refer to the actual amount of greenhouse gas emissions produced by a company or entity, as reported in its emissions report and verified by an accredited verifier by 31 March of the following year. 
Based on verified emissions, companies must surrender a corresponding number of emission allowances by the end of April of that year. 
If a company's verified emissions exceed the number of allowances it holds, it may need to purchase additional allowances from the market to cover the excess emissions. 
Figure \ref{fig:Fig02Allocated} highlights that Phases I and II are characterized by an over-allocation of permits, with verified emissions being systematically lower than allocated emissions. 
The cap on emissions has become tighter during Phases III and IV when verified emissions exceed allocated emissions for most of the sample period.

\section{Data and methods}\label{sec:datamethods}
\subsection{Data}\label{sec:data}

\begin{figure}[!ht]
    \centering
        \caption{Real EU ETS price (left), actual and interpolated verified emissions for EU-19 countries (right) from March 2005 to September 2023.}    
  \includegraphics[width = .495\textwidth]{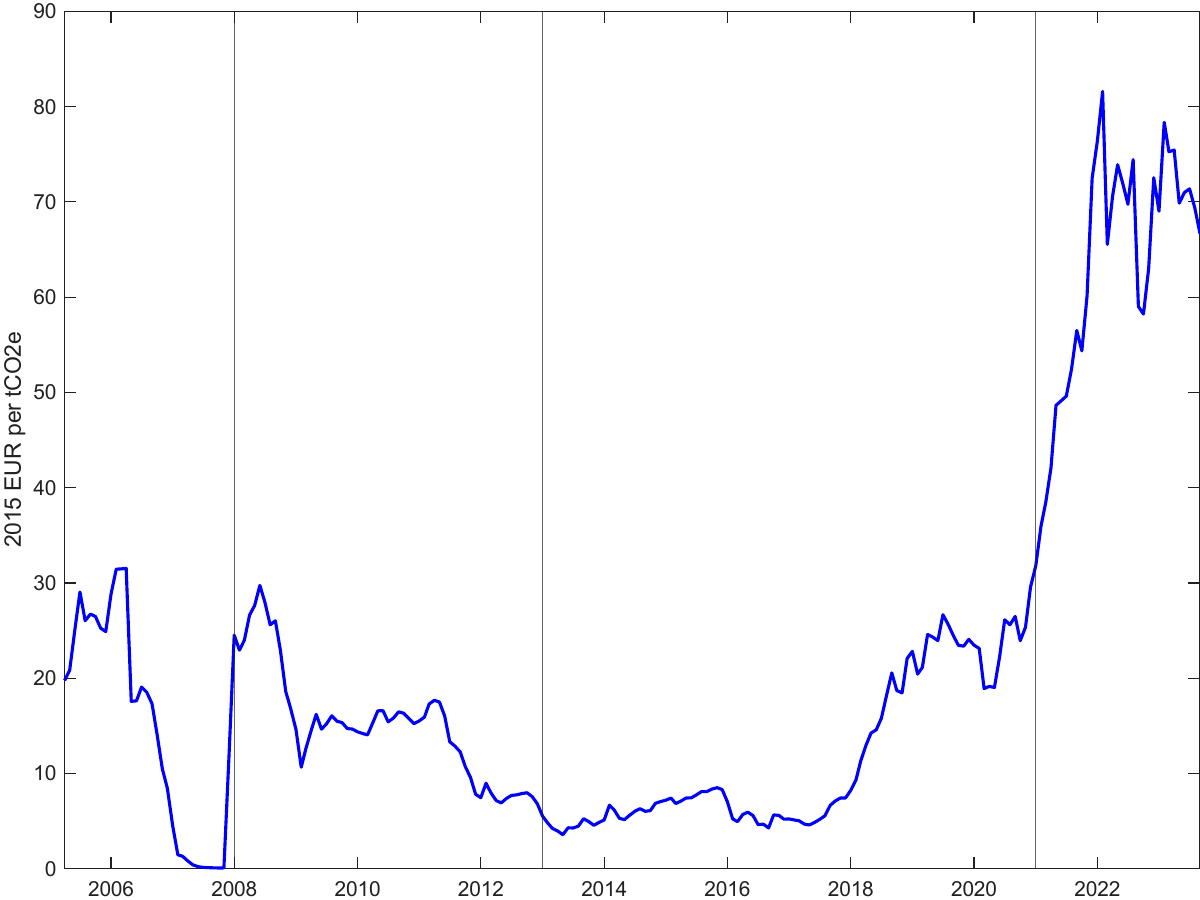}%
  \includegraphics[width = .495\textwidth]{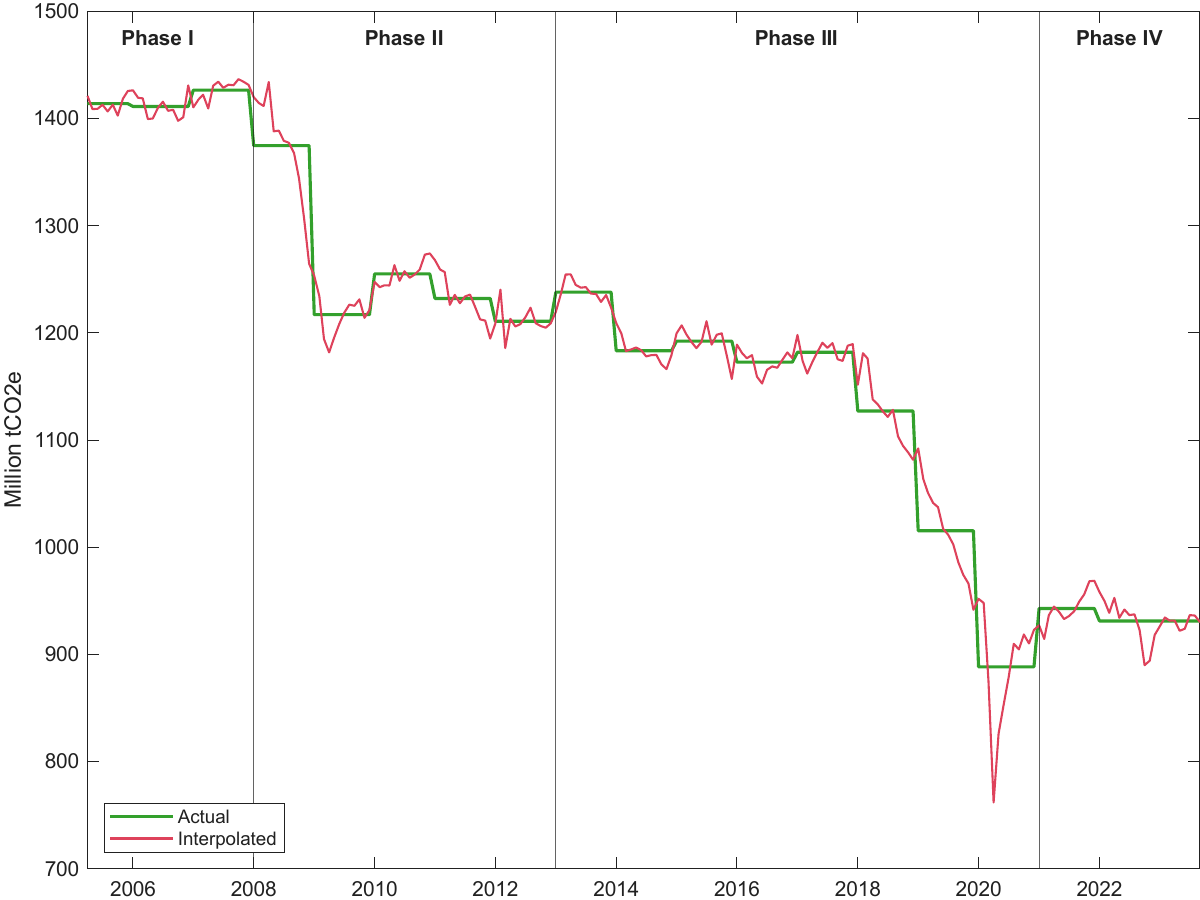}
    \label{fig:Fig02PandEmiss}
        \caption*{\scriptsize{\textit{Notes:} Interpolated emissions are obtained using IP indices of six sectors covered by the EU ETS.}}
\end{figure}

Following \citet{bjornland2023}, we focus on forecasting the one-month ahead futures contract traded on ICE that represents the most closely watched series by practitioners. 
We deflate the nominal futures price -- computed as the monthly average of daily prices -- by using the Euro area Harmonized Index of Consumer Prices (HICP) (see the left panel of Figure \ref{fig:Fig02PandEmiss}).

As for the predictors, we follow the approach of \citet{boivin2006more} and \citet{baumeister2022energy}. 
Therefore, instead of collecting a large number of series, we carefully select 20 predictors that capture demand and supply-side forces driving the price of carbon. 
More precisely, we concentrate on macroeconomic, energy market, and financial variables within the following categories:
\begin{itemize}
    \item \textit{Economic activity} (7 series): we collect data on aggregate industrial production (IP) for the 19 Eurozone countries (EU-19 area), as well as indices for sectors covered by the EU ETS (i.e. electricity, gas, steam, and air conditioning supply, basic metals, manufacture of paper and its products, coke and refined petroleum products, chemical products, non-metallic mineral products) from Eurostat.

    \item \textit{Real energy prices} (7 series): we consider the prices of Brent crude oil, TTF natural gas (front-month and front-year), ARA API-2 coal (front-year), German power price (front-year), clean dark, and clean spark spreads (front month). These variables are sourced from LSEG and deflated using the Euro Area HICP.
    
    \item \textit{Technical and financial indicators} (4 series): we select some of the variables that practitioners use to track the functioning of the EU ETS market \citep{stateofeuets2023}. The auction coverage ratio, defined as the total number of bids in an auction divided by the number of available EUA, is a proxy for the actual auction demand relative to supply on the primary market. As a rule of thumb, a value greater (lower) than two indicates a high (low) auction demand relative to supply. The auction clearing price (deflated using the Euro Area HICP), and a volatility proxy based on the monthly auction price range. These variables are sourced from the EUA Primary Market Auction Report maintained by EEX. Moreover, we consider the Euro Stoxx 50 stock price index, sourced from LSEG and deflated using the Euro Area HICP.
    
    \item \textit{Weather conditions} (2 series): temperature and precipitation anomalies for EU-19 countries are constructed as differences from long-term moving averages using data sourced from the Weather for Energy Tracker maintained by the International Energy Agency (IEA).
  
\end{itemize}

\noindent Moreover, we also collect information on annual verified emissions for all stationary installations in six sectors covered by the EU ETS in the EU-19 countries. 
Following \citet{bjornland2023} and \citet{Kanzig2023}, we temporally disaggregate annual data with the \citet{chowlin} method.\footnote{We consider IP indices and emissions for all EU ETS sectors (except aviation) and construct an emission-weighted IP index to interpolate annual verified emissions via the \citet{chowlin} method, implemented using the MATLAB library of \cite{quilis}. Aviation is excluded because it joined the EU ETS only in 2012, whereas our interpolation starts in 2006; the 2022 value is used for 2023, as the latest verified emissions were unavailable at the time of writing. Results excluding 2023 are qualitatively unchanged and are available from the authors upon request. An alternative proxy for demand-side pressure would be Eurostat's Quarterly Greenhouse Gas Emissions Accounts, but this presents several drawbacks: they rely on temporal disaggregation of annual Air Emissions Accounts, cover total national emissions rather than ETS-verified ones, and are quarterly rather than monthly, so the mixed-frequency and look-ahead problems would remain. That said, as shown in Supplement S.4.2, the two emission measures are highly correlated ($\approx 0.95$) and react similarly to key events -- such as COVID-19 -- confirming that our results are not sensitive to this choice. Higher-frequency alternatives exist, such as daily EU CO2 emission tracker maintained by the Center for Research on Energy and Clean Air (CREA) estimates, which are available with a one-week delay but these also rely on indicator-based extrapolation rather than verified data.}  
Annual and interpolated monthly verified emissions are shown in the right panel of Figure \ref{fig:Fig02PandEmiss}.

\subsection{The forecasting framework}
The forecasting exercise is based on data spanning from June 2012 to September 2023, comprising a total of 136 monthly observations. 
The start date of the sample is dictated by the availability of data on auctions, which are particularly relevant for explaining the allocation mechanism of emission allowances during Phases III and IV of the EU ETS (see Figure \ref{fig:Fig02Allocated}). 
Forecasts are generated using an expanding window approach: each time a new forecast is produced, the estimation sample is updated by adding a new observation. 
The first estimation sample ends in December 2017, and the last forecast is issued in September 2022. We analyze forecast horizons ranging from one month to one year ahead. The evaluation sample is consistent across all horizons, covering December 2018 to September 2023, which includes the late part of Phase III and the beginning of Phase IV of the EU ETS. %

We denote the level of the real price of carbon in month $t$ as $R_t$ and the log price as $r_t = \log R_t$. 
Models are estimated using the first difference of the logarithm of the real price of carbon, $\Delta r_t$, and forecasts are constructed iteratively from the estimated models and converted into levels as follows:
\begin{align*}
\hat{R}_{t+h|t}  & = \exp\left(r_{t} + \sum_{\ell=1}^h \Delta \hat{r}_{t+\ell|t}\right), %
\end{align*}
where $\Delta \hat{r}_{t+\ell|t}$ is the $\ell$-step-ahead forecasted value (for ease of notation, we omit the superscript $s$ referring to the draw). Point forecasts are obtained as the posterior mean of $\hat{R}_{T+h|T}$ across draws. The evaluation of point forecasts relies on the relative Root Mean Squared Forecast Error (RMSFE) that represents the ratio of the RMSFE of a model to the RMSFE of the benchmark, such as the Random Walk (RW). 
Therefore, a relative RMSFE lower than unity is taken as evidence that a certain model is more accurate than the benchmark.

Sign, or direction-of-change forecasts, are defined as: $\text{sign}\left(\hat{R}_{t+h|t} - R_t \right)$, where $\mathrm{sign}(x)$ equals -1 if $x < 0$, 0 if $x = 0$, and 1 if $x > 0$. 
We use the Success Ratio (SR), defined as the proportion of correctly predicted signs, to gauge the directional accuracy. 
A SR greater than 0.5 indicates a gain in accuracy relative to the RW model, which implies a no-change forecast.

Density forecasts are obtained from the empirical distribution of the draws of $\hat{R}_{T+h|T}$, which integrates over parameter uncertainty and forecast innovations at all horizons $h = 1, \ldots, H$. As an additional measure of predictive performance, we use the quantile-based continuous ranked probability score (qCRPS) of \citet{gneiting2011comparing}, a proper scoring rule for evaluating density forecasts, denoted by
\begin{equation}
    \widehat{QS}_{t}=\frac{1}{J-1} \sum_{j=1}^{J-1} \widehat{QS}_{t}^{\alpha_j}= \frac{1}{J-1} \sum_{j=1}^{J-1} 2\left[ \mathbb{I}\left(R_{t+h}\leq \hat{q}^{\alpha_j}_{t+h|t}\right) - \alpha_j \right]\times\left(\hat{q}^{\alpha_j}_{t+h|t}-R_{t+h}\right),
    \label{eq:crps}
\end{equation}
where $\mathbb{I}(\cdot)$ denotes the indicator function and $\hat{q}^{\alpha_j}_{t+h|t}$ is the $h$-step ahead quantile forecast for $R_{t+h}$ at level $\alpha_j=j/J$ with $J=20$, which corresponds to $\alpha_j = 0.05, 0.10, \ldots, 0.95$. 
We construct weighted versions of the qCRPS, where the weights are selected to emphasize specific regions, such as the center or one of the tails of the distribution:
\begin{align}
    \widehat{wQS}_t & = \frac{1}{J-1} \sum_{j=1}^{J-1} \nu(\alpha_j) \widehat{QS}_{t}^{\alpha_j}, \quad \text{with }%
    \nu(\alpha_j)=\begin{cases}
    \alpha_j(1-\alpha_j) &  \text{(center)},\\
    \alpha_j^2  & \text{(right tail)},\\
    (1-\alpha_j)^2 &  \text{(left tail)}.
    \end{cases}
\end{align}
In a pairwise comparison, the model with the lowest score is ranked as the most accurate. 

\subsection{VAR models of the EU ETS carbon market}\label{sec:varmodel}
To link the real price of carbon to its determinants, a natural starting point is a small-scale VAR(p) model of the EU ETS market:
\begin{equation}
\mathbf{y}_t = \mathbf{a} + \sum_{j=1}^{p} \mathbf{A}_j \mathbf{y}_{t-j} + \mathbf{u}_t,
    \label{eq:varm1}
\end{equation}
where $\mathbf{a}$ is a $n\times1$ vector of intercepts, $\mathbf{A}_j$ are $n\times n$ matrices of coefficients for $j=1,\ldots,p$, and $\mathbf{u}_t$ is an $n\times1$ vector of zero-mean innovations Normally distributed with covariance matrix $\boldsymbol{\Sigma}$. 
Following \citet{bjornland2023}, we consider a baseline VAR specification in which $\mathbf{y}_t$ is a $3\times1$ vector including $\Delta r_t$ (or alternatively, $r_t$), the first difference of the logarithm of interpolated verified emissions, $emis_t$, and first difference of the logarithm of aggregate industrial production for EU-19 countries, $\Delta ip_t$.

The baseline model excludes several predictors that are closely monitored by practitioners, including energy prices, technical and financial indicators related to EU ETS allowance auctions, and weather anomalies that may affect electricity and gas demand. We also extend the information set relative to \citet{bjornland2023} in two directions. First, we include the Euro Stoxx stock price index as an additional financial variable. Second, we broaden the measure of real economic activity by adding IP indices for the main sectors covered by the EU ETS, with the aim of better capturing demand-side pressures on the real price of carbon.

We exploit this enlarged information set in two ways.\footnote{Before the analysis, all variables are transformed to induce stationarity. IP indices, the Euro Stoxx price index, real energy prices, and the real auction price are transformed into monthly growth rates, computed as first differences of logarithms. Clean dark and clean spark spreads are first-differenced, while the monthly real auction price range is log-transformed. No transformation is applied to the auction cover ratio, temperature anomalies, or precipitation anomalies. We also carry out a preliminary screening for outliers. Since only a few extreme observations, defined as observations exceeding 20 times the interquartile range from the median, are detected during the COVID-19 pandemic, we retain them in the sample.} First, we estimate factor-augmented VAR models, replacing the aggregate EU-19 IP index in the baseline VAR with up to three factors extracted from the 20 predictors. The factors are obtained by principal component analysis and are re-estimated recursively at each forecast origin using only the available estimation sample. Second, we estimate a large BVAR that includes the 20 predictors individually, after standardizing them to have zero mean and unit variance at each forecast origin.

All Bayesian models are estimated under a Normal-inverse-Wishart prior with a Minnesota-style specification. The prior shrinks coefficients toward a parsimonious autoregressive benchmark while allowing relatively diffuse beliefs on the intercepts. The posterior is simulated using 5,200 draws, with the first 200 discarded as burn-in. Full details on the prior specification, posterior distribution, factor extraction, data transformations, and estimation algorithm are provided in Supplement S.1.1.

\begin{figure}[!ht]
    \centering
    \caption{Contribution of predictors to Factor 1}
    \includegraphics[width=0.8\linewidth]{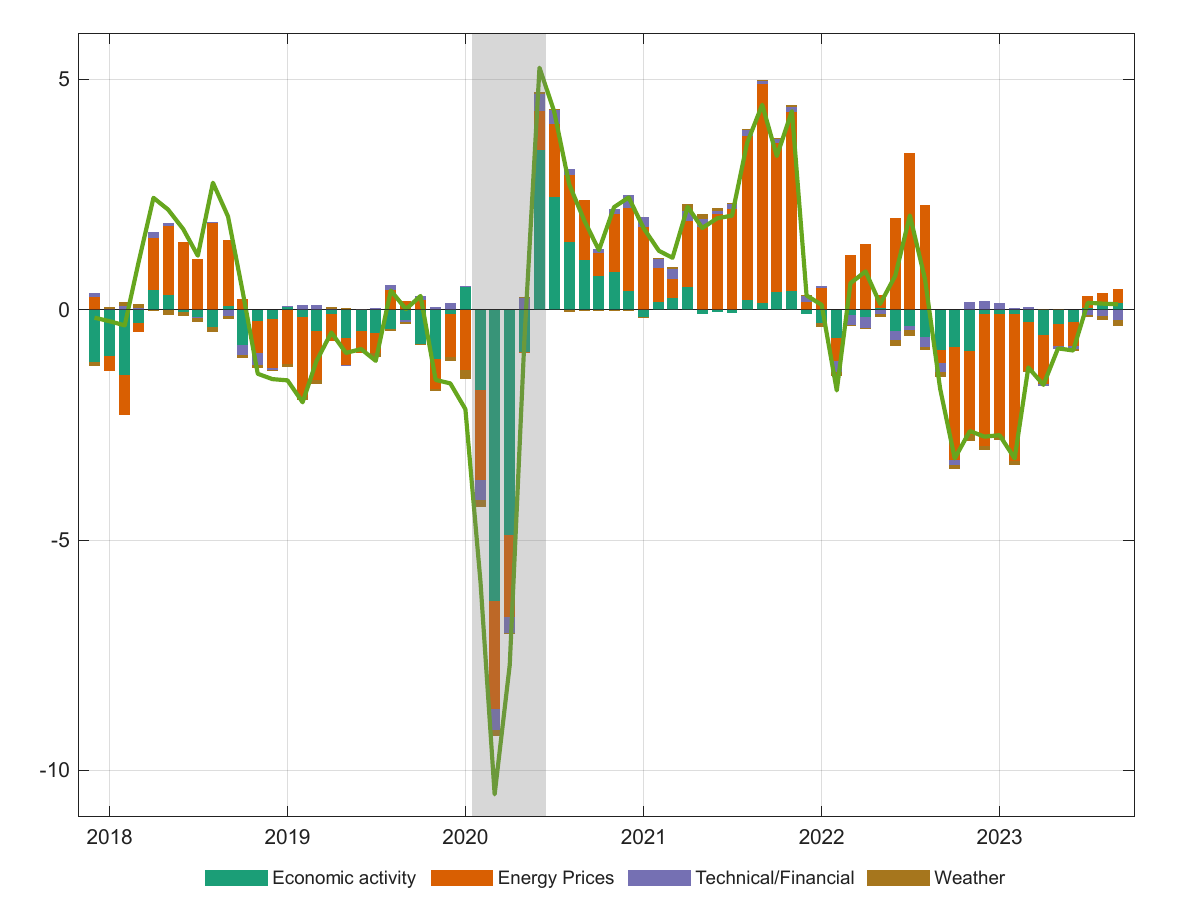}
    \label{fig:Factor1LoadsFeb2025}
    \caption*{\scriptsize{\textit{Notes}: we depict the centered 3-month moving average of both Factor 1 and the predictors' contributions to it. Shaded areas represent recessions in the Euro Area, as determined by the CEPR-EABCN Euro Area Business Cycle Dating Committee (\url{https://eabcn.org}).}}
\end{figure}

\paragraph{Economic interpretation of factors.}
The first three factors explain 22.0\%, 17.8\%, and 8.6\% of the total variance of the 20 predictors, respectively, for a cumulative share of about 48\%. Although this share is moderate, the forecasting results below show that a BVAR(1) augmented with a single factor performs well in predicting both the real price of carbon and verified emissions, and generally outperforms less parsimonious alternatives.\footnote{See Supplement S.3 for more detailed results.}

Our framework does not give the factors a structural interpretation. Nevertheless, it is informative to characterize the variables underlying their dynamics by examining their correlations with observable predictors, as is commonly done in the literature \citep[see, e.g.,][]{baumeister2022energy,bjornland2018shale}. Figure \ref{fig:Factor1LoadsFeb2025} shows the evolution of the first factor over time, together with bars representing the contribution of different predictors grouped by class: economic activity, energy prices, technical and financial indicators, and weather anomalies.

The variables underlying movements in the first factor change over time, but its dynamics are consistently dominated by proxies for economic activity and energy prices. By contrast, technical and financial indicators and weather anomalies play a more limited role. Measures of real economic activity account for most of the downward movement of the factor in 2020, when European economies contracted sharply during the COVID-19 lockdowns, and for the subsequent recovery. The upward movement of the factor in 2022, followed by its decline in 2023, instead reflects pressures related to the war in Ukraine and the subsequent easing of conditions in European energy markets.

Consistent with this interpretation, the first factor is strongly correlated with IP growth, with a correlation of 0.77, and with TTF natural gas price growth, with a correlation of 0.74. It therefore tracks both the European business cycle and energy-market conditions closely. This interpretation is in line with the structural VAR literature on carbon markets: \citet{bjornland2023} identify macroeconomic demand shocks as the primary driver of carbon-price dynamics, and our first factor captures this channel in reduced form. More broadly, this result connects to the literature on global commodity-price co-movement \citep{alquist2020commodity,delle2022common}, which shows that demand-driven fluctuations in economic activity account for a large share of co-movement across commodity prices. It is also consistent with \citet{baumeister2022energy}, who show that indicators of global economic activity contain substantial information about energy demand and prices.

\section{Forecasting the real price of carbon}\label{sec:results}

\subsection{Univariate time series models}\label{sec:unviariate}
Before turning to the multivariate forecasting results, we discuss the choice of the univariate benchmark used in the forecast evaluation. We compare the RW with several standard univariate alternatives: the RW with drift, an Unobserved Component model, and three Autoregressive Integrated Moving Average (ARIMA) specifications, namely an ARIMA(1,1,1), an ARIMA(0,1,1), and an ARIMA(1,0,1). The first two ARIMA models impose a unit root, while the third is estimated on the log real carbon price, $r_t$. Moreover, the RW is also compared to Bayesian AR (BAR) models with lag order $p \in \{1, 3, 12\}$ and AIC-based selection, estimated on both 
$\Delta r_t$ and $r_t$. The comparison, reported in Section S.4.1 of the Supplement, shows that no univariate specification systematically improves upon the RW across forecast horizons. We therefore retain the RW as the main benchmark. This choice is also in line with the commodity-price forecasting literature -- and with crude-oil forecasting in particular -- where the RW is commonly used as the main benchmark.

\subsection{Can VAR models forecast the real price of carbon?}\label{sec:varresults}
We then evaluate Bayesian multivariate forecasting models for the real carbon price. We consider the baseline Bayesian VAR (BVAR) specification of \citet{bjornland2023}, factor-augmented VAR models, and a large BVAR including all 20 predictors used to extract the factors. All specifications reported in Table \ref{tab:Tab01PaperBVARS} are estimated using the first difference of log real carbon prices, $\Delta r_t$.\footnote{In this and in the remaining sections of the paper, we focus on selected forecast horizons, $h=1,2,3,6,9,12$. Supplement S.4.2 reports the full set of results for all horizons from one to twelve months ahead.}

Table \ref{tab:Tab01PaperBVARS} reports the forecasting performance of selected specifications. The first two model columns report the baseline BVARs with one and three lags; the next six columns report factor-augmented BVARs with one, two, and three factors, estimated with either one or three lags; and the last two columns report large BVARs that include the full set of predictors individually. Panel~(a) shows that Bayesian multivariate models provide little improvement over the RW at very short horizons. At horizons of one to three months, relative RMSFEs are generally above one, indicating that the no-change benchmark remains hard to beat.

Forecast gains emerge more clearly at medium and long horizons. At six and nine months ahead, the one-factor BFAVAR with one lag delivers the lowest RMSFE, reducing forecast errors relative to the RW by about 0.8\% and 3.9\%, respectively. At the one-year horizon, however, the baseline BVAR(1) becomes the best-performing point-forecast model, with a relative RMSFE of 0.9372, corresponding to a reduction of about 6.3\% relative to the RW. The one-factor BFAVAR(1) remains very close, with a relative RMSFE of 0.9416. The results therefore suggest that factor augmentation can be useful, but mainly when the information set is summarized parsimoniously. Adding a second or third factor does not systematically improve performance and often deteriorates accuracy, especially in specifications with three lags. The large BVARs, reported in the last two columns, perform particularly poorly in terms of RMSFE: their relative RMSFEs are above one at all horizons, suggesting that including all predictors individually adds estimation noise that is not offset by additional predictive information.

Panel~(b) shows a more balanced picture for directional accuracy. Success ratios are above 0.5 for most models and increase with the forecast horizon. At medium and long horizons, BVAR(1) and the one-lag BFAVAR specifications often share the highest success ratio. At six, nine, and twelve months ahead, for example, BVAR(1) and the one-lag BFAVARs all attain the best success ratio in the table. Large BVARs also predict the direction of price changes better than chance at most horizons, but they rarely match the best parsimonious specifications and are clearly dominated in terms of point forecast accuracy.

Overall, the evidence favours parsimonious Bayesian specifications estimated on $\Delta r_t$. The choice between the baseline BVAR(1) and the one-factor BFAVAR(1) is horizon-dependent: the factor-augmented model performs best at medium horizons, while the baseline BVAR(1) is marginally more accurate at the one-year horizon. By contrast, increasing the number of factors or estimating a large BVAR with all predictors does not improve forecast accuracy.

All tables comparing point and density forecasts report Diebold--Mariano tests of the null of equal predictive accuracy relative to the random walk. The null is never rejected at conventional significance levels. This finding should be interpreted with caution, however, since the test compares average losses over the full evaluation sample and may have low power when relative forecast performance varies over time, as documented below using the fluctuation test of \citet{giacomini2010forecast}.

\paragraph{Sensitivity checks.}
We assess the sensitivity of these findings along several dimensions in Section S.4.2 of the Supplement. First, models estimated on the log level of the real carbon price, $r_t$, perform poorly relative to models based on $\Delta r_t$, supporting our use of differenced log prices in the main analysis. Second, VAR(12) specifications, commonly used with monthly data, do not improve forecasting performance. Third, specifications that recursively select the VAR lag order using the Akaike information criterion also fail to deliver systematic gains. These checks confirm that the main results are not driven by the specific lag choices reported in Table \ref{tab:Tab01PaperBVARS}, and reinforce the conclusion that parsimonious Bayesian models provide the most reliable forecasting performance.

We also allow the number of factors in the BFAVAR to vary over time, selecting it recursively using the \citet{bai2002determining} information criteria. The resulting Bai--Ng-based BFAVARs do not improve upon the fixed-factor specifications, suggesting that simple fixed-factor models capture the relevant forecasting gains more effectively.

As an additional sensitivity check, we re-estimate the Bayesian VAR models using the Horseshoe prior\footnote{Methodological details are provided in Section S.1.} The results closely confirm the baseline evidence. Forecast gains remain concentrated among parsimonious specifications estimated on $\Delta r_t$, while models with additional factors or all predictors included individually do not deliver systematic improvements. The one-factor BFAVAR with one lag delivers the lowest relative RMSFE at most horizons from two months ahead onward, and directional accuracy is also robust, with success ratios generally above 0.5 and increasing with the forecast horizon. Overall, the Horseshoe-prior results reinforce the conclusion that real carbon price predictability is best captured by compact Bayesian VAR specifications rather than large unrestricted predictor systems.

Finally, we assess whether the results depend on our focus on real rather than nominal carbon prices. We therefore replicate the previous analysis in using nominal carbon prices. The nominal-price results closely align with those based on real prices and, if anything, provide even stronger support for our main findings: parsimonious specifications continue to dominate, and the factor-augmented BVAR(1) remains the best model at intermediate forecast horizons.
\begin{table}[ht]
\centering
\caption{Relative RMSFE and Success Ratio of selected Bayesian forecasting models}
\newcolumntype{C}[1]{>{\scriptsize\hsize=#1\hsize\centering\arraybackslash}X}
\renewcommand{\arraystretch}{0.8}
\begin{tabularx}{\linewidth}{C{.5}|C{.9}C{.9}|C{.9}C{.9}|C{.9}C{.9}|C{.9}C{.9}|C{.9}C{.9}}
\hline
\multicolumn{11}{l}{\scriptsize (a) Relative RMSFE}\\\hline
 & \multicolumn{2}{c|}{\scriptsize BVAR} & \multicolumn{2}{c|}{\scriptsize 1 Factor} & \multicolumn{2}{c|}{\scriptsize 2 Factors} & \multicolumn{2}{c|}{\scriptsize 3 Factors} & \multicolumn{2}{c}{\scriptsize Large BVAR} \\\cline{2-11}
h & BVAR(1) & BVAR(3) & BVAR(1) & BVAR(3) & BVAR(1) & BVAR(3) & BVAR(1) & BVAR(3) & BVAR(1) & BVAR(3) \\\hline
1 & 1.0505 & 1.0752 & 1.0604 & 1.0739 & 1.0500 & 1.0950 & 1.0577 & 1.1239 & 1.2143 & 1.2711 \\
2 & 1.0337 & 1.0433 & 1.0407 & 1.0562 & 1.0647 & 1.1343 & 1.0816 & 1.1643 & 1.1360 & 1.2415 \\
3 & 1.0366 & 1.0313 & 1.0499 & 1.0605 & 1.0468 & 1.0997 & 1.0725 & 1.1151 & 1.3368 & 1.3904 \\
6 & \textbf{0.9997} & 1.0008 & \underline{\textbf{0.9921}} & 1.0106 & 1.0036 & 1.0364 & 1.0184 & 1.0435 & 1.2851 & 1.4535 \\
9 & \textbf{0.9667} & \textbf{0.9735} & \underline{\textbf{0.9610}} & \textbf{0.9904} & \textbf{0.9695} & 1.0417 & \textbf{0.9843} & 1.0428 & 1.2898 & 1.5440 \\
12 & \underline{\textbf{0.9372}} & \textbf{0.9554} & \textbf{0.9416} & \textbf{0.9808} & \textbf{0.9492} & 1.0406 & \textbf{0.9558} & 1.0463 & 1.2098 & 1.6891 \\\hline
\multicolumn{11}{l}{\scriptsize (b) Success Ratio}\\\hline
 & \multicolumn{2}{c|}{\scriptsize BVAR} & \multicolumn{2}{c|}{\scriptsize 1 Factor} & \multicolumn{2}{c|}{\scriptsize 2 Factors} & \multicolumn{2}{c|}{\scriptsize 3 Factors} & \multicolumn{2}{c}{\scriptsize Large BVAR} \\\cline{2-11}
h & BVAR(1) & BVAR(3) & BVAR(1) & BVAR(3) & BVAR(1) & BVAR(3) & BVAR(1) & BVAR(3) & BVAR(1) & BVAR(3) \\\hline
1 & \textbf{0.5517} & \textbf{0.5690} & \textbf{0.5517} & \textbf{0.5517} & \textbf{0.5517} & \underline{\textbf{0.6207}} & \textbf{0.5690} & \textbf{0.5690} & \textbf{0.5690} & \underline{\textbf{0.6207}}$^{\ast}$ \\
2 & \textbf{0.5517} & \textbf{0.5690} & \textbf{0.5517} & \textbf{0.5690} & \textbf{0.5517} & \textbf{0.5517} & \textbf{0.5517} & 0.5000 & \underline{\textbf{0.6207}} & \textbf{0.5690} \\
3 & \textbf{0.5690} & \textbf{0.5862} & \textbf{0.5517} & \textbf{0.5690} & \textbf{0.5690} & \underline{\textbf{0.6034}} & \textbf{0.5517} & \textbf{0.5690} & \textbf{0.5862} & \underline{\textbf{0.6034}} \\
6 & \underline{\textbf{0.6897}} & \textbf{0.6724} & \underline{\textbf{0.6897}} & \textbf{0.6552} & \underline{\textbf{0.6897}} & \textbf{0.6379} & \underline{\textbf{0.6897}} & \textbf{0.6552} & \textbf{0.6207} & \textbf{0.6034} \\
9 & \underline{\textbf{0.7931}} & \textbf{0.7759} & \underline{\textbf{0.7931}} & \textbf{0.7586} & \underline{\textbf{0.7931}} & \textbf{0.7586} & \underline{\textbf{0.7931}} & \textbf{0.7759} & \textbf{0.7586} & \textbf{0.7069} \\
12 & \underline{\textbf{0.7931}} & \textbf{0.7759} & \underline{\textbf{0.7931}} & \textbf{0.7414} & \underline{\textbf{0.7931}} & \textbf{0.7586} & \underline{\textbf{0.7931}} & \textbf{0.7586} & \textbf{0.7759} & \textbf{0.7241} \\\hline
\end{tabularx}
\label{tab:Tab01PaperBVARS}
\caption*{\scriptsize\textit{Notes}: Panel (a) shows relative RMSFE (ratio to RW). Values below one in bold; best model per row underlined. $^\ast$ ($^{\ast\ast}$) denotes rejection of the one-sided Diebold-Mariano test at 10\% (5\%). Panel (b) reports success ratios; values above 0.5 in bold; best model per row underlined. $^\ast$ indicates p-value of Pesaran-Timmermann (2009) test below 0.1. All models estimated in first differences. Lag orders in parentheses.}
\end{table}

\FloatBarrier

\subsection{Density forecasts and the role of stochastic volatility}\label{sec:densfore}
To quantify the uncertainty surrounding point forecasts, and in light of evidence that stochastic volatility (SV) improves density forecasts of macroeconomic aggregates \citep{clark2015macroeconomic, chan2023comparing}, we compare the performance of the baseline BVAR models with that of otherwise identical specifications augmented with SV. We evaluate both point and density forecasts for future values of the real price of carbon.
 
We augment the BVAR models with a Cholesky multivariate SV process \citep[][for details]{chan2023comparing}\footnote{The Cholesky-SV specification is not invariant to the ordering of the variables, since the Cholesky decomposition imposes a recursive structure on the reduced-form covariance matrix. This issue is well known in the literature: \citet{arias2023macroeconomic} show that variable ordering can affect density and interval forecasts in multivariate SV models, while \citet{chan2024large} discuss order-invariant alternatives. See also \citet{carriero2019large} and \citet{primiceri2005time}.}. The model is estimated in a Bayesian framework using Minnesota-style shrinkage priors for the VAR coefficients and hierarchical priors for the stochastic-volatility parameters. Full details of the prior specification are provided in the Supplement S.1.2.

Point and direction-of-change forecast comparisons, together with the full tables for all forecast horizons, are reported in Supplement S.4.3. The main message is that allowing for SV does not improve point-forecast accuracy. For sign forecasts, the evidence is more mixed: SV improves directional accuracy at short horizons, while from horizon six onward differences across models become negligible, with most specifications delivering identical or nearly identical success ratios.

We therefore focus here on density-forecast accuracy. To this end, we rely on qCRPS \citep{gneiting2011comparing}, where the model with the lowest score is ranked as the most accurate.\footnote{Contrary to the point-forecast tables, these numbers are not ratios to a benchmark but are expressed in the same scale as real prices.} Panel (a) of Table \ref{tab:Tab03DensPaper} focuses on the center of the predictive distribution, while Panels (b) and (c) evaluate accuracy in the right and left tails, respectively. These tail forecasts are relevant for assessing the probability of extreme price movements.

With very few exceptions, homoskedastic factor-augmented BVAR models -- especially the specification with one factor -- are more accurate than the SV alternatives in forecasting both the center and the tails of the predictive distribution. In Panel (a), SV delivers the lowest qCRPS only at the one-month-ahead horizon; from horizon two onward, the best-performing specification is always homoskedastic. A similar pattern emerges for the right tail in Panel (b): although the BVAR-SV specification performs best at horizon two, the lowest qCRPS is otherwise obtained by homoskedastic models, most often the one-factor BVAR. The evidence for the left tail in Panel (c) is consistent with this conclusion. Apart from horizons one and six, where the BVAR-SV without factors records the lowest score, the best-performing models are again homoskedastic.

In the case of the real carbon price, incorporating SV therefore does not seem to offer any meaningful advantage. As shown in Figure \ref{fig:Fig02PandEmiss} the 2018--2023 evaluation period is dominated by abrupt repricing episodes -- the COVID-19 recession and rebound, and the sharp increase during 2021--2022 -- that are better interpreted as level adjustments driven by policy, macroeconomic, and energy-market shocks than as persistent changes in conditional volatility. Since SV is expected to improve density forecasts primarily when time variation in volatility is persistent \citep{clark2015macroeconomic}, these results suggest that most of the predictable variation in this sample is already captured by the mean equation, especially once common information from the larger predictor set is summarized through factors. This leaves limited scope for SV to further improve the forecast density.

\begin{table}[ht]
\centering
\caption{Quantile-weighted Continuous Ranked Probability Score (qCRPS)}
\newcolumntype{C}[1]{>{\scriptsize\hsize=#1\hsize\centering\arraybackslash}X}
\renewcommand{\arraystretch}{0.8}
\begin{tabularx}{\textwidth}{C{.4}C{1.1}|C{1.1}|C{1.1}|C{1.1}|C{1.1}|C{1.1}}
\multicolumn{7}{l}{\scriptsize (a) Quantile-weighted CRPS: center}\\\hline
 & & 1 Factor & 2 Factors & & 1 Factor & 2 Factors  \\\cline{3-4}\cline{6-7}
h & BVAR(1) & BVAR(1) & BVAR(1) &  BVAR-SV(1) & BVAR-SV(1) & BVAR-SV(1) \\\hline
1 & 0.5025 & 0.4973 & 0.4919 & \underline{\textbf{0.4900}} & 0.4936 & 0.4925\\
2 & \underline{\textbf{0.7263}} & 0.7276 & 0.7322 & 0.7277 & 0.7315 & 0.7373\\
3 & 0.9037 & 0.9085 & \underline{\textbf{0.9028}} & 0.9173 & 0.9226 & 0.9243\\
6 & 1.2594 & \underline{\textbf{1.2506}} & 1.2662 & 1.2703 & 1.2762 & 1.2880\\
9 & 1.6191 & \underline{\textbf{1.6149}} & 1.6221 & 1.6503 & 1.6524 & 1.6648\\
12 & \underline{\textbf{1.9409}} & 1.9645 & 1.9573 & 2.0465 & 2.0520 & 2.0628\\\hline
\multicolumn{7}{l}{\scriptsize (b) Quantile-weighted CRPS: right tail}\\\hline
 & & 1 Factor & 2 Factors & & 1 Factor & 2 Factors  \\\cline{3-4}\cline{6-7}
h & BVAR(1) & BVAR(1) & BVAR(1) &  BVAR-SV(1) & BVAR-SV(1) & BVAR-SV(1) \\\hline
1 & 0.7592 & 0.7542 & \underline{\textbf{0.7410}} & 0.7457 & 0.7500 & 0.7441\\
2 & 1.1810 & 1.1756 & 1.1829 & \underline{\textbf{1.1724}} & 1.1815 & 1.1835\\
3 & 1.4730 & \underline{\textbf{1.4646}} & 1.4662 & 1.4938 & 1.4976 & 1.5065\\
6 & 2.1408 & \underline{\textbf{2.1228}} & 2.1487 & 2.1814 & 2.1988 & 2.2061\\
9 & 2.7457 & \underline{\textbf{2.7269}} & 2.7309 & 2.8172 & 2.8234 & 2.8425\\
12 & 3.5462 & 3.5620 & \underline{\textbf{3.5432}} & 3.7103 & 3.7282 & 3.7499\\\hline
\multicolumn{7}{l}{\scriptsize (c) Quantile-weighted CRPS: left tail}\\\hline
 & & 1 Factor & 2 Factors & & 1 Factor & 2 Factors  \\\cline{3-4}\cline{6-7}
h & BVAR(1) & BVAR(1) & BVAR(1) &  BVAR-SV(1) & BVAR-SV(1) & BVAR-SV(1) \\\hline
1 & 0.8016 & 0.7983 & 0.7968 & \underline{\textbf{0.7965}} & 0.8024 & 0.8033\\
2 & \underline{\textbf{1.0537}} & 1.0667 & 1.0734 & 1.0876 & 1.0939 & 1.1067\\
3 & \underline{\textbf{1.2809}} & 1.3010 & 1.2883 & 1.3118 & 1.3269 & 1.3230\\
6 & 1.7339 & 1.7304 & 1.7511 & \underline{\textbf{1.7287}} & 1.7342 & 1.7498\\
9 & 2.2823 & \underline{\textbf{2.2821}} & 2.2989 & 2.3107 & 2.3174 & 2.3185\\
12 & \underline{\textbf{2.5460}} & 2.5767 & 2.5805 & 2.6116 & 2.6200 & 2.6258\\\hline
\end{tabularx}
\label{tab:Tab03DensPaper}
\caption*{\scriptsize\textit{Notes}: Panel (\textit{a}), (\textit{b}) and (\textit{c}) show quantile-weighted Continuous Ranked Probability Scores (CRPS) with weighting that emphasizes the center and the right or the left tail of the distribution, respectively. The best forecasts, associated with the lowest scores, are underlined. $^\ast$ ($^{\ast\ast}$) denotes rejection of the one-sided Diebold-Mariano test at 10\% (5\%).}
\end{table}

\FloatBarrier

\subsection{Forecast instability and pooling}\label{sec:instabil}
The forecast comparisons reported so far evaluate models using average losses over the full evaluation sample. However, relative forecast accuracy may vary over time because of parameter instability, changes in shock volatility, or shifts in the predictive content of the regressors \citep{giacomini2010forecast,RossiJEL}. This concern is particularly relevant in our application, since the evaluation period includes the COVID-19 recession, the 2021--2022 energy crisis, and the subsequent period of lower volatility in carbon prices.

In Supplement S.4.4, we investigate this issue using the fluctuation test of \citet{giacomini2010forecast}.

\begin{figure}[!ht]
    \centering
    \caption{Fluctuation test statistic for a BFAVAR with 1 factor (left) and 2 factors (right) against different benchmarks for forecast horizons $h=1,3,12$ months from September 2019 to December 2022.}
   \includegraphics[width=1\linewidth]{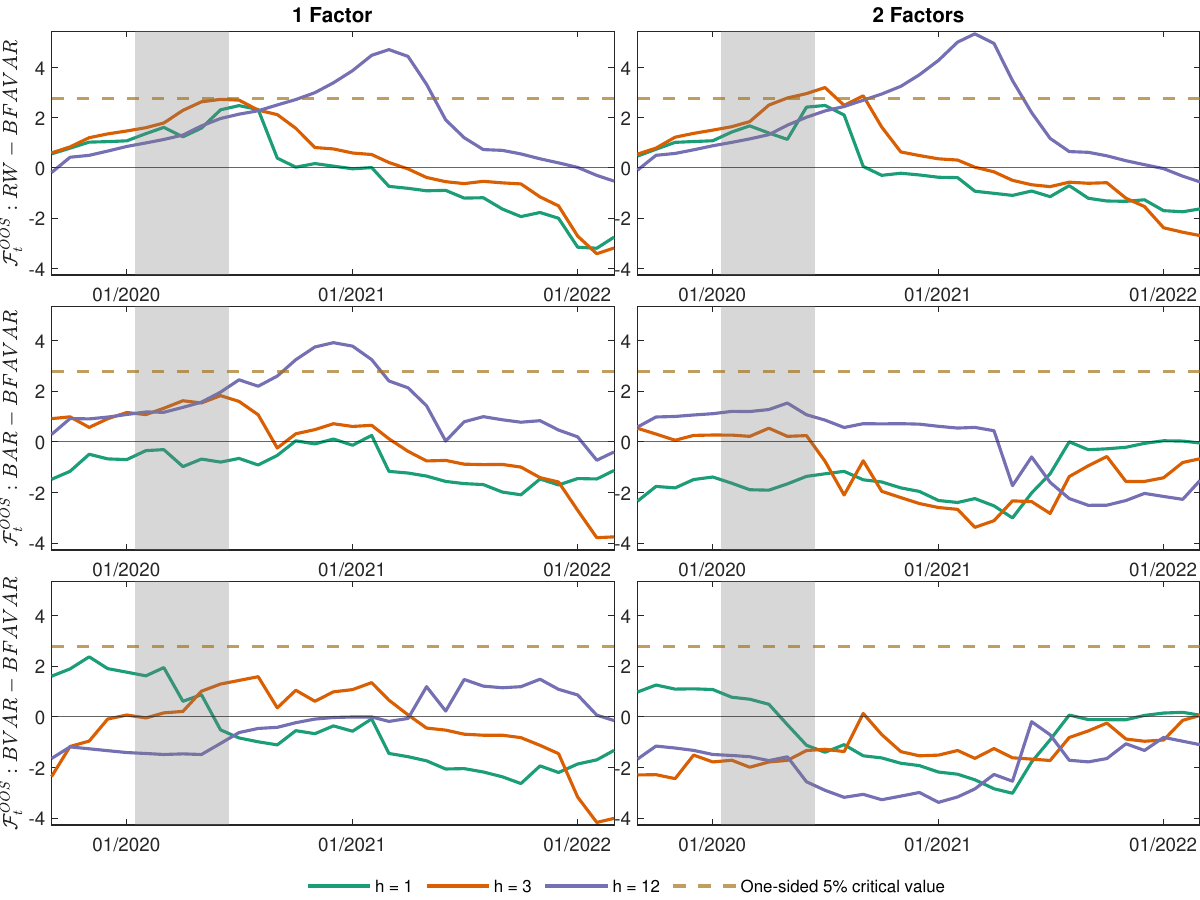}
    \label{fig:FigGW18}
   \caption*{\scriptsize{\textit{Notes:} the fluctuation test statistic, $\mathcal{F}_t^{OOS}$, of \citet{giacomini2010forecast}, is calculated with a 19-month centered rolling window. Positive values indicate that the BFAVAR model is better than the benchmark. All tests are one-sided, with the null hypothesis being that the BFAVAR(1) model has the same MSFE as the benchmark; the alternative is that the BFAVAR(1) model forecasts better than the benchmark. The dashed line indicates the one-sided 5\% critical value, $CV_{0.05}$. The null hypothesis is rejected when $\max \mathcal{F}_t^{OOS}>CV_{0.05}$. The shaded area represents the COVID recession, as determined by the CEPR-EABCN Euro Area Business Cycle Dating Committee (\url{https://eabcn.org}).}}
\end{figure}

Figure \ref{fig:FigGW18} reports fluctuation-test statistics for BFAVAR(1) specifications with one and two factors, based on a 19-month centered rolling window. Forecast performance is clearly time-varying. The one-factor specification generally outperforms the two-factor model and performs well relative to the random walk at the one-month and one-quarter horizons until early 2022, although it rarely improves on the BVAR(1) at the one-month horizon. At longer horizons, particularly $h=12$, factor-augmented models outperform the benchmarks over most of the evaluation sample, including the COVID-19 recession. Against the random walk, equal predictive accuracy is rejected in late 2020 and part of 2021 for $h=3$ and $h=12$. However, relative accuracy deteriorates toward the end of 2022, especially at shorter horizons. Similar time variation emerges for density forecasts, including right-tail qCRPS. Overall, the fluctuation tests indicate that factor models contain useful predictive information, particularly at medium and long horizons, but that their relative performance is not stable over time.

Similar evidence emerges when forecast performance is evaluated using density losses, including right-tail qCRPS. Thus, the fluctuation tests indicate that factor models contain useful predictive information, but also that their performance is not stable over time. See Supplement S.4.4 for further details and results.

We therefore go beyond documenting forecast instability and ask whether it can be mitigated in practice. Forecast averaging is a natural device in this setting because it reduces dependence on a single specification when model rankings change over time \citep[see e.g.][]{bates1969combination,aag2025forecasting,baumeister2015forecasting,stock1999comparison}.

We implement the pooling exercise using the parsimonious lag-one specifications considered above. Here we focus on the simplest and best-performing combination, which averages the forecasts from the AR(1), the BVAR(1), and the one-factor BFAVAR(1). We use a simple equal-weight average, which is transparent, feasible in real time, and avoids estimating additional weights from our short evaluation sample. Additional pooling schemes, including median combinations and pools based on BFAVAR models with one, two, and three factors, are reported in Supplement S.4.5.

Results in Supplement S.4.5 show that pooling does not improve forecast accuracy at very short horizons, where the RW remains difficult to beat. Forecast gains emerge from medium horizons onward. From $h=6$, the pooled forecast delivers RMSFE ratios below one, and at horizons $h=7$ and $h=10$ it is the best-performing specification. At the one-year horizon, the pooled forecast remains very close to the best individual models, preserving most of the gains obtained by the BVAR(1) and the one-factor BFAVAR(1).

Plots of cumulative RMSFE ratios relative to the RW confirm that relative forecast accuracy changes substantially over the evaluation sample. The one-factor BFAVAR(1) performs well before and after the COVID-19 recession, but its cumulative performance deteriorates around 2022. The pooled forecast does not offset this deterioration, but it remains competitive and broadly tracks the best-performing individual specifications. Hence, pooling in this case does not mitigate the instability documented by the fluctuation tests.

Overall, the exercise reinforces the main message of the paper. The useful information in the broader predictor set is concentrated in the parsimonious one-factor BFAVAR.

\subsection{How important are verified emissions when forecasting prices?}\label{sec:robck}
Our results show that simple VAR models of the EU ETS yield sizable accuracy gains when forecasting the real price of carbon. Since both the baseline VAR model of \citet{bjornland2023} and our preferred factor-augmented specifications include verified emissions as an endogenous variable, we might want to assess their role in forecasting real prices. 

As a simple robustness check, we consider a stripped-down version of VAR models where we omit interpolated verified emissions, with the aim of assessing whether the interpolation process induces a look-ahead bias that artificially inflates forecast gains. We focus on point forecasts from the baseline BVAR(1) and the BFAVAR(1) specifications with one and two factors. Results in Supplement S.4.6 show that omitting verified emissions leaves point forecast accuracy largely unchanged: differences in relative RMSFE between models with and without emissions are small across horizons, with no systematic pattern favoring either specification. If anything, models without emissions perform marginally better at some short horizons, while differences become negligible at longer horizons. Results for sign and density forecasts point in the same direction and are reported in Supplement S.4.6.

This finding alleviates concerns about look-ahead bias: were the interpolation artificially inflating forecast performance, removing emissions would be expected to worsen accuracy rather than leave it unchanged or improve it. We therefore retain verified emissions in the main specifications, since including them allows us to produce forecast scenarios conditional on changes in the emissions cap, which are directly relevant for EU ETS market monitoring.

\section{Further results}\label{sec:robustness}
\subsection{Expert forecasts of the nominal price of carbon}
As a further step, we provide a qualitative comparison of model-based nominal price forecasts against the fixed-event forecasts issued by the Carbon Team at the London Stock Exchange Group (LSEG, formerly Refinitiv) and its survey forecasts, converted to a fixed-horizon basis following \citet{dovern2012disagreement}. Full details and figures are provided in Supplement S.4.7.

\bigskip

\noindent\textit{LSEG Carbon Team's forecasts.} The Carbon Team at LSEG produces forecasts of nominal EU ETS price with an irregular cadence for the period 2014-2023 ranging from 3 to 6 times per year, where the forecast horizon can be either the current year or several years in the future. There are 19 one-year-ahead forecasts irregularly spaced over the period December 2018 to September 2023 that overlap with those in our evaluation period.

The comparison between LSEG, RW, and model-based forecasts indicates that forecast performance is not stable over time. LSEG and RW forecasts are very close over most of the sample, with a correlation of forecast errors equal to 0.95 computed over the period January 2015 to May 2023. However, the relative performance of the different approaches changes across market regimes. During the high-volatility period of 2021--2022, the inclusion of macroeconomic variables improves the ability of the  BFAVAR(1) model-based forecasts to track the sharp increase in nominal EU ETS prices. By contrast, in the more stable environment of late 2022 and 2023, LSEG forecasts appear to follow realized prices more closely than the model-based forecasts. Overall, these results are consistent with the evidence for the real carbon price in Section~\ref{sec:instabil} and suggest that the relative performance of LSEG and model-based forecasts of the nominal EU ETS price is time-varying.

\bigskip

\noindent\textit{Survey forecasts.} Carbon Market Surveys run by LSEG each year from 2020 to 2022, conducted between February and April (March in 2020), capture the market sentiment of respondents from a multitude of countries who are mostly stakeholders with tangible and financial interests in carbon markets (e.g., traders, firms covered by an ETS). 
Participants who have answered the section of the survey concerning EU ETS price expectations are 60 in 2020, 119 in 2021, and 88 in 2022. 

\begin{figure}[t]
    \centering
    \caption{\small{Survey and model-based one-year-ahead density forecasts of the nominal EU ETS price from 2021 to 2023.}}
    \includegraphics[width=1\linewidth]{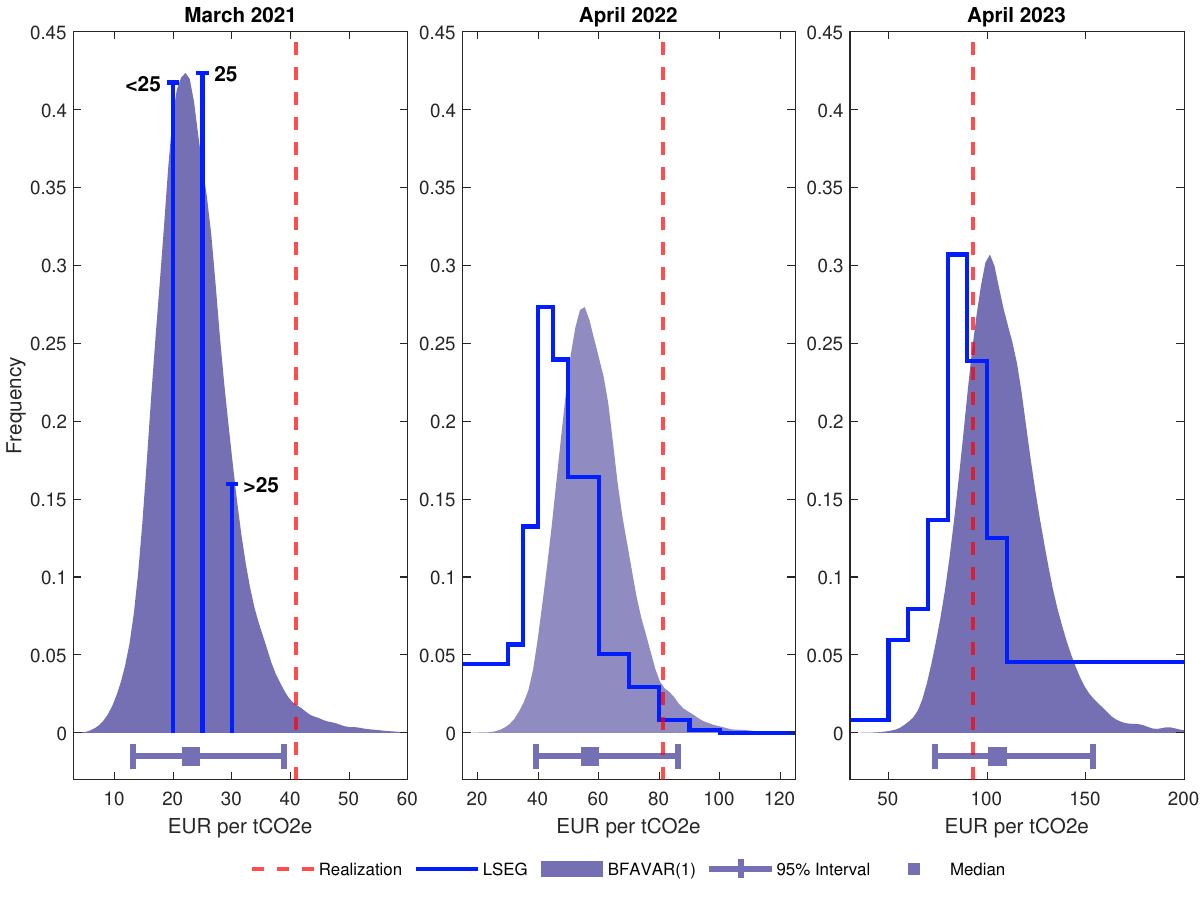}
   \label{fig:FigLSEGDens_nominal}
   \caption*{\scriptsize{\textit{Notes}: each year the survey provides a different set of price ranges across which respondents can choose. The title indicates the date of the forecast, while the data come from the survey of the previous year. Model based forecasts are here produced using nominal price data. The 2020 survey -- leftmost plot -- provided only three categories ``$< 25$'', ``About 25'' and ``$>25$'' Euro per tCO2e.}}
\end{figure}

Figure \ref{fig:FigLSEGDens_nominal} reports LSEG survey forecasts and BFAVAR one-year-ahead predictive densities for the nominal EU ETS price. The comparison confirms that forecast performance differs markedly across periods. In 2021 and 2022, both survey respondents and the single-factor BFAVAR failed to anticipate the rapid run-up in nominal EU ETS prices. Nevertheless, realized prices generally fall within the BFAVAR predictive distribution, mostly in its right tail, with March 2021 the only realization outside the 95\% prediction interval. This suggests that the inclusion of macroeconomic information helps capture part of the upside risk, even when point forecasts substantially underpredict realized prices. By contrast, forecasts for 2023 are much closer to realized values. The exceptional macroeconomic conditions surrounding the COVID-19 pandemic also make forecast evaluation for 2021 and 2022 challenging.

\subsection{Verified emissions forecasts}\label{sec:emissions}
Forecasts of verified emissions are as crucial as price forecasts. Entities participating in the EU ETS need reliable emission forecasts to plan emission reduction strategies and ensure compliance with regulatory requirements, while environmental agencies rely on these forecasts to assess the impact of emissions reduction initiatives. We now turn to assessing whether the BVAR models considered in this paper can serve as a reliable tool for this purpose.

\begin{table}[ht]
\centering
\caption{Relative RMSFE: verified emissions}
\newcolumntype{C}[1]{>{\scriptsize\hsize=#1\hsize\centering\arraybackslash}X}
\renewcommand{\arraystretch}{0.8}
\begin{tabularx}{\textwidth}{C{.2}|C{.9}|C{.9}|C{.9}|C{.9}|C{1.3}|C{1.3}|C{1.3}|C{1.3}}\hline
 &  & &  1 Factor &  2 Factors   & & &  1 Factor &  2 Factors  \\\cline{4-5}\cline{8-9}
h & BAR(1) & BVAR(1) & BVAR(1) & BVAR(1) & BAR(1)-SV & BVAR(1)-SV & BVAR(1)-SV & BVAR(1)-SV \\\hline
1 & 1.1050 & 1.2382 & 1.1805 & 1.2370 & 1.0064 & 1.0206 & 1.0320 & 1.0134\\
2 & 1.0567 & 1.2274 & 1.1630 & 1.2034 & 1.0020 & 1.0003 & 1.0017 & 1.0002\\
3 & 1.0470 & 1.2477 & 1.1716 & 1.2162 & \textbf{0.9999} & \textbf{0.9882} & \underline{\textbf{0.9867}} & \textbf{0.9915}\\
6 & 1.0296 & 1.2332 & 1.1379 & 1.1881 & \textbf{0.9967} & \textbf{0.9593} & \underline{\textbf{0.9576}} & \textbf{0.9584}\\
9 & 1.0179 & 1.1705 & 1.0836 & 1.1264 & \textbf{0.9964} & \underline{\textbf{0.9411}} & \textbf{0.9454} & \textbf{0.9441}\\
12 & 1.0124 & 1.1280 & 1.0502 & 1.0878 & \textbf{0.9962} & \underline{\textbf{0.9295}} & \textbf{0.9347} & \textbf{0.9330}\\\hline
\end{tabularx}
\label{tab:Tab06EmissRMSEPaper}
\caption*{\scriptsize\textit{Notes}: see notes to Table \ref{tab:Tab01PaperBVARS}}
\end{table}

We report full results in Supplement S.4.8. Table \ref{tab:Tab06EmissRMSEPaper} illustrates that for point forecasts, no specification improves upon the RW at horizons of one and two months, but SV models deliver systematic accuracy gains from $h=3$ onward. Among SV specifications, the one-factor BFAVAR(1)-SV performs best at medium horizons, while the baseline BVAR(1)-SV dominates at longer ones. At the one-year horizon -- the most policy-relevant given the annual publication of verified emissions data -- SV models reduce RMSFE relative to the RW by between 6.5\% and 7.1\%. The AR(1) is uniformly worse than the RW at all horizons, providing evidence against its use as a univariate benchmark. For density forecasts, SV models also dominate for the center of the predictive distribution, with the one-factor BFAVAR(1)-SV performing best at short and medium horizons and the BVAR(1)-SV at longer ones. The contrast with the real carbon price results -- where SV yields no systematic gains -- reflects the different time-series properties of the two variables. Verified emissions follow a smoother path with volatility concentrated around the COVID-19 episode; SV can identify this as a persistent high-uncertainty regime and improve density accuracy accordingly \citep{clark2015macroeconomic}. By contrast, real carbon prices mainly reflect abrupt policy, macroeconomic, and energy-market repricing, largely captured by the mean equation.

\subsection{Market monitoring}\label{sec:monitoring}
Following \citet{baumeister2022energy}, we construct indices of \textit{demand pressure}, \textit{upward}, and \textit{downward price pressure} for the EU ETS market. 
For approximating demand pressure, we take the difference between one-year and one-month-ahead verified emission forecasts from the BFAVAR(1)-SV model. 
A negative value of the proxy signals expectations of loosening market conditions over the next year.

Using the predictive densities delivered by Bayesian estimation of the single-factor BVAR(1) model, upward and downward price pressure indices are defined as follows:
\begin{align}
    PP^{+}_t & = \frac{1}{12}\sum_{h=1}^{12} \mathbb{I}\left[\hat{R}_{t+h|t}>\max\left(R_t,R_{t-1},\ldots,R_{t-11}\right)\right],\\
        PP^{-}_t & = \frac{1}{12}\sum_{h=1}^{12} \mathbb{I}\left[\hat{R}_{t+h|t}<\min\left(R_t,R_{t-1},\ldots,R_{t-11}\right)\right].
\end{align}
These proxies estimate the probability that over the next 12 months, the real price of carbon is above (below) the maximum (minimum) value observed in the previous year.

\begin{figure}[!ht]
    \centering
        \caption{Demand and price pressure indices for the EU ETS market from March 2018 to August 2023.}    
        \centering
               \includegraphics[width=0.8\textwidth, keepaspectratio]{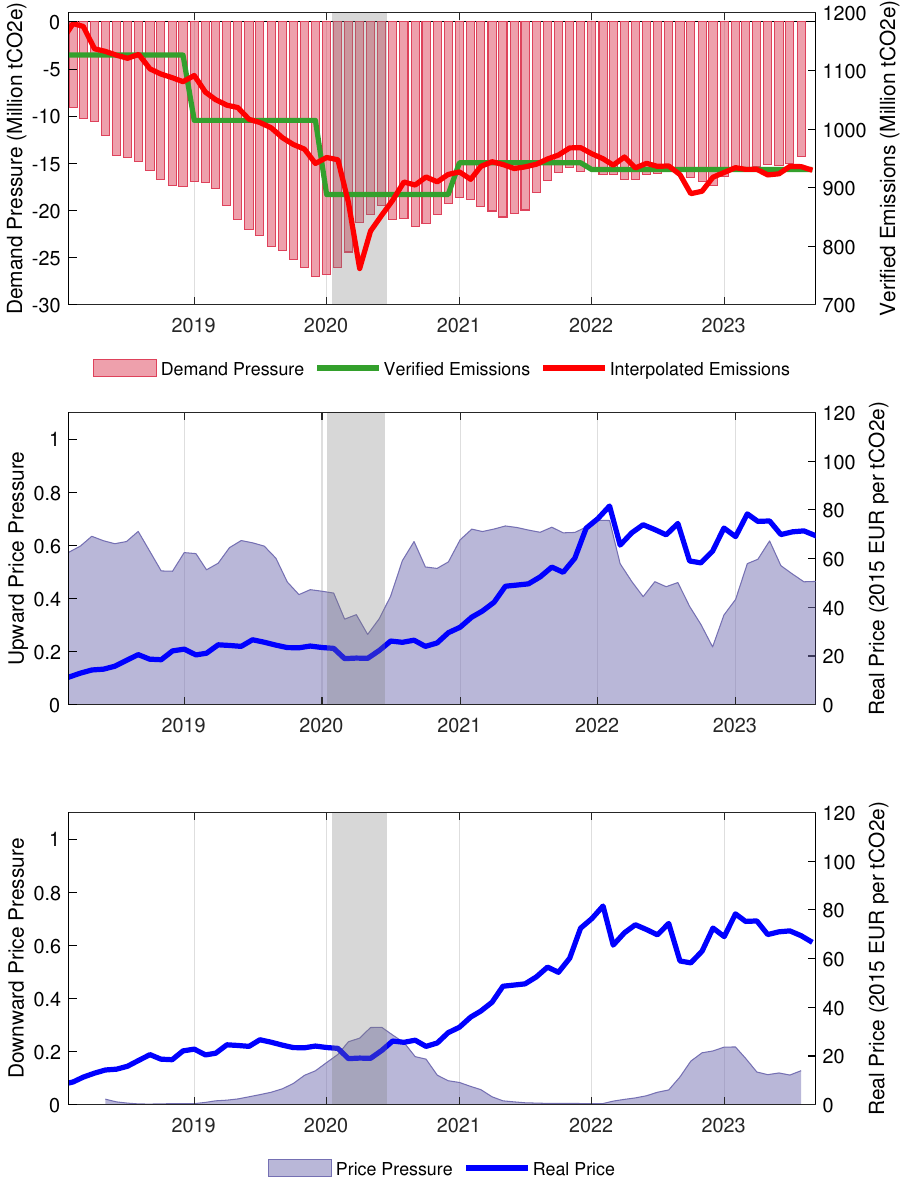}
        \label{fig:FigPpress}
        \caption*{\scriptsize{\textit{Notes}: we plot the backward 3-month moving average of price and demand pressure indices. Indices are aligned with the forecast origin.}}
\end{figure}

The upper panel of Figure \ref{fig:FigPpress} displays the demand pressure proxy as a set of bars along with lines representing (actual and interpolated) verified emissions. 
Note that, although verified emissions are available at an annual sampling frequency, BVAR models are estimated with monthly data; therefore, the demand pressure index allows monitoring expectations about the EU ETS market in real time each month. 
The index is always negative to track the long-term decline of verified emissions, interrupted by an increase in 2021 in the aftermath of the COVID-19 recession.

As shown by \citet{bjornland2023}, the dynamics of verified emissions are elicited by three main forces: a supply shock and two distinct demand shocks. %
The two demand shocks are related to economic activity and transition demand. 
The first essentially depends on the business cycle and, therefore, can be associated with positive or negative changes in the demand pressure index. 
For instance, during the COVID pandemic, expectations of a future quick recovery associated with an increase in industrial activity are captured by a rebound of the demand pressure index. 
The transition-demand shock -- capturing, among other things, the increased usage of renewables -- contributes to the steady decline in verified emissions and therefore keeps the pressure index in negative territory.

The two price pressure indices, along with the evolution of the real EU ETS prices, are shown in the middle and bottom panels of Figure \ref{fig:FigPpress}. 
Given that between 2018 and 2023, the real price has steadily increased -- peaking at almost 120 Euros in early 2023 -- the upward pressure index in the middle panel always signals expectations of soaring prices. 
Two exceptions are recorded in 2020 and in early 2023 when the downward price pressure index rises temporarily, capturing expectations of price decreases.

\section{Conclusions}\label{sec:concl}
The fact that the EU ETS regulation is strengthening over time makes further analysis of the macroeconomic effects of carbon price shocks increasingly warranted. 
Indeed, the ECB already embeds technical assumptions on carbon pricing in its projections \citep{ECB2021press}. 
Technical assumptions amount to setting the trajectory for key variables entering the ECB's macroeconomic models over the projection horizon and are derived in a variety of ways, including univariate and multivariate econometric models and using the price of futures contracts \citep{ECBprojections}. 

In this paper, we have identified predictors and methodological choices that can directly inform projections and scenario analyses of the EU ETS carbon price. 
Our results show that EU ETS prices and verified emissions can be forecast with relatively simple BVAR models estimated with monthly data to include macroeconomic predictors.

There are at least two aspects of our analysis that deserve further investigation. 
First, the use of time series sampled at different frequencies. Mixed Frequency and Mixed-data sampling (MIDAS) models could be used to further improve monthly forecasts relying on data sampled at a higher frequency, such as weather conditions, energy prices, and financial variables. 
Moreover, given that several key predictors of real carbon prices -- such as the level of verified emissions and many macroeconomic aggregates -- are available only at the lower sampling frequency, it would also be informative to exploit reverse-MIDAS approaches \citep{foroni2023low}. Lastly, the comparison between model-based, survey, and LSEG forecasts could open a debate regarding the usage of tilting techniques \citep[see e.g.][]{kruger2017using}. 

\section*{Supplementary Materials}
An online Supplement provides details on the Bayesian estimation of the models, describes the temporal disaggregation of verified emissions data, reports descriptive statistics for the factor models. Moreover, the Supplement shows further robustness checks on prior specification, number of factors and nominal carbon prices, while it provides different fluctuation tests for various models, forecast pooling and the importance of verified emissions.

\clearpage
\setcounter{footnote}{0}
\singlespacing
\arxivRestoreMakeTitle

\title{\vspace{-60pt} \textbf{Supplementary Material for \\
``Forecasting the Price of Carbon with Macroeconomic and Financial variables''}}
\author{
Andrea Bastianin\thanks{University of Milan, Italy and Fondazione Eni Enrico Mattei (FEEM). \color{blue}\texttt{andrea.bastianin@unimi.it}}
\and
Elisabetta Mirto\thanks{Study Center Gerzensee, Foundation of the Swiss National Bank. Switzerland. {\color{blue}\texttt{elisabetta.mirto@szgerzensee.ch}}}
\and
Yan Qin\thanks{ClearBlue Markets, Norway. \color{blue}\texttt{qinmaria@gmail.com}}
\and
Luca Rossini\thanks{University of Milan, Italy and Fondazione Eni Enrico Mattei (FEEM). \color{blue}\texttt{luca.rossini@unimi.it}}
}
\date{}
\maketitle
\vspace{.25cm}

\doublespacing
\normalsize
\appendix
\renewcommand*{\theHequation}{supp.\arabic{equation}}
\renewcommand{\thesection}{S.\arabic{section}}
\renewcommand{\theequation}{S.\arabic{equation}}
\renewcommand{\thefigure}{S.\arabic{figure}}
\renewcommand{\thetable}{S.\arabic{table}}
\setcounter{section}{0}
\setcounter{table}{0}
\setcounter{figure}{0}
\setcounter{equation}{0}

\noindent\textbf{Abstract:} This Supplement provides additional methodological details, descriptive statistics, and further forecasting results. Section \ref{sec:methods} details the Bayesian estimation of the models, including a discussion of the prior distributions. Section \ref{sec:tempdisag} describes the temporal disaggregation of verified emissions data, while Section \ref{sec:facdesc} reports descriptive statistics for the factor models. Additional forecasting results are discussed in Section \ref{sec:appForecasting}. Section~\ref{sec:data_sources} provides details on the data sources.

\newpage
\section{Bayesian estimation of models}\label{sec:methods}
\subsection{Bayesian Estimation of AR, VAR and factor VAR models}\label{sec:bvarest}
We provide additional details about the estimation of VAR(p) models considered in the paper. We begin by rewriting the model and establishing the notation:
\begin{equation}
\mathbf{y}_t = \mathbf{a} + \sum_{j=1}^{p} \mathbf{A}_j \mathbf{y}_{t-j} + \mathbf{u}_t,
    \label{eq:supp-varm1}
\end{equation}
where $\mathbf{a}$ is a $n\times1$ vector of intercepts, $\mathbf{A}_j$ are $n\times n$ matrices of coefficients for $j=1,\ldots,p$, and $\mathbf{u}_t$ is an $n\times1$ vector of zero-mean innovations Normally distributed with covariance matrix $\boldsymbol{\Sigma}$. Let $k=np+1$ be the number of regressors and $nk=n^2p+n$ the number of coefficients.

It is useful to rewrite models \eqref{eq:supp-varm1} as:
\begin{equation}
\mathbf{Y} = \mathbf{Z}\mathbf{A}+\mathbf{U}
    \label{eq:varm1mat}
\end{equation}
where the dependent variables are stacked in the $T\times n$ matrix $\mathbf{Y}$ with $t$-th row being $\mathbf{y}_t^{\prime}$. The $T\times k$ matrix $\mathbf{Z}$ collects regressors, with $t$-th row being $\mathbf{x}_t^{\prime}=\left(1,\mathbf{y}_{t-1}^{\prime},\ldots,\mathbf{y}_{t-p}^{\prime}\right)$. The $k\times n$ matrix $\mathbf{A} = \left(\mathbf{a},\mathbf{A}_1,\ldots,\mathbf{A}_p\right)^{\prime}$ collects VAR coefficients, and $\mathbf{U}$ is a $T\times n$ matrix of innovations with $t$-th row being $\mathbf{u}_t^{\prime}$.

From this representation, it follows that $\text{vec}(\mathbf{U})\sim N(\mathbf{0},\boldsymbol{\Sigma}\otimes \mathbf{I}_T)$. Moreover, the likelihood function can be written as:
\begin{equation}
p\left(\mathbf{Y}\left|\mathbf{A},\boldsymbol{\Sigma}\right.\right) = (2\pi)^{-0.5Tn} \ \left|\boldsymbol{\Sigma}\right|^{-0.5T} \ \exp\left\{-0.5 \ \text{tr}\left[\boldsymbol{\Sigma}^{-1}\left(\mathbf{Y} - \mathbf{Z}\mathbf{A}\right)^\prime \left(\mathbf{Y} - \mathbf{Z}\mathbf{A}\right)\right]\right\}
    \label{eq:varlik}
\end{equation}
where $|\cdot|$ denotes the determinant and $\text{tr}(\cdot)$ the trace function.

\bigskip

\noindent\textit{The natural conjugate prior.} In all the baseline results estimation of VAR models relies on the natural conjugate prior in which we consider a joint distribution for $\left(\text{vec}(\mathbf{A}),\boldsymbol{\Sigma}\right)$. More precisely, we assume that $\boldsymbol{\Sigma}$ has the following inverse-Wishart prior:
\begin{equation}
\boldsymbol{\Sigma}\sim \mathcal{W}^{-1}({\nu}_0,\mathbf{S}_0)
    \label{eq:iwish}
\end{equation}
where ${\nu}_0>0$ is a shape parameter and $\mathbf{S}_0$ a scale matrix. In addition we have the following conditional multivariate Normal prior for $(\text{vec}(\mathbf{A})|\boldsymbol{\Sigma})$:
\begin{equation}
\left(\text{vec}(\mathbf{A})\left|\right. \boldsymbol{\Sigma}\right)\sim N\left(\text{vec}(\mathbf{A}_0),\boldsymbol{\Sigma}\otimes\mathbf{V}_\mathbf{A}\right)
    \label{eq:condNAmat}
\end{equation}
Then $\left(\text{vec}(\mathbf{A}), \boldsymbol{\Sigma}\right)$ has the following Normal-inverse-Wishart (NIW) prior:
\begin{equation}
\left(\text{vec}(\mathbf{A}), \boldsymbol{\Sigma}\right)\sim\text{NIW}\left(\text{vec}(\mathbf{A}_0),\mathbf{V}_\mathbf{A},\nu_0,\mathbf{S}_0\right)
    \label{eq:niw}
\end{equation}
The joint density function is:
\begin{equation}
p(\mathbf{A}, \boldsymbol{\Sigma}) \propto %
|\boldsymbol{\Sigma}|^{-0.5(\nu_0 + n + k + 1)} %
\exp \left\{-0.5 \ \text{tr}\left[\boldsymbol{\Sigma}^{-1}\mathbf{S}_0\right]\right\} %
\exp \left\{-0.5 \ \text{tr}\left[\boldsymbol{\Sigma}^{-1} (\mathbf{A} - \mathbf{A}_0)'\mathbf{V}_{\mathbf{A}}^{-1}(\mathbf{A} - \mathbf{A}_0)\right]\right\}
\label{eq:niwpdf}
\end{equation}

\bigskip

\noindent\textit{Selection of the hyperparameters}.  The hyperparameters of the NIW prior are: $\text{vec}(\mathbf{A}_0)$, $\mathbf{V}_\mathbf{A}$, $\nu_0$, and $\mathbf{S}_0$. In the empirical implementation we adopt a Minnesota-style specification embedded in the natural conjugate prior.

The prior mean $\mathbf{A}_0$ is set to shrink the VAR dynamics towards a parsimonious benchmark: the intercepts and all coefficients are centered at zero, except for the coefficients on the own first lag. Specifically, diagonal elements of the first lag matrix are set to zero when variables are modeled in first differences, and to unity when they are modeled in levels or log-levels (unit-root prior). In our application, diagonal elements of the first lag matrix are set to zero also for estimated factors. This choice reflects the fact that factors are extracted via principal component analysis from a set of variables that have been transformed to induce stationarity prior to estimation.

The prior covariance matrix $\mathbf{V}_\mathbf{A}$ is assumed to be diagonal. Diagonal elements for coefficients associated with the $j$-th lag of variable $r$ are set equal to $c_1/(j^2\hat{\sigma}_r^2)$ where $\hat{\sigma}_r^2$ is the residual variance of an AR($p$) model for variable $r$. The intercepts prior variance are set separately to be equal to $c_2$.
The hyperparameters $(c_1,c_2)$ regulate the strength of shrinkage: smaller $c_1$ tightens the prior around $\mathbf{A}_0$ (i.e. stronger regularization). The lag-decay $j^{-2}$ progressively shrinks longer lags. Scaling by $\hat{\sigma}_r^2$ ensures that shrinkage is comparable across variables measured in different units. In our application, we set $c_1=0.04$ and $c_2 = 100$: the former imposes moderate shrinkage on the autoregressive dynamics, while the latter reflects diffuse prior beliefs about the intercepts.

The prior for the degrees of freedom parameter is $\nu_0=n+3$ which implies a relatively diffuse inverse-Wishart prior. The scale matrix $\mathbf{S}_0$ is diagonal with elements $\hat{\sigma}_r^2$, that is the residual variance of an AR($p$) model for variable $r$ with lag-order $p$ that matches the lag-order of the VAR($p$) model.

\bigskip

\noindent\textit{Estimation}. For the derivation of the posterior distribution and the estimation, we follow \citet{supp:chanlargebvar}, whose replication code we adapt for the purposes of this paper. In all cases, we take a total of 5200 draws, discarding the first 
200 as burn-in.

The factors entering the BFAVAR models are extracted via principal component analysis from the set of 20 predictors described in the paper. Prior to extraction, all variables are transformed to induce stationarity and standardized. Specifically, IP indices, the Euro Stoxx price index, energy prices, and the auction price are transformed into monthly growth rates (first differences of logarithms). Clean dark and clean spark spreads are first-differenced, while the monthly auction price range is log-transformed. We do not apply any transformation to the auction cover ratio, temperature, and precipitation anomalies. A preliminary screening for outliers was also carried out; however, given that only a few extreme observations (i.e. observations exceeding 20 times the interquartile range from the median) were detected during the COVID pandemic, we decided to keep them in the sample. At each forecast origin, the principal components are re-estimated as described above on the available sample only, so as to avoid any look-ahead bias. In the large-VAR estimated in the paper, we include the 20 individual predictors transformed as described above. The same code is also used for Bayesian AR models.

\subsubsection{The Horseshoe prior}\label{sec:hsprior}
As a robustness check, we define a global-local shrinkage prior for the vectorized matrix of coefficients. 
In particular, we rely on the Horseshoe prior (HS) introduced by \cite{supp:Carvalho2010} and adapted to VAR by \cite{supp:Follett2019}, which, unlike other global-local shrinkage priors, does not require hyperparameter selection. 
Let us denote $\boldsymbol{\alpha} = \text{vec}(\mathbf{A})$, then the general HS prior takes the form
\begin{align*}
    \alpha_j| \lambda_j^\alpha, \tau^\alpha &\sim \mathcal{N}(0,\lambda_j^\alpha \tau^\alpha) \\
    \lambda_j^\alpha &\sim \mathcal{C}^+(0,1), \\
    \tau^\alpha &\sim \mathcal{C}^+(0,1),
\end{align*}
where $\mathcal{C}^+(\cdot,\cdot)$ denotes the half-Cauchy distribution, $\lambda_j^\alpha$ and $\tau^\alpha$ are the local and global shrinkage parameter, respectively.

Following \cite{supp:Cross2020Shrink}, we consider the scale mixture representation of the half-Cauchy distribution as
\begin{align*}
    \lambda_j^\alpha| \nu_j^\alpha &\sim \mathcal{I}G\left(\frac{1}{2},\frac{1}{\nu_j^\alpha}\right) \\
    \tau^\alpha|\xi^\alpha &\sim \mathcal{I}G\left(\frac{1}{2},\frac{1}{\xi^\alpha}\right) \\
    \nu_j^\alpha &\sim \mathcal{I}G\left(\frac{1}{2},1\right) \\
    \xi^\alpha &\sim \mathcal{I}G\left(\frac{1}{2},1\right), 
    \end{align*}
    where $\nu_j^\alpha$ and $\xi^\alpha$ are independent auxiliary random variables. 
    
Inference for the local and global shrinkage parameters follows the one provided by \cite{supp:Cross2020Shrink}, while inference for $\boldsymbol{\alpha}$ and for $\Sigma$ follows the same steps as in the conjugate case.

\subsection{Bayesian Estimation of VAR and factor VAR Models with Stochastic 
Volatility}\label{sec:bvarsvestsv}
We provide additional details about the estimation of VAR($p$) models with stochastic volatility considered in the paper. The BVAR model with Cholesky multivariate SV process \citep[][for details]{supp:chan2023comparing}  can be represented as follows:
\begin{align}
\mathbf{y}_t &= \mathbf{a} + \sum_{j=1}^{p} \mathbf{A}_j 
\mathbf{y}_{t-j} + \mathbf{u}_t \\
\mathbf{u}_t &\sim N(\mathbf{0},\boldsymbol{\Sigma}_t), \quad 
\boldsymbol{\Sigma}_t^{-1}=\mathbf{B}_0^{\prime}\mathbf{D}_t^{-1}
\mathbf{B}_0, \label{eq:svvar}\\
h_{i,t} &= \mu_i + \phi_i\left(h_{i,t-1}-\mu_i\right)+\varepsilon_{i,t}, 
\quad \varepsilon_{i,t}\sim N(0,\sigma^2_{\varepsilon_i}),
\label{eq:svola}
\end{align}
where $\mathbf{a}$ is a $n\times1$ vector of intercepts, $\mathbf{A}_j$ are $n\times n$ matrices of coefficients for $j=1,\ldots,p$, and $\mathbf{u}_t$ is an $n\times1$ vector of reduced-form idiosyncratic errors. Next, we have: $\mathbf{B}_0$, a lower triangular matrix of size $n\times n$ with ones along the main diagonal and $\mathbf{D}_t = \mathrm{diag}(e^{h_{1,t}},\ldots,e^{h_{n,t}})$. Equation \eqref{eq:svola} specifies an independent autoregressive volatility process for each variable in the model for $t=2,\ldots,T$ with initial condition $h_{i,1}\sim N(\mu_i,\sigma^2_{\varepsilon_i}/(1-\phi_i^2))$. The model parameters are:
\begin{itemize}
    \item the equation-specific VAR coefficient vectors $\boldsymbol{\alpha}_i = (a_i, \mathbf{A}_{i,1},\ldots,\mathbf{A}_{i,p})^{\prime}$ of dimension $k\times 1$ where $k=np+1$, for $i=1,\ldots,n$;
    \item the free below-diagonal elements $\boldsymbol{\beta}_i=(B_{0,i,1},\ldots,B_{0,i,i-1})^{\prime}$ of $\mathbf{B}_0$, for $i=2,\ldots,n$;
    \item and the stochastic volatility parameters $(\mu_i,\phi_i,\sigma^2_{\varepsilon_i})$ for $i=1,\ldots,n$.
\end{itemize}

\bigskip

\noindent\textit{Prior on VAR coefficients.} We assume that the intercept and the VAR coefficients are independent across equations and assign each $\boldsymbol{\alpha}_i$ a hierarchical Normal prior \citep{supp:chan2023comparing}:
\begin{equation}
\boldsymbol{\alpha}_i \sim N\!\left(\boldsymbol{\alpha}_{0,i}, 
\mathbf{V}_{\boldsymbol{\alpha}_i}\right), \quad i=1,\ldots,n,
    \label{eq:alphaprior}
\end{equation}
with prior mean $\boldsymbol{\alpha}_{0,i} = \mathbf{0}$. The prior covariance $\mathbf{V}_{\boldsymbol{\alpha}_i}$ is diagonal with elements:
\begin{equation}
V_{\alpha_i,r}^{(l)} = \begin{cases}
\kappa_0 & \text{intercept}\\[4pt]
\dfrac{\kappa_1}{l^2} & \text{own lag } l \text{ in equation } i 
\\[8pt]
\dfrac{\kappa_2\,\hat{\sigma}^2_i}{l^2\,\hat{\sigma}^2_r} & \text{lag } l 
\text{ of variable } r\neq i \text{ in equation } i
\end{cases}
    \label{eq:minnesotasv}
\end{equation}
where $l=1,\ldots,p$ and $\hat{\sigma}^2_r$ denotes the residual variance of a univariate AR($p$) regression for variable $r$. As in the NIW case described in Section \ref{sec:bvarest}, scaling by $\hat{\sigma}^2_r$ ensures unit-free shrinkage across variables, and the intercept variance $\kappa_0=100$ is diffuse.

A key feature of this specification is that $\kappa_1$ and $\kappa_2$ are hyperparameters controlling, respectively, the 
overall shrinkage strength on own-lag and cross-variable coefficients \citep{supp:chan2023comparing}. Both $\kappa_1$ and $\kappa_2$ are treated as unknown parameters to be estimated from the data.

\bigskip

\noindent\textit{Prior on the impact matrix.} The free elements $\boldsymbol{\beta}_i$ of $\mathbf{B}_0$ have an independent 
hierarchical Normal prior:
\begin{equation}
\boldsymbol{\beta}_i \sim N\!\left(\mathbf{0}, 
\mathbf{V}_{\boldsymbol{\beta}_i}\right), \quad i=2,\ldots,n,
    \label{eq:betaprior}
\end{equation}
where $\mathbf{V}_{\boldsymbol{\beta}_i}$ is diagonal with elements: $V_{\beta_i,j} = \kappa_3\,\frac{\hat{\sigma}^2_i}{\hat{\sigma}^2_j}$, $ j = 1,\ldots,i-1$, which shrinks contemporaneous off-diagonal relationships toward zero. The tightness parameter $\kappa_3$ is also treated as unknown and estimated from the data.

\bigskip

\noindent\textit{Prior on stochastic volatility parameters.} 
For the parameters governing the state equation we adopt the priors below. The innovation variance $\sigma^2_{\varepsilon_i}$ has an Inverse-Gamma prior:
\begin{equation}
\sigma^2_{\varepsilon_i} \sim \mathcal{IG}(\nu_\sigma, S_\sigma), 
\qquad \nu_\sigma = 3,\quad S_\sigma = 0.2,
    \label{eq:sig2prior}
\end{equation}
implying a prior mean of $S_\sigma/(\nu_\sigma-1)=0.1$. The unconditional log-volatility mean $\mu_i$ has a diffuse Normal prior:
\begin{equation}
\mu_i \sim N(0,\,100).
    \label{eq:muprior}
\end{equation}
The persistence parameter $\phi_i$ is assigned a Normal prior truncated to the stationarity region $|\phi_i|<1$:
\begin{equation}
\phi_i \sim N(0.98,\;0.0025)\;\mathbf{1}(|\phi_i|<1),
    \label{eq:phiprior}
\end{equation}
In the sampler, stationarity is further enforced by retaining Metropolis--Hastings proposals only when $|\phi_i|<0.998$.

\bigskip

\noindent\textit{Hierarchical GIG priors on shrinkage 
hyperparameters.} The tightness parameters $\kappa_1$, $\kappa_2$ and $\kappa_3$ are treated as latent variables and sampled within 
the MCMC. Each is assigned a Generalised Inverse Gaussian (GIG) hyperprior \citep{supp:chan2023comparing}:
\begin{equation}
\kappa_s \sim \mathcal{GIG}(a_s,\,b_s), \qquad s=1,2,3,
    \label{eq:gigprior}
\end{equation}
The hyperparameters $(a_s,b_s)$ are set following \citet{supp:chan2023comparing}.

\bigskip

\noindent\textit{Special case.} To make the comparison as close as possible to models without stochastic volatility, we consider an alternative specification of the prior for this model. Specifically, (1) we modify Equation \eqref{eq:minnesotasv} so that the prior variance of the VAR coefficients is specified exactly as in Section \ref{sec:bvarest}. In particular, we set $V_{\alpha_{i,r}}^{(l)}=\kappa_1/(l^2\hat{\sigma}_r^2)$, where $i=1,\dots,n$ indexes the equation (i.e., the dependent variable), $r=1,\dots,n$ indexes the variable whose lag appears as a regressor, and $l=1,\dots,p$ denotes the lag order. Here $\hat{\sigma}_r^2$ is the estimated innovation variance of variable $r$, obtained from a univariate autoregression. Moreover, (2) we do not estimate the hyperparameters $\kappa_s$ for $s\in\{1,3\}$ from the data, but we set them to: $\kappa_1=0.04$, and $\kappa_3=0.04$.

\bigskip

\noindent\textit{Estimation.} Posterior inference is conducted via MCMC, using a Gibbs sampler as in \cite{supp:chan2023comparing} whose replication code we adapt for the purposes of this paper. We take a total of $5{,}200$ draws, discarding the first $200$ as burn-in and retaining $5{,}000$ draws. The same code is extended to BFAVAR with SV models by augmenting $\mathbf{y}_t$ with principal components as described in Section \ref{sec:bvarest}.

\section{Temporal disaggregation of verified emissions data}\label{sec:tempdisag}
For the temporal disaggregation process, we follow \citet{supp:bjornland2023} and \citet{supp:Kanzig2023} in applying Chow-Lin temporal disaggregation of verified emissions from annual to monthly data, implemented via the MATLAB library of \citet{supp:quilis}. The Chow-Lin formula
\[
\hat{y} = X\hat{\beta} + VC'(CVC')^{-1}(Y - CX\hat{\beta})
\]
is applied, where $Y$ is the vector of verified annual emissions, $y$ denotes the unobserved monthly emissions, $X$ is the matrix of monthly indicators --the logarithm of the emission-weighted IP index and the logarithm of the HICP--, $C$ is the aggregation matrix converting months into years and ensuring that annual totals are preserved, and $V$ is the
covariance matrix of the monthly residuals. The procedure first estimates $\hat{\beta}$ via Generalized Least Squares (GLS) from the regression of annual verified emissions on the annualised monthly indicator, computes annual residuals $\hat{u} = Y - CX\hat{\beta}$, and recovers monthly estimates as $\hat{y} = X\hat{\beta} + VC'(CVC')^{-1}\hat{u}$. The term $X\hat{\beta}$ provides the monthly profile driven by the IP indicator, while the correction component distributes the annual residuals across months according to the assumed covariance structure. By construction, $Y = C\hat{y}$.

\section{Factor Models -- Descriptive Statistics}\label{sec:facdesc}
\begin{figure}[h!]
    \centering
    \caption{Percentage of total variance explained by factors 1, 2, and 3 from December 2017 to September 2023.}
   \includegraphics[width=0.95\linewidth]{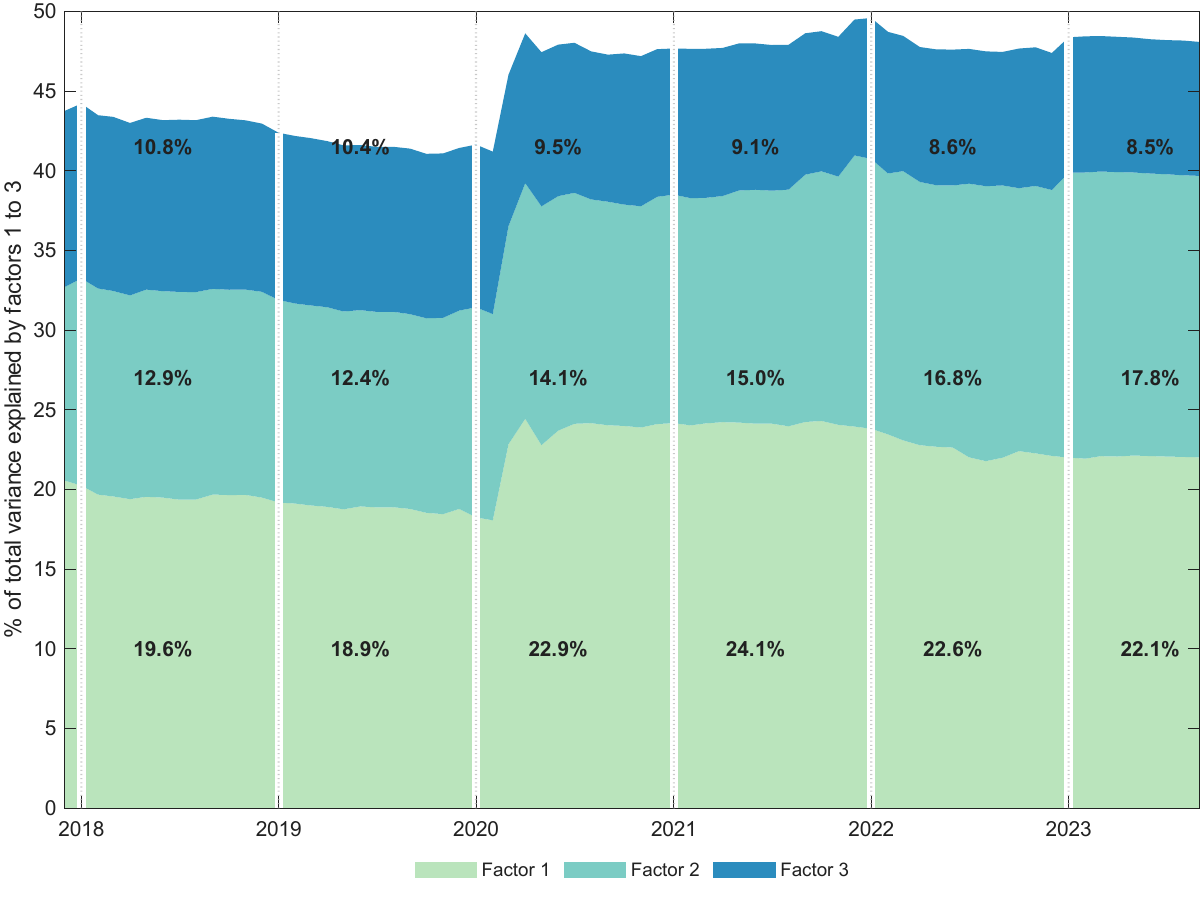}
    \label{fig:PctVar}
\end{figure}

\noindent\textit{Explained variance.} Figure \ref{fig:PctVar} shows the share of the total variance explained by factors from December 2017 to September 2023. 

\bigskip

\begin{figure}[!ht]
    \centering
    \caption{Characterization of the First Three Factors: $R^2$ from Regressing Factors on Individual Predictors (June 2012-December 2017)}
    \includegraphics[width=\linewidth]{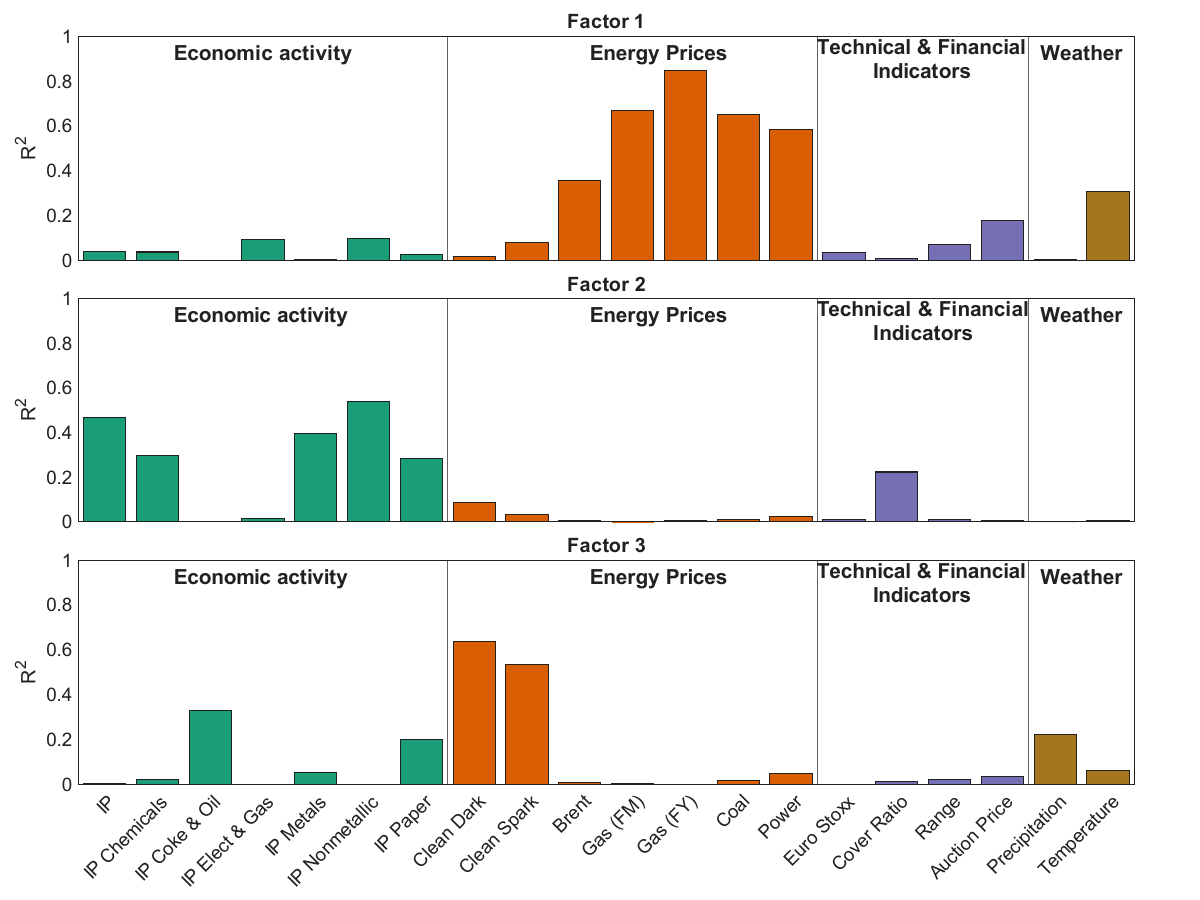}
    \label{fig:FactorR2}
\end{figure}

\begin{figure}[!ht]
    \centering
    \caption{Dynamics of the First Factor and Its Correlation with IP and Natural Gas Prices}
    \includegraphics[width=\linewidth]{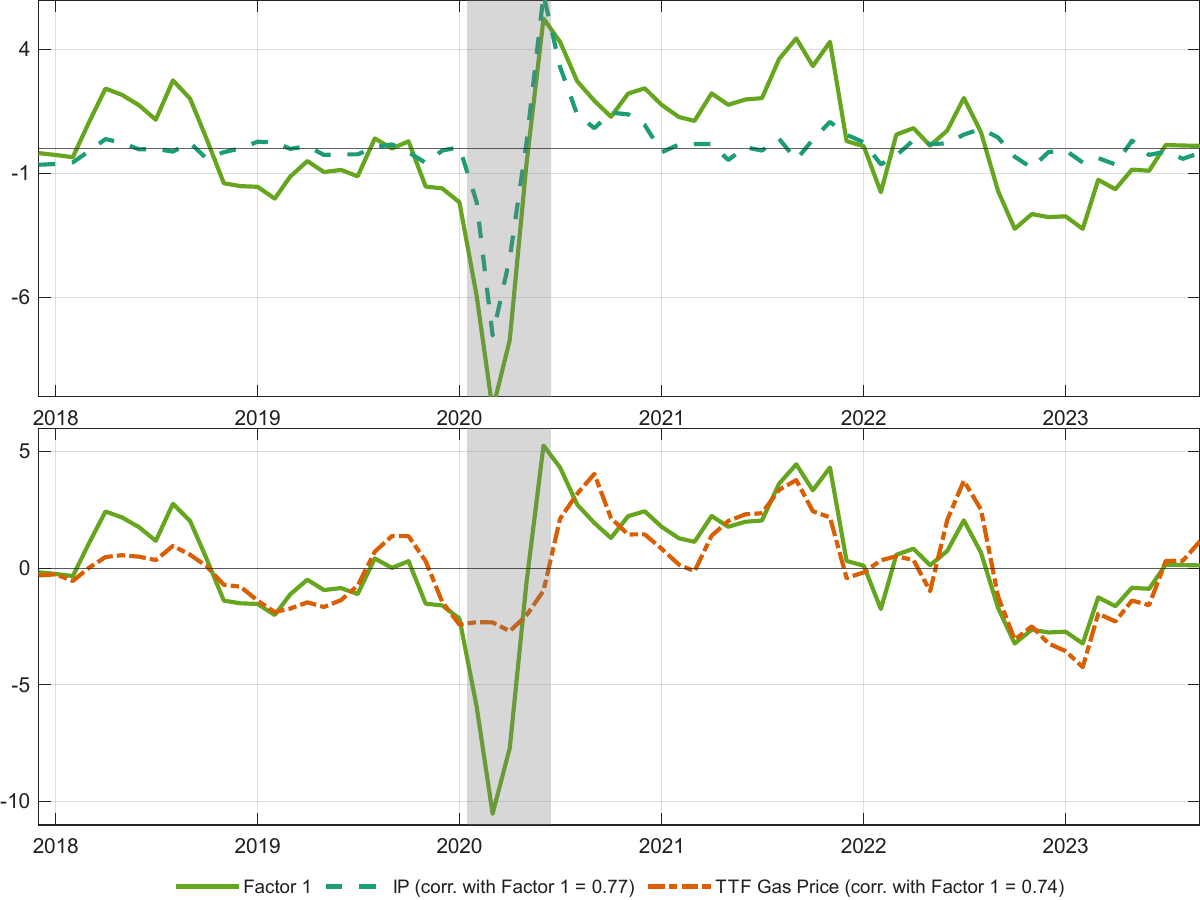}
    \label{fig:FigFactor1Corr}
    {\scriptsize{\textit{Notes}: IP and TTF gas price series are standardized to have zero mean and unit variance.}}
\end{figure}

\noindent\textit{Economic interpretation of factors.} Figure \ref{fig:FactorR2} reports the $R^2$ from regressing each factor on individual predictors
using data from June 2012 to December 2017, corresponding to the information
available at the beginning of the forecasting exercise. The first factor loads
primarily on energy prices and IP indices. The second factor loads on a combination
of several IP indices and the cover ratio, while the third factor loads mainly on
financial spreads, precipitation anomalies, and the remaining IP indices. The
relative contribution of predictors to the factors evolves over time, so the
composition of the variables underlying factor movements is not constant throughout
the sample.

Figure \ref{fig:FigFactor1Corr} shows that the first factor closely tracks the European business cycle and
the evolution of energy prices, displaying a strong correlation with both IP growth
(0.77) and the growth of TTF natural gas prices (0.74). The first factor turns
negative and falls well below its mean during the COVID-19 recession, recovers
during the subsequent expansion, and remains elevated for most of the first half of
2022, when natural gas prices surged following the Russian invasion of Ukraine.

\FloatBarrier
\section{Additional results} \label{sec:appForecasting}

Tables in the following subsections share a common structure. Panel (a) 
reports relative Root Mean Squared Forecast Errors (RMSFE), computed as 
the ratio of the RMSFE of model $m$ over the RMSFE of the Random Walk 
(RW) model, for forecast horizons $h = 1, \ldots, 12$ months ahead. 
Models are estimated on both the log real carbon price, $r_t$, and its 
first difference, $\Delta r_t$. Forecast evaluation is carried out on the real price in level, that is after exponentiating the forecasts. Relative RMSFEs below one, shown in 
bold, indicate superior point forecast accuracy relative to the RW. 
Panel (b) reports Success Ratios (SR); entries in bold exceed 0.5, 
indicating that the model predicts the direction of change correctly more 
than half of the time. %
In each table, the model yielding the lowest RMSFE or the highest 
SR is underlined. Since most existing studies rely on daily or weekly 
nominal prices, the choice of benchmark for monthly real carbon prices is 
not straightforward. Following the commodity price forecasting literature 
--- and crude oil forecasting in particular --- we adopt the simple RW 
model as the primary benchmark.

\subsection{Univariate time series models}\label{sec:univmodels}
Table \ref{tab:Tab01UnivApp} reports RMSFE and SR for the full set of 
univariate time series models. These models include: the Random 
Walk with Drift (RWD); three ARIMA specifications --- ARIMA(1,1,1), 
ARIMA(0,1,1), and ARIMA(1,0,1), where the first two impose a unit root 
and the third is estimated on $r_t$; an Unobserved Component (UC) 
model;\footnote{We specify the UC model as $r_t = \tau_t + \varepsilon_t$ 
with $\varepsilon_t \overset{iid}{\sim} N(0,\sigma^2_{\varepsilon})$. 
The stochastic trend evolves as $\tau_t = \tau_{t-1} + \eta_t$ with 
$\eta_t \overset{iid}{\sim} N(0,\sigma^2_{\eta})$ and 
$\text{cov}(\varepsilon_t, \eta_s) = 0$ for all $t,s$. Estimation 
follows the Markov chain Monte Carlo sampler of \citet{supp:chan2013}.} and Bayesian AR (BAR) models with lag order $p \in \{1, 3, 12\}$ and AIC-based selection, estimated on both 
$\Delta r_t$ and $r_t$. Prior specification and estimation details are 
provided in Section \ref{sec:methods}.

Panel (a) shows that no model improves upon the RW at $h = 1$ and 
$h = 2$. From $h = 3$ onward, the RWD and several ARIMA and BAR 
specifications in first differences generate RMSFE reductions, with 
gains reaching up to 5.5\% at $h = 10$. The UC model underperforms the 
RW at all horizons. BAR models in levels deteriorate monotonically with 
the horizon, with RMSFE ratios well above 1.5 at $h = 12$, and are 
dominated by their first-difference counterparts. Among first-difference 
BAR models, small fixed lag orders ($p = 1, 3$) consistently outperform 
both $p = 12$ and AIC-based selection.

Panel (b) shows that the RWD displays SR above 0.5 at all horizons, 
exceeding 0.8 from $h = 9$ onward. ARIMA models in first differences 
and BAR models in first differences exhibit comparable directional 
accuracy. The UC model performs poorly at all horizons and is discarded 
as a competitor to the RW. BAR models in level show SR declining below 
0.5 from $h = 7$ onward, confirming the unsuitability of level 
specifications for this series.

Since no model improves upon the RW at short horizons, the RW cannot be 
definitively ruled out as a benchmark. Differences in point 
forecast accuracy are evaluated with the \citet{supp:dmtest} test, applied 
following \citet{supp:baumeister2022energy} and \citet{supp:carriero2015bayesian} 
also to nested models. Possibly due to the limited evaluation sample or 
to instabilities in forecasting performance, no statistically significant 
improvements relative to the RW are detected. Forecast instabilities are 
investigated in Section \ref{sec:appFlcutuationTest}.

\begin{landscape}
\begin{table}[ht]
\centering
\caption{Relative RMSFE and Success Ratio of univariate time series models}
\newcolumntype{C}[1]{>{\scriptsize\hsize=#1\hsize\centering\arraybackslash}X}
\renewcommand{\arraystretch}{0.8}
\begin{tabularx}{\linewidth}{C{.4}|C{.7}C{1.3}C{1.3}C{1.3}C{1}|C{1}C{1}C{1}C{1}|C{1}C{1}C{1}C{1}}
\multicolumn{14}{l}{\scriptsize (a) Relative RMSFE}\\\hline
 & \multicolumn{5}{l|}{ } & \multicolumn{4}{c|}{\scriptsize First-difference} & \multicolumn{4}{c}{\scriptsize Level} \\\cline{7-14}
h & RWD & ARIMA(1,1,1) & ARIMA(0,1,1) & ARIMA(1,0,1) & UC & BAR(1) & BAR(3) & BAR(12) & BAR(aic) & BAR(1) & BAR(3) & BAR(12) & BAR(aic)\\\hline
1 & 1.0039 & 1.0668 & 1.0658 & 1.0671 & 1.0159 & 1.0437 & 1.0490 & 1.0419 & 1.0487 & 1.0554 & 1.0585 & 1.0369 & 1.0657\\
2 & 1.0030 & 1.0362 & 1.0297 & 1.0337 & 1.0104 & 1.0324 & 1.0294 & 1.0294 & 1.0316 & 1.1087 & 1.0947 & 1.0528 & 1.0990\\
3 & \underline{\textbf{0.9999}} & 1.0377 & 1.0330 & 1.0384 & 1.0040 & 1.0355 & 1.0283 & 1.0327 & 1.0249 & 1.1581 & 1.1369 & 1.0708 & 1.1381\\
4 & \underline{\textbf{0.9930}} & 1.0090 & 1.0049 & 1.0126 & 1.0126 & 1.0127 & 1.0052 & 1.0276 & 1.0022 & 1.2109 & 1.1668 & 1.0814 & 1.1643\\
5 & \underline{\textbf{0.9913}} & 1.0201 & 1.0160 & 1.0279 & 1.0063 & 1.0193 & 1.0162 & 1.0437 & 1.0152 & 1.2692 & 1.2193 & 1.1172 & 1.2149\\
6 & \underline{\textbf{0.9737}} & \textbf{0.9871} & \textbf{0.9841} & 1.0036 & 1.0151 & \textbf{0.9870} & \textbf{0.9887} & 1.0268 & \textbf{0.9900} & 1.3208 & 1.2436 & 1.1239 & 1.2435\\
7 & \underline{\textbf{0.9652}} & \textbf{0.9794} & \textbf{0.9780} & 1.0024 & 1.0151 & \textbf{0.9813} & \textbf{0.9815} & 1.0215 & \textbf{0.9796} & 1.3790 & 1.2911 & 1.1457 & 1.2960\\
8 & \textbf{0.9515} & \textbf{0.9494} & \underline{\textbf{0.9490}} & \textbf{0.9804} & 1.0155 & \textbf{0.9566} & \textbf{0.9569} & \textbf{0.9996} & \textbf{0.9525} & 1.4350 & 1.3268 & 1.1598 & 1.3324\\
9 & \underline{\textbf{0.9554}} & \textbf{0.9636} & \textbf{0.9623} & \textbf{0.9993} & 1.0108 & \textbf{0.9695} & \textbf{0.9725} & 1.0129 & \textbf{0.9672} & 1.5100 & 1.3903 & 1.1978 & 1.4012\\
10 & \underline{\textbf{0.9450}} & \textbf{0.9489} & \textbf{0.9477} & \textbf{0.9906} & 1.0115 & \textbf{0.9514} & \textbf{0.9608} & 1.0061 & \textbf{0.9538} & 1.5723 & 1.4285 & 1.2108 & 1.4375\\
11 & \underline{\textbf{0.9464}} & \textbf{0.9502} & \textbf{0.9492} & \textbf{0.9962} & 1.0112 & \textbf{0.9499} & \textbf{0.9604} & 1.0137 & \textbf{0.9545} & 1.6586 & 1.4848 & 1.2354 & 1.4911\\
12 & \textbf{0.9521} & \textbf{0.9472} & \underline{\textbf{0.9462}} & \textbf{0.9973} & 1.0146 & \textbf{0.9495} & \textbf{0.9595} & 1.0221 & \textbf{0.9541} & 1.7525 & 1.5438 & 1.2577 & 1.5470\\\hline
\multicolumn{14}{l}{\scriptsize (b) Success Ratio}\\\hline
 & \multicolumn{5}{l|}{ } & \multicolumn{4}{c|}{\scriptsize First-difference} & \multicolumn{4}{c}{\scriptsize Level} \\\cline{7-14}
h & RWD & ARIMA(1,1,1) & ARIMA(0,1,1) & ARIMA(1,0,1) & UC & BAR(1) & BAR(3) & BAR(12) & BAR(aic) & BAR(1) & BAR(3) & BAR(12) & BAR(aic)\\\hline
1 & \underline{\textbf{0.5862}} & 0.5000 & 0.5000 & 0.5000 & 0.5000 & \textbf{0.5172} & \textbf{0.5345} & \textbf{0.5172} & \textbf{0.5172} & \underline{\textbf{0.5862}} & \textbf{0.5345} & \textbf{0.5345} & \textbf{0.5172}\\
2 & \underline{\textbf{0.6379}} & \textbf{0.5517} & \textbf{0.5517} & \textbf{0.5690} & \textbf{0.5345} & \textbf{0.5345} & \textbf{0.5345} & \underline{\textbf{0.6379}} & \textbf{0.5345} & \underline{\textbf{0.6379}} & \textbf{0.5345} & \textbf{0.5862} & \textbf{0.5172}\\
3 & \underline{\textbf{0.6207}} & \textbf{0.5172} & \textbf{0.5172} & \textbf{0.5172} & 0.4655 & \textbf{0.5172} & \textbf{0.5345} & \underline{\textbf{0.6207}} & \textbf{0.5690} & \underline{\textbf{0.6207}} & \textbf{0.5172} & \textbf{0.5862} & 0.5000\\
4 & \underline{\textbf{0.6379}} & \textbf{0.5690} & \textbf{0.5690} & \textbf{0.5172} & 0.3621 & \textbf{0.5690} & \textbf{0.6207} & \underline{\textbf{0.6379}} & \textbf{0.6034} & \underline{\textbf{0.6379}} & \textbf{0.5690} & \textbf{0.5862} & \textbf{0.5345}\\
5 & \underline{\textbf{0.6379}} & \textbf{0.6207} & \textbf{0.6207} & \textbf{0.5345} & 0.4483 & \textbf{0.6207} & \textbf{0.6207} & \underline{\textbf{0.6379}} & \textbf{0.6207} & \underline{\textbf{0.6379}} & \textbf{0.5517} & \textbf{0.5172} & \textbf{0.5172}\\
6 & \underline{\textbf{0.6897}} & \textbf{0.6724} & \textbf{0.6724} & \textbf{0.5690} & 0.4483 & \textbf{0.6724} & \textbf{0.6724} & \underline{\textbf{0.6897}} & \textbf{0.6724} & \underline{\textbf{0.6897}} & \textbf{0.6034} & 0.5000 & \textbf{0.5517}\\
7 & \underline{\textbf{0.7069}} & \textbf{0.6897} & \textbf{0.6897} & \textbf{0.5690} & 0.4310 & \underline{\textbf{0.7069}} & \underline{\textbf{0.7069}} & \underline{\textbf{0.7069}} & \underline{\textbf{0.7069}} & \underline{\textbf{0.7069}} & \textbf{0.6034} & 0.4655 & \textbf{0.5517}\\
8 & \underline{\textbf{0.7759}} & \textbf{0.7586} & \textbf{0.7586} & \textbf{0.6207} & 0.3966 & \underline{\textbf{0.7759}} & \underline{\textbf{0.7759}} & \underline{\textbf{0.7759}} & \underline{\textbf{0.7759}} & \textbf{0.7586} & \textbf{0.6552} & 0.4828 & \textbf{0.6034}\\
9 & \underline{\textbf{0.7931}} & \underline{\textbf{0.7931}} & \underline{\textbf{0.7931}} & \textbf{0.6207} & 0.4310 & \underline{\textbf{0.7931}} & \underline{\textbf{0.7931}} & \underline{\textbf{0.7931}} & \underline{\textbf{0.7931}} & \textbf{0.7586} & \textbf{0.6897} & 0.4655 & \textbf{0.6034}\\
10 & \underline{\textbf{0.8103}} & \underline{\textbf{0.8103}} & \underline{\textbf{0.8103}} & \textbf{0.6207} & 0.4655 & \underline{\textbf{0.8103}} & \underline{\textbf{0.8103}} & \underline{\textbf{0.8103}} & \underline{\textbf{0.8103}} & \textbf{0.7586} & \textbf{0.6897} & 0.5000 & \textbf{0.6034}\\
11 & \underline{\textbf{0.8276}} & \underline{\textbf{0.8276}} & \underline{\textbf{0.8276}} & \textbf{0.6207} & 0.4310 & \underline{\textbf{0.8276}} & \underline{\textbf{0.8276}} & \underline{\textbf{0.8276}} & \underline{\textbf{0.8276}} & \textbf{0.7586} & \textbf{0.6897} & 0.4828 & \textbf{0.6034}\\
12 & \underline{\textbf{0.8103}} & \textbf{0.7931} & \textbf{0.7931} & \textbf{0.5862} & 0.3966 & \textbf{0.7931} & \textbf{0.7931} & \underline{\textbf{0.8103}} & \textbf{0.7931} & \textbf{0.7241} & \textbf{0.6552} & 0.4828 & \textbf{0.5690}\\\hline
\end{tabularx}
\label{tab:Tab01UnivApp}
\caption*{\scriptsize\textit{Notes}: Panel (a) shows relative Root Mean Squared Forecast Errors (RMSFE) computed as the ratio of RMSFE of model $m$ over the RMSFE of the Random Walk (RW) model. Relative RMSFE below one are in bold and suggest superior forecast performance of model $m$ with respect to the RW. $^\ast$ ($^\ast$) denotes that the null hypothesis of the Diebold-Mariano test (i.e. equal predictive accuracy) is rejected at the 90\% (95\%) confidence level.Panel (b) reports the success ratios; entries in bold suggest that the model can accurately predict the direction-of-change over 50\% of the times. $^\ast$ indicates that the p-value of the Pesaran and Timmermann (2009) test of the null of no directional accuracy is below 0.1 and hence provides evidence of statistical accuracy at the 10\% significance level.}
\end{table}

\end{landscape}

\FloatBarrier

\subsection{Bayesian VAR models}\label{sec:bvars}
Table \ref{tab:Tab02AppendixBVARS} reports the forecasting performance of
the full set of Bayesian VAR specifications considered in the analysis.
Panel (a) reports relative RMSFEs, while Panel (b) reports success
ratios. The table compares four groups of models: BVARs estimated on the
first difference of log real carbon prices, $\Delta r_t$ (columns 2--5);
BVARs estimated on the log level, $r_t$ (columns 6--9); BFAVAR models
with one, two, and three factors (columns 10--15); and large BVARs
including the full set of predictors (columns 16--17).

Three main results emerge from Panel (a). First, estimating the baseline BVAR models on
$\Delta r_t$ is clearly preferable to estimating them on $r_t$ in terms
of point forecast accuracy. The level specifications in columns 6--9
display relative RMSFEs above one at all horizons, and their performance
deteriorates markedly as the forecast horizon increases. By contrast,
the first-difference BVARs in columns 2--5 start improving upon the RW
benchmark at medium horizons, with gains becoming larger at longer
horizons.

Second, focusing again on the baseline BVAR model, parsimonious lag structures perform best among the BVAR
specifications. Within columns 2--5, BVAR(1) delivers the lowest RMSFE at
almost all horizons. The BVAR(12)
specification, corresponding to the lag length used in
\citet{supp:bjornland2023}, is systematically less accurate than the shorter
lag specifications in terms of RMSFE. The AIC-selected specification in column (5)
performs close to the short-lag models, but it does not systematically
improve upon BVAR(1) or BVAR(3).

Third, factor augmentation is useful when implemented parsimoniously. The
one-factor BFAVAR specifications in columns 10--11 are among the best
models at medium and long horizons, and the BFAVAR(1) model in column 10
delivers the lowest RMSFE from horizons six to ten. Adding more
factors does not systematically improve point forecast accuracy. The
two-factor and three-factor specifications in columns 12--15 generally
perform worse than the one-factor specification, especially when combined
with three lags.

The large BVARs in columns 16--17 perform poorly in
terms of RMSFE: their relative RMSFEs are above one at all horizons and
are substantially larger than those of the parsimonious BVAR and BFAVAR
models. This suggests that including all predictors individually adds
estimation noise that is not offset by additional forecasting
information.

Panel (b) shows that the evidence on directional accuracy is more
balanced. Success ratios generally increase with the forecast horizon and
are above 0.5 for most specifications. At medium and long horizons, the
best success ratios are often shared by several models, including
BVAR(1), BVAR(12), and the BFAVAR(1) specifications. Large BVARs also
deliver success ratios above 0.5 at most horizons, but they rarely match
the best-performing parsimonious specifications at medium and long
horizons. Hence, while directional accuracy does not identify a single
dominant model, it broadly confirms that parsimonious BVAR and BFAVAR
specifications contain useful predictive information.

Overall, Table \ref{tab:Tab02AppendixBVARS} supports three conclusions.
Forecasting performance is substantially better when models are estimated
on $\Delta r_t$ rather than on $r_t$; short lag structures dominate
longer ones in terms of point forecast accuracy; and summarizing the
large information set through a small number of factors is preferable to
including all predictors individually. The choice between BVAR(1) and the
one-factor BFAVAR(1) remains horizon-dependent: the latter performs best
at medium horizons, while the former becomes marginally more accurate at
the longest horizons.

\paragraph{Robustness check on the number of factors.} We test whether a data-driven approach to the selection of the number of factors in the BFAVAR could further improve forecast performance. For this purpose, we rely on the information criteria proposed by \citet{supp:bai2002determining}. The predictors are treated as an approximate factor model,

\[
X_{it} = \lambda_i' F_t + e_{it},
\]

where \(X_{it}\) denotes the standardized set of predictors, \(F_t\) is a vector of latent common factors, and \(e_{it}\) is the idiosyncratic component.

At each recursive forecast origin, the set of predictors is standardized and factors are estimated by principal components. For each candidate number of factors \(k\), the residual variance is computed as

\[
V(k)
=
\frac{1}{NT}
\left\|
X - \widehat{X}(k)
\right\|_F^2,
\]

where \(N\) is the number of predictors and \(T\) is the length of the estimation sample. The number of factors is then selected by minimizing the \citet{supp:bai2002determining} information criteria. We consider two criteria,

\[
IC_{p1}(k)
=
\log V(k)
+
k
\left(
\frac{N+T}{NT}
\right)
\log
\left(
\frac{NT}{N+T}
\right),
\]

and

\[
IC_{p2}(k)
=
\log V(k)
+
k
\left(
\frac{N+T}{NT}
\right)
\log
\left(
\min\{N,T\}
\right).
\]

The two criteria differ only in the penalty term for adding an additional factor, with \(IC_{p2}\) being slightly more conservative.

Given the relatively small cross-sectional dimension of the predictor panel, we set the maximum number of candidate factors to \(k_{\max}=6\).

Thus, at each forecast origin and for each criterion \(c \in \{IC_{p1},IC_{p2}\}\), the number of factors is selected by minimizing the corresponding information criterion.

\begin{figure}[ht]
\caption{Number of factors selected by the Bai-Ng information criteria at each forecast origin.}
    \centering
    \includegraphics[width=1\linewidth]{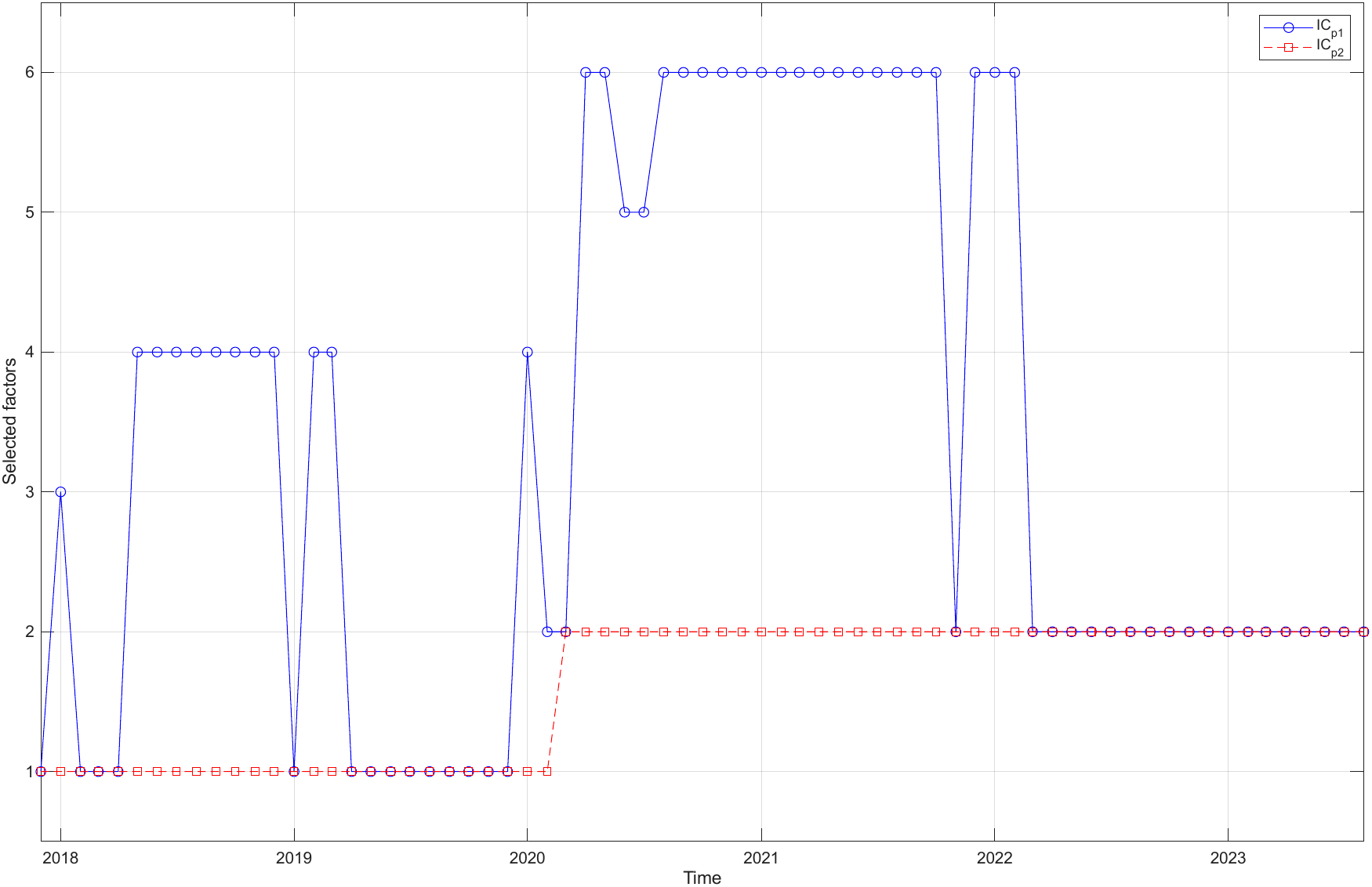}
    \label{fig:nofactorsbaing}
\end{figure}

The selected factors are then included together with the endogenous variables in a BFAVAR with one lag. The forecast evaluation reports two Bai--Ng-based BFAVAR specifications, one using the number of factors selected by \(IC_{p1}\) and one using the number selected by \(IC_{p2}\), alongside the fixed-factor BFAVAR models. Figure \ref{fig:nofactorsbaing} reports the number of factors selected by the two criteria. Consistent with expectations, \(IC_{p2}\) tends to select one or two factors, while \(IC_{p1}\) often selects a larger number of factors. Moreover, both criteria tend to select a larger number of factors from the onset of the COVID-19 pandemic onwards.

Table \ref{tab:Tab_BFAVAR_BaiNg_1Lag_ICp1_ICp2} shows that relying on the Bai-Ng criteria for selecting the number of factors does not lead to improvements with respect to the BVAR models based on a predetermined number of factors.

\paragraph{Robustness check on the prior.} Table \ref{tab:Tab02BVARHS} reports the corresponding results when the Minnesota prior is replaced by the Horseshoe prior. The main conclusions are unchanged. In Panel (a), relative RMSFEs are close to those obtained under the baseline prior and remain generally below one at medium and long horizons for the more parsimonious specifications. The strongest point-forecast gains are obtained by the one-factor BFAVAR with one lag: this specification delivers the lowest RMSFE at almost all horizons from $h=2$ onward, with gains relative to the RW benchmark of about 5--7 percent at longer horizons. Baseline BVARs with short lag structures also
perform well, especially from horizons four to twelve, whereas adding more factors or using three lags within the factor-augmented models tends to reduce forecast accuracy.

The results also confirm that including a large number of predictors individually is not systematically beneficial. Although the large BVAR with three lags improves upon the RW benchmark at several medium and long horizons, its performance is generally weaker than that of the best parsimonious BVAR and BFAVAR specifications. The large BVAR with one lag performs particularly poorly at short horizons and only becomes competitive at longer horizons. This again suggests that shrinkage alone does not fully offset the estimation uncertainty generated by a large predictor set, whereas extracting a small number of factors provides a more effective way of summarizing the information contained in the data.

Panel (b) shows that the evidence on directional accuracy is consistent with the RMSFE results. Success ratios are above 0.5 for most specifications and increase substantially with the forecast horizon. At longer horizons, several parsimonious models achieve success ratios close to or above 0.8, with the BVAR(1), BVAR(12), one-factor BFAVAR(1), and two-factor BFAVAR(1) specifications often sharing the best performance. Overall, the Horseshoe-prior results reinforce the baseline evidence: forecast improvements are concentrated in parsimonious BVAR and BFAVAR models, factor augmentation is useful when limited to a small number of factors, and large unrestricted predictor sets do not deliver systematic forecasting gains.

\paragraph{Nominal price forecasts.} Although the use of real commodity prices is widespread in empirical macroeconomics \citep{supp:baumeister2022energy,supp:bjornland2023,supp:issler2014using,supp:kilian2011responses,supp:wang2023commodity}, forecasts of nominal carbon prices also deserve attention for three reasons. First, practitioners often refer to the nominal price of carbon \citep{supp:icap,supp:stateofeuets2023}. Second, much of the carbon-price forecasting literature focuses on nominal prices \citep[see, e.g.,][]{supp:koop2013JRSS,supp:lei2022probability,supp:tan2022IJF}. Third, one may be concerned that the forecasting gains documented above reflect the ability of VAR models to forecast inflation rather than carbon-price movements \citep{supp:baumeister2014central}. This concern appears limited, however, since most of the variability in nominal carbon prices is accounted for by real prices rather than inflation. A bivariate regression of the first difference of log nominal carbon prices on the first difference of inflation yields an $R^2$ of 0.04, suggesting that inflation explains only a small fraction of nominal carbon-price variability. Consistently, the variance of the nominal log price of carbon, $p_t$, can be decomposed as
\[
\operatorname{var}(p_t)
=
\operatorname{var}(rp_t)
+
\operatorname{var}(hcpi_t)
+
2\operatorname{cov}(rp_t,hcpi_t),
\]
where $p_t=rp_t+hcpi_t$ and all terms, including $HCPI_t/100$, are expressed in logs. The share of variance accounted for by the real-price component, $\operatorname{var}(rp_t)/\operatorname{var}(p_t)$, is 82.06\%, whereas the corresponding share for the inflation component, $\operatorname{var}(hcpi_t)/\operatorname{var}(p_t)$, is only 1.19\%.

We next replicate the analysis in Section 4.2 using nominal carbon prices. The results, reported in Table \ref{tab:Tab02BVAR_nominal}, closely align with those based on real prices and, if anything, provide even stronger support for the main findings. Panel (a) shows that Bayesian VAR models provide little or no improvement over the RW at very short horizons: relative RMSFEs are above one for all specifications from one to three months ahead. Forecast gains emerge from four months ahead and become sizeable at longer horizons. At intermediate horizons, the one-factor BFAVAR with one lag is typically the best-performing model, delivering the lowest RMSFE between five and nine months ahead. At longer horizons, the baseline BVAR(1) becomes marginally more accurate, reaching a relative RMSFE of 0.8774 at twelve months ahead, corresponding to a reduction in forecast errors of about 12.3\% relative to the RW.

Panel (b) shows a similar pattern for directional accuracy. Success ratios are above 0.5 for all models and horizons, and increase markedly with the forecast horizon. From six months ahead onward, several parsimonious specifications attain success ratios above 0.7, rising to almost 0.9 at eleven months ahead. As in the baseline results, the best directional forecasts are generally delivered by compact models, especially BVAR(1), BVAR(12), and the one-lag factor-augmented specifications. Overall, the nominal-price exercise confirms that the forecasting gains are not driven by the deflation procedure. It also reinforces the conclusion that carbon-price predictability is concentrated at medium and long horizons and is best captured by parsimonious Bayesian VAR models rather than by specifications with many factors or richer lag structures.

\begin{figure}[H] \centering 
\caption{\href{https://ec.europa.eu/eurostat/databrowser/view/env_ac_aigg_q__custom_21764548/default/table?lang=en}{Eurostat quarterly GHG emissions} vs monthly interpolated \href{https://www.eea.europa.eu/en/analysis/maps-and-charts/emissions-trading-viewer-1-dashboards}{ETS Verified Emissions} in EU-19 countries}
\label{fig:GHGvsVerifiedEmiss}
\includegraphics[width=0.85\linewidth]{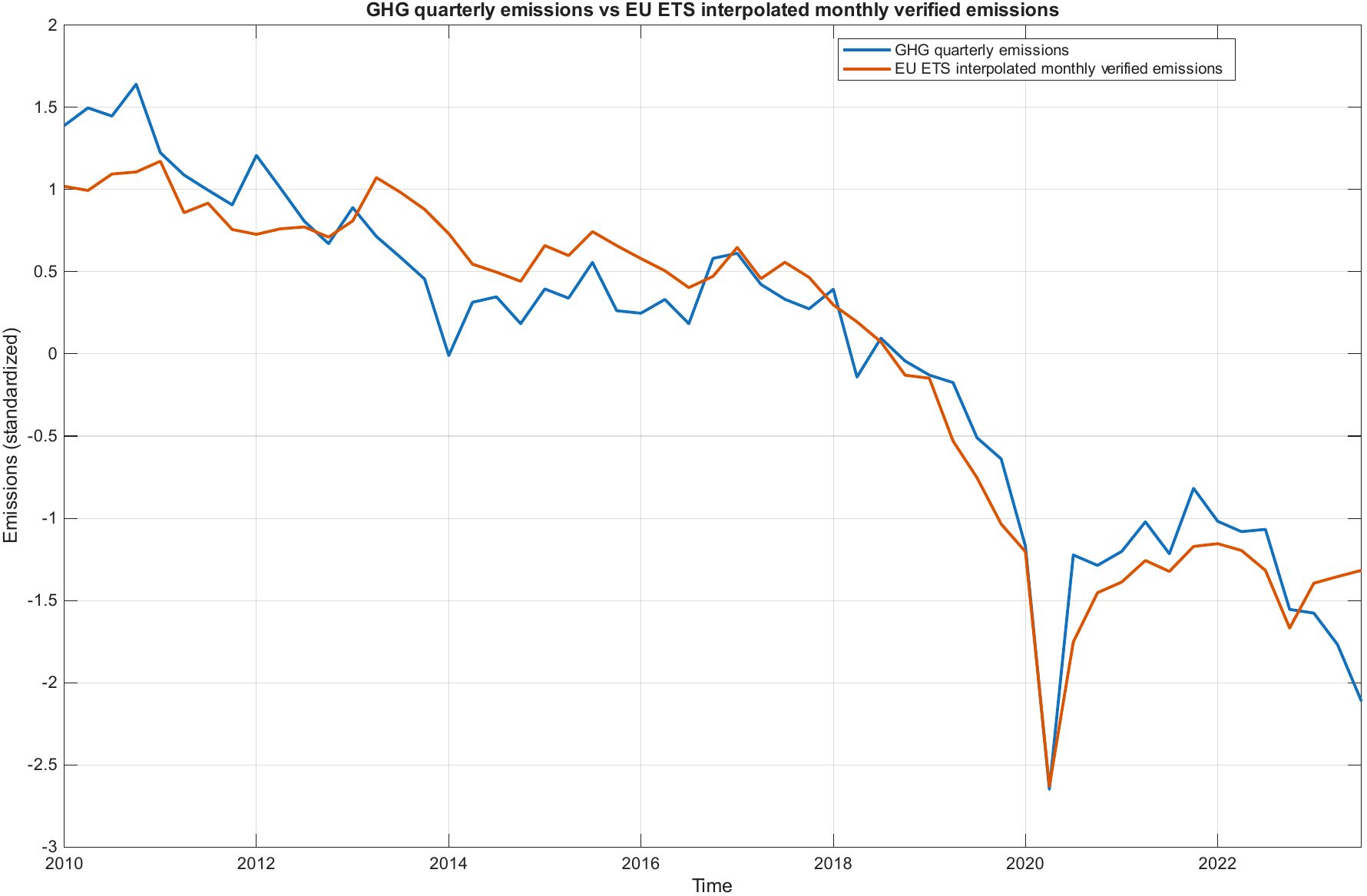} 
\scriptsize{\mbox{\textit{Notes:} The figure compares the series after they have been standardized to have zero mean and unit variance. }}
\end{figure}

\paragraph{Alternative emissions proxies.} As discussed in Section \ref{sec:tempdisag}, our monthly verified-emissions series is constructed following the approach used in the recent carbon-market literature, which applies temporal disaggregation to annual verified emissions. A potential concern with this procedure is that the interpolated series may introduce some degree of look-ahead bias. To assess this issue, we compare our monthly proxy with the main available alternatives. Figure \ref{fig:GHGvsVerifiedEmiss} compares our interpolated series with the one based on Eurostat's Quarterly Greenhouse Gas Emissions Accounts. The two series are highly correlated ($\approx 0.95$) and show very similar dynamics around the main episodes in our sample, including the post-2014 decline, the COVID-19 crisis, and the 2022 increase. Moreover, as discussed in Section 3.1, the available alternatives do not appear to provide a clear improvement over the interpolated verified-emissions series, since they also suffer from limitations in terms of timeliness, frequency, or reliance on indicator-based extrapolation rather than verified data.

\begin{landscape}
\begin{table}[ht]
\centering
\caption{Relative RMSFE and Success Ratio of Bayesian VAR models: full model set}
\newcolumntype{C}[1]{>{\scriptsize\hsize=#1\hsize\centering\arraybackslash}X}
\renewcommand{\arraystretch}{0.8}
\begin{tabularx}{\linewidth}{C{.5}|C{.775}C{.775}C{.85}C{1}|C{.775}C{.775}C{.85}C{1}|C{.85}C{.85}|C{.85}C{.85}|C{.85}C{.85}|C{.85}C{.85}}
\hline
\multicolumn{17}{l}{\scriptsize (a) Relative RMSFE}\\\hline
 & \multicolumn{4}{c|}{\scriptsize First-diff.} & \multicolumn{4}{c|}{\scriptsize Log-Level} & \multicolumn{2}{c|}{\scriptsize 1 Factor} & \multicolumn{2}{c|}{\scriptsize 2 Factors} & \multicolumn{2}{c|}{\scriptsize 3 Factors} & \multicolumn{2}{c}{\scriptsize Large BVAR} \\\cline{2-17}
h & BVAR(1) & BVAR(3) & BVAR(12) & BVAR(aic) & BVAR(1) & BVAR(3) & BVAR(12) & BVAR(aic) & BVAR(1) & BVAR(3) & BVAR(1) & BVAR(3) & BVAR(1) & BVAR(3) & BVAR(1) & BVAR(3)  \\\hline
1 & 1.0505 & 1.0752 & 1.0721 & 1.0726 & 1.0738 & 1.0876 & 1.0675 & 1.0802 & 1.0604 & 1.0739 & 1.0500 & 1.0950 & 1.0577 & 1.1239 & 1.2143 & 1.2711 \\
2 & 1.0337 & 1.0433 & 1.0483 & 1.0438 & 1.1057 & 1.1139 & 1.0774 & 1.1230 & 1.0407 & 1.0562 & 1.0647 & 1.1343 & 1.0816 & 1.1643 & 1.1360 & 1.2415 \\
3 & 1.0366 & 1.0313 & 1.0361 & 1.0352 & 1.1617 & 1.1448 & 1.0928 & 1.1680 & 1.0499 & 1.0605 & 1.0468 & 1.0997 & 1.0725 & 1.1151 & 1.3368 & 1.3904 \\
4 & 1.0178 & 1.0113 & 1.0264 & 1.0128 & 1.2141 & 1.1813 & 1.1119 & 1.2262 & 1.0130 & 1.0182 & 1.0301 & 1.0392 & 1.0481 & 1.0541 & 1.3023 & 1.3992 \\
5 & 1.0259 & 1.0230 & 1.0487 & 1.0203 & 1.2711 & 1.2350 & 1.1566 & 1.2807 & 1.0187 & 1.0271 & 1.0320 & 1.0512 & 1.0492 & 1.0588 & 1.2930 & 1.4373 \\
6 & \textbf{0.9997} & 1.0008 & 1.0363 & 1.0041 & 1.3228 & 1.2669 & 1.1766 & 1.3391 & \underline{\textbf{0.9921}} & 1.0106 & 1.0036 & 1.0364 & 1.0184 & 1.0435 & 1.2851 & 1.4535 \\
7 & \textbf{0.9838} & \textbf{0.9920} & 1.0341 & \textbf{0.9920} & 1.3803 & 1.3129 & 1.2075 & 1.4009 & \underline{\textbf{0.9806}} & \textbf{0.9985} & \textbf{0.9940} & 1.0263 & 1.0093 & 1.0278 & 1.2942 & 1.4695 \\
8 & \textbf{0.9533} & \textbf{0.9606} & 1.0022 & \textbf{0.9583} & 1.4265 & 1.3445 & 1.2140 & 1.4503 & \underline{\textbf{0.9398}} & \textbf{0.9654} & \textbf{0.9642} & 1.0032 & \textbf{0.9783} & 1.0046 & 1.2322 & 1.4493 \\
9 & \textbf{0.9667} & \textbf{0.9735} & 1.0177 & \textbf{0.9689} & 1.5019 & 1.4076 & 1.2610 & 1.5201 & \underline{\textbf{0.9610}} & \textbf{0.9904} & \textbf{0.9695} & 1.0417 & \textbf{0.9843} & 1.0428 & 1.2898 & 1.5440 \\
10 & \textbf{0.9455} & \textbf{0.9554} & 1.0029 & \textbf{0.9513} & 1.5598 & 1.4410 & 1.2713 & 1.5718 & \underline{\textbf{0.9448}} & \textbf{0.9788} & \textbf{0.9539} & 1.0373 & \textbf{0.9637} & 1.0417 & 1.2693 & 1.6097 \\
11 & \underline{\textbf{0.9412}} & \textbf{0.9564} & 1.0123 & \textbf{0.9499} & 1.6449 & 1.4966 & 1.2978 & 1.6472 & \textbf{0.9433} & \textbf{0.9796} & \textbf{0.9509} & 1.0350 & \textbf{0.9574} & 1.0406 & 1.2561 & 1.6527 \\
12 & \underline{\textbf{0.9372}} & \textbf{0.9554} & 1.0227 & \textbf{0.9482} & 1.7360 & 1.5559 & 1.3172 & 1.7298 & \textbf{0.9416} & \textbf{0.9808} & \textbf{0.9492} & 1.0406 & \textbf{0.9558} & 1.0463 & 1.2098 & 1.6891 \\\hline
\multicolumn{17}{l}{\scriptsize (b) Success Ratio}\\\hline
 & \multicolumn{4}{c|}{\scriptsize First-diff.} & \multicolumn{4}{c|}{\scriptsize Log-Level} & \multicolumn{2}{c|}{\scriptsize 1 Factor} & \multicolumn{2}{c|}{\scriptsize 2 Factors} & \multicolumn{2}{c|}{\scriptsize 3 Factors} & \multicolumn{2}{c}{\scriptsize Large BVAR} \\\cline{2-17}
h & BVAR(1) & BVAR(3) & BVAR(12) & BVAR(aic) & BVAR(1) & BVAR(3) & BVAR(12) & BVAR(aic) & BVAR(1) & BVAR(3) & BVAR(1) & BVAR(3) & BVAR(1) & BVAR(3) & BVAR(1) & BVAR(3)  \\\hline
1 & \textbf{0.5517} & \textbf{0.5690} & \textbf{0.5517} & 0.5000 & \textbf{0.5862} & \textbf{0.5690} & \textbf{0.6034} & \textbf{0.5517} & \textbf{0.5517} & \textbf{0.5517} & \textbf{0.5517} & \underline{\textbf{0.6207}} & \textbf{0.5690} & \textbf{0.5690} & \textbf{0.5690} & \underline{\textbf{0.6207}}$^{\ast}$ \\
2 & \textbf{0.5517} & \textbf{0.5690} & \textbf{0.5862} & \textbf{0.5517} & \underline{\textbf{0.6207}} & \textbf{0.5690} & \textbf{0.5517} & \textbf{0.5862} & \textbf{0.5517} & \textbf{0.5690} & \textbf{0.5517} & \textbf{0.5517} & \textbf{0.5517} & 0.5000 & \underline{\textbf{0.6207}} & \textbf{0.5690} \\
3 & \textbf{0.5690} & \textbf{0.5862} & \underline{\textbf{0.6207}} & \textbf{0.5690} & \textbf{0.6034} & \textbf{0.5690} & \textbf{0.5690} & \textbf{0.5690} & \textbf{0.5517} & \textbf{0.5690} & \textbf{0.5690} & \textbf{0.6034} & \textbf{0.5517} & \textbf{0.5690} & \textbf{0.5862} & \textbf{0.6034} \\
4 & \textbf{0.6034} & \underline{\textbf{0.6207}} & \underline{\textbf{0.6207}} & \textbf{0.5862} & \underline{\textbf{0.6207}} & \textbf{0.5517} & \textbf{0.5862} & \textbf{0.5517} & \underline{\textbf{0.6207}} & \textbf{0.6034} & \textbf{0.6034} & \underline{\textbf{0.6207}} & \textbf{0.5690} & \textbf{0.6034} & \textbf{0.5862} & \underline{\textbf{0.6207}} \\
5 & \underline{\textbf{0.6379}} & \underline{\textbf{0.6379}} & \textbf{0.6207} & \underline{\textbf{0.6379}} & \underline{\textbf{0.6379}} & \textbf{0.5345} & \textbf{0.5172} & \textbf{0.5345} & \underline{\textbf{0.6379}} & \textbf{0.6034} & \textbf{0.6207} & \underline{\textbf{0.6379}} & \underline{\textbf{0.6379}} & \underline{\textbf{0.6379}} & \textbf{0.6034} & \textbf{0.5862} \\
6 & \underline{\textbf{0.6897}} & \textbf{0.6724} & \textbf{0.6724} & \textbf{0.6552} & \underline{\textbf{0.6897}} & \textbf{0.5517} & 0.5000 & \textbf{0.5690} & \underline{\textbf{0.6897}} & \textbf{0.6552} & \underline{\textbf{0.6897}} & \textbf{0.6379} & \underline{\textbf{0.6897}} & \textbf{0.6552} & \textbf{0.6207} & \textbf{0.6034} \\
7 & \underline{\textbf{0.7069}} & \textbf{0.6897} & \underline{\textbf{0.7069}} & \textbf{0.6897} & \textbf{0.6897} & \textbf{0.5862} & 0.5000 & \textbf{0.5690} & \underline{\textbf{0.7069}} & \textbf{0.6724} & \underline{\textbf{0.7069}} & \textbf{0.6724} & \underline{\textbf{0.7069}} & \textbf{0.6897} & \textbf{0.6552} & \textbf{0.6207} \\
8 & \underline{\textbf{0.7759}} & \textbf{0.7586} & \underline{\textbf{0.7759}} & \textbf{0.7586} & \textbf{0.7586} & \textbf{0.6379} & 0.5000 & \textbf{0.6207} & \underline{\textbf{0.7759}} & \textbf{0.7414} & \underline{\textbf{0.7759}} & \textbf{0.7414} & \underline{\textbf{0.7759}} & \textbf{0.7586} & \textbf{0.7241} & \textbf{0.7069} \\
9 & \underline{\textbf{0.7931}} & \textbf{0.7759} & \underline{\textbf{0.7931}} & \textbf{0.7759} & \textbf{0.7586} & \textbf{0.6724} & 0.4655 & \textbf{0.6379} & \underline{\textbf{0.7931}} & \textbf{0.7586} & \underline{\textbf{0.7931}} & \textbf{0.7586} & \underline{\textbf{0.7931}} & \textbf{0.7759} & \textbf{0.7586} & \textbf{0.7069} \\
10 & \underline{\textbf{0.8103}} & \textbf{0.7759} & \underline{\textbf{0.8103}} & \textbf{0.7931} & \textbf{0.7586} & \textbf{0.6897} & 0.4655 & \textbf{0.6552} & \underline{\textbf{0.8103}} & \textbf{0.7586} & \underline{\textbf{0.8103}} & \textbf{0.7586} & \underline{\textbf{0.8103}} & \textbf{0.7586} & \textbf{0.7759} & \textbf{0.6897} \\
11 & \underline{\textbf{0.8276}} & \textbf{0.7931} & \underline{\textbf{0.8276}} & \textbf{0.8103} & \textbf{0.7586} & \textbf{0.6897} & 0.4310 & \textbf{0.6552} & \underline{\textbf{0.8276}} & \textbf{0.7759} & \underline{\textbf{0.8276}} & \textbf{0.7759} & \underline{\textbf{0.8276}} & \textbf{0.7759} & \textbf{0.7931} & \textbf{0.7241} \\
12 & \textbf{0.7931} & \textbf{0.7759} & \underline{\textbf{0.8103}} & \textbf{0.7759} & \textbf{0.7241} & \textbf{0.6552} & 0.4655 & \textbf{0.6207} & \textbf{0.7931} & \textbf{0.7414} & \textbf{0.7931} & \textbf{0.7586} & \textbf{0.7931} & \textbf{0.7586} & \textbf{0.7759} & \textbf{0.7241} \\\hline
\end{tabularx}
\label{tab:Tab02AppendixBVARS}
\caption*{\scriptsize\textit{Notes}: Panel (a) shows relative RMSFE (ratio to RW). Values below one in bold; best model per row underlined. $^\ast$ ($^{\ast\ast}$) denotes rejection of the one-sided Diebold-Mariano test at 10\% (5\%). Panel (b) reports success ratios; values above 0.5 in bold; best model per row underlined. $^\ast$ indicates p-value of Pesaran-Timmermann (2009) test below 0.1. Lag orders in parentheses.}
\end{table}

\end{landscape}

\begin{table}[ht]
\centering
\caption{Relative RMSFE and Success Ratio: RW, BFAVAR fixed factors and Bai--Ng selected factors, 1 lag}
\newcolumntype{C}[1]{>{\scriptsize\hsize=#1\hsize\centering\arraybackslash}X}
\begin{tabularx}{\linewidth}{C{.45}|C{.75}C{.75}C{.75}C{.75}C{.95}C{.95}}
\hline
\multicolumn{7}{l}{\scriptsize (a) Relative RMSFE}\\\hline
 & \multicolumn{1}{c}{} & \multicolumn{3}{c}{\scriptsize BVAR(1) Fixed no. factors} & \multicolumn{2}{c}{\scriptsize BVAR(1) Bai--Ng selected} \\\cline{2-7}
h & RW & 1 Factor & 2 Factors & 3 Factors & $IC_{p1}$ & $IC_{p2}$ \\\hline
1 & 1.0000 & 1.0604 & 1.0500 & 1.0577 & 1.0412 & 1.0466 \\
2 & 1.0000 & 1.0407 & 1.0647 & 1.0816 & 1.0486 & 1.0567 \\
3 & 1.0000 & 1.0499 & 1.0468 & 1.0725 & 1.0694 & 1.0446 \\
4 & 1.0000 & 1.0130 & 1.0301 & 1.0481 & 1.0483 & 1.0290 \\
5 & 1.0000 & 1.0187 & 1.0320 & 1.0492 & 1.0446 & 1.0342 \\
6 & 1.0000 & \underline{\textbf{0.9921}} & 1.0036 & 1.0184 & 1.0225 & 1.0050 \\
7 & 1.0000 & \underline{\textbf{0.9806}} & \textbf{0.9940} & 1.0093 & 1.0132 & \textbf{0.9956} \\
8 & 1.0000 & \underline{\textbf{0.9398}} & \textbf{0.9642} & \textbf{0.9783} & \textbf{0.9775} & \textbf{0.9640} \\
9 & 1.0000 & \underline{\textbf{0.9610}} & \textbf{0.9695} & \textbf{0.9843} & \textbf{0.9894} & \textbf{0.9690} \\
10 & 1.0000 & \underline{\textbf{0.9448}} & \textbf{0.9539} & \textbf{0.9637} & \textbf{0.9757} & \textbf{0.9513} \\
11 & 1.0000 & \underline{\textbf{0.9433}} & \textbf{0.9509} & \textbf{0.9574} & \textbf{0.9667} & \textbf{0.9473} \\
12 & 1.0000 & \underline{\textbf{0.9416}} & \textbf{0.9492} & \textbf{0.9558} & \textbf{0.9622} & \textbf{0.9463} \\\hline
\multicolumn{7}{l}{\scriptsize (b) Success Ratio}\\\hline
 & \multicolumn{1}{c}{\scriptsize} & \multicolumn{3}{c}{\scriptsize Fixed no. factors} & \multicolumn{2}{c}{\scriptsize Bai--Ng selected} \\\cline{2-7}
h & RW & 1F & 2F & 3F & ICp1 & ICp2 \\\hline
1 & 0.0000 & \textbf{0.5517} & \textbf{0.5517} & \underline{\textbf{0.5690}} & \textbf{0.5517} & \textbf{0.5517} \\
2 & 0.0000 & \textbf{0.5517} & \textbf{0.5517} & \textbf{0.5517} & \underline{\textbf{0.5862}} & \textbf{0.5345} \\
3 & 0.0000 & \textbf{0.5517} & \underline{\textbf{0.5690}} & \textbf{0.5517} & \underline{\textbf{0.5690}} & \underline{\textbf{0.5690}} \\
4 & 0.0000 & \underline{\textbf{0.6207}} & \textbf{0.6034} & \textbf{0.5690} & \textbf{0.5690} & \textbf{0.6034} \\
5 & 0.0000 & \underline{\textbf{0.6379}} & \textbf{0.6207} & \underline{\textbf{0.6379}} & \textbf{0.6207} & \textbf{0.6207} \\
6 & 0.0000 & \underline{\textbf{0.6897}} & \underline{\textbf{0.6897}} & \underline{\textbf{0.6897}} & \textbf{0.6724} & \underline{\textbf{0.6897}} \\
7 & 0.0000 & \underline{\textbf{0.7069}} & \underline{\textbf{0.7069}} & \underline{\textbf{0.7069}} & \underline{\textbf{0.7069}} & \underline{\textbf{0.7069}} \\
8 & 0.0000 & \underline{\textbf{0.7759}} & \underline{\textbf{0.7759}} & \underline{\textbf{0.7759}} & \underline{\textbf{0.7759}} & \underline{\textbf{0.7759}} \\
9 & 0.0000 & \underline{\textbf{0.7931}} & \underline{\textbf{0.7931}} & \underline{\textbf{0.7931}} & \underline{\textbf{0.7931}} & \underline{\textbf{0.7931}} \\
10 & 0.0000 & \underline{\textbf{0.8103}} & \underline{\textbf{0.8103}} & \underline{\textbf{0.8103}} & \underline{\textbf{0.8103}} & \underline{\textbf{0.8103}} \\
11 & 0.0000 & \underline{\textbf{0.8276}} & \underline{\textbf{0.8276}} & \underline{\textbf{0.8276}} & \underline{\textbf{0.8276}} & \underline{\textbf{0.8276}} \\
12 & 0.0000 & \underline{\textbf{0.7931}} & \underline{\textbf{0.7931}} & \underline{\textbf{0.7931}} & \underline{\textbf{0.7931}} & \underline{\textbf{0.7931}} \\\hline
\end{tabularx}
\label{tab:Tab_BFAVAR_BaiNg_1Lag_ICp1_ICp2}
\caption*{\scriptsize\textit{Notes}: Panel (a) shows relative RMSFE, ratio to RW. Values below one are in bold; the best model in each row is underlined. $^\ast$ ($^{\ast\ast}$) denotes rejection of the one-sided Diebold--Mariano test at 10\% (5\%). Panel (b) reports success ratios; values above 0.5 are in bold; the best model in each row is underlined. $^\ast$ indicates Pesaran--Timmermann test p-value below 0.1. All BFAVAR models use one lag. Bai--Ng ICp1 and ICp2 select the number of factors recursively.}
\end{table}

\begin{table}[ht]
\centering
\caption{Relative RMSFE and Success Ratio of Bayesian Vector Autoregressive models -- Horseshoe Prior}
\newcolumntype{C}[1]{>{\scriptsize\hsize=#1\hsize\centering\arraybackslash}X}
\begin{tabularx}{\textwidth}{C{.1}C{1}C{1}C{1.3}|C{1.05}C{1.1}|C{1.05}C{1.1}|C{1.05}C{1.1}|C{1.05}C{1.1}}
\multicolumn{12}{l}{\scriptsize (a) Relative RMSFE}\\\hline
 & \multicolumn{3}{c}{} & \multicolumn{2}{|c}{\scriptsize 1 Factor} & \multicolumn{2}{|c}{\scriptsize 2 Factors} & \multicolumn{2}{|c}{\scriptsize 3 Factors} & \multicolumn{2}{|c}{\scriptsize Large-BVAR} \\\cline{2-12}
h & BVAR(1) & BVAR(3) & BVAR(12) & BVAR(1) & BVAR(3) & BVAR(1) & BVAR(3) & BVAR(1) & BVAR(3) & BVAR(1) & BVAR(3) \\\hline
1 & 1.0116 & 1.0091 & 1.0047 & 1.0034 & 1.0064 & 1.0044 & 1.0308 & 1.0273 & 1.0228 & 1.0437 & 1.0036\\
2 & 1.0017 & 1.0002 & 1.0012 & \underline{\textbf{0.9766}} & 1.0118 & 1.0068 & 1.0738 & 1.0725 & 1.0527 & 1.0273 & 1.0108\\
3 & \textbf{0.9986} & \textbf{0.9955} & \textbf{0.9929} & \underline{\textbf{0.9785}} & 1.0129 & 1.0045 & 1.0680 & 1.0655 & 1.0461 & 1.0290 & \textbf{0.9905}\\
4 & \textbf{0.9891} & \textbf{0.9857} & \textbf{0.9866} & \underline{\textbf{0.9581}} & 1.0084 & 1.0007 & 1.0536 & 1.0561 & 1.0418 & 1.0137 & \textbf{0.9758}\\
5 & \textbf{0.9892} & \textbf{0.9849} & \textbf{0.9880} & \underline{\textbf{0.9552}} & 1.0069 & \textbf{0.9793} & 1.0337 & 1.0332 & 1.0209 & 1.0127 & \textbf{0.9778}\\
6 & \textbf{0.9748} & \textbf{0.9686} & \textbf{0.9725} & \underline{\textbf{0.9454}} & 1.0016 & \textbf{0.9648} & 1.0073 & 1.0086 & 1.0160 & \textbf{0.9933} & \textbf{0.9624}\\
7 & \textbf{0.9720} & \textbf{0.9693} & \textbf{0.9688} & \underline{\textbf{0.9429}} & 1.0007 & \textbf{0.9620} & \textbf{0.9999} & 1.0033 & 1.0146 & \textbf{0.9872} & \textbf{0.9544}\\
8 & \textbf{0.9613} & \textbf{0.9650} & \textbf{0.9640} & \underline{\textbf{0.9330}} & 1.0006 & \textbf{0.9553} & \textbf{0.9917} & \textbf{0.9951} & 1.0124 & \textbf{0.9659} & \textbf{0.9471}\\
9 & \textbf{0.9653} & \textbf{0.9709} & \textbf{0.9693} & \underline{\textbf{0.9458}} & 1.0030 & \textbf{0.9572} & \textbf{0.9945} & \textbf{0.9953} & 1.0139 & \textbf{0.9738} & \textbf{0.9565}\\
10 & \textbf{0.9572} & \textbf{0.9651} & \textbf{0.9630} & \underline{\textbf{0.9456}} & 1.0034 & \textbf{0.9600} & \textbf{0.9887} & \textbf{0.9896} & 1.0150 & \textbf{0.9621} & \textbf{0.9500}\\
11 & \textbf{0.9553} & \textbf{0.9651} & \textbf{0.9639} & \underline{\textbf{0.9443}} & 1.0038 & \textbf{0.9575} & \textbf{0.9801} & \textbf{0.9800} & 1.0088 & \textbf{0.9649} & \textbf{0.9542}\\
12 & \textbf{0.9510} & \textbf{0.9632} & \textbf{0.9660} & \underline{\textbf{0.9441}} & 1.0050 & \textbf{0.9554} & \textbf{0.9765} & \textbf{0.9762} & 1.0058 & \textbf{0.9703} & \textbf{0.9641}\\\hline
\multicolumn{12}{l}{\scriptsize (b) Success Ratio}\\\hline
& \multicolumn{3}{c}{} & \multicolumn{2}{|c}{\scriptsize 1 Factor} & \multicolumn{2}{|c}{\scriptsize 2 Factors} & \multicolumn{2}{|c}{\scriptsize 3 Factors} & \multicolumn{2}{|c}{\scriptsize Large-BVAR} \\\cline{2-12}
h & BVAR(1) & BVAR(3) & BVAR(12) & BVAR(1) & BVAR(3) & BVAR(1) & BVAR(3) & BVAR(1) & BVAR(3) & BVAR(1) & BVAR(3) \\\hline
1 & \textbf{0.5345} & \textbf{0.5172} & \textbf{0.5862} & \underline{\textbf{0.6207}}$^{\ast}$ & \textbf{0.6034} & \textbf{0.5690} & \textbf{0.6034} & \textbf{0.6034} & \textbf{0.5690} & \textbf{0.5690} & \textbf{0.5345}\\
2 & \textbf{0.5345} & \textbf{0.5345} & \underline{\textbf{0.6379}} & \underline{\textbf{0.6379}} & \textbf{0.5690} & \textbf{0.5690} & \textbf{0.5345} & \textbf{0.5172} & 0.4483 & \textbf{0.6207} & \textbf{0.5862}\\
3 & \textbf{0.5345} & \textbf{0.5345} & \underline{\textbf{0.6207}} & \underline{\textbf{0.6207}} & \textbf{0.5517} & \textbf{0.5862} & \textbf{0.5345} & \textbf{0.5345} & 0.4310 & \underline{\textbf{0.6207}} & \textbf{0.5690}\\
4 & \textbf{0.5862} & \textbf{0.5862} & \underline{\textbf{0.6379}} & \underline{\textbf{0.6379}} & \textbf{0.6034} & \textbf{0.6207} & \textbf{0.5862} & \textbf{0.5862} & \textbf{0.5172} & \underline{\textbf{0.6379}} & \textbf{0.5862}\\
5 & \textbf{0.6207} & \textbf{0.5862} & \underline{\textbf{0.6379}} & \underline{\textbf{0.6379}} & \textbf{0.6034} & \textbf{0.6207} & \textbf{0.6034} & \textbf{0.6034} & \textbf{0.5690} & \underline{\textbf{0.6379}} & \textbf{0.6034}\\
6 & \underline{\textbf{0.6897}} & \textbf{0.6379} & \underline{\textbf{0.6897}} & \underline{\textbf{0.6897}} & \textbf{0.6379} & \textbf{0.6552} & \textbf{0.6552} & \textbf{0.6552} & 0.5000 & \underline{\textbf{0.6897}} & \textbf{0.6552}\\
7 & \underline{\textbf{0.7069}} & \textbf{0.6724} & \underline{\textbf{0.7069}} & \underline{\textbf{0.7069}} & \textbf{0.6552} & \textbf{0.6724} & \textbf{0.6724} & \textbf{0.6724} & \textbf{0.5517} & \underline{\textbf{0.7069}} & \textbf{0.6724}\\
8 & \underline{\textbf{0.7759}} & \textbf{0.7586} & \textbf{0.7586} & \underline{\textbf{0.7759}} & \textbf{0.7241} & \textbf{0.7414} & \textbf{0.7414} & \textbf{0.7414} & \textbf{0.5862} & \underline{\textbf{0.7759}} & \textbf{0.7414}\\
9 & \underline{\textbf{0.7931}} & \textbf{0.7759} & \textbf{0.7759} & \underline{\textbf{0.7931}} & \textbf{0.7414} & \textbf{0.7759} & \textbf{0.7586} & \textbf{0.7586} & \textbf{0.5862} & \underline{\textbf{0.7931}} & \textbf{0.7586}\\
10 & \underline{\textbf{0.8103}} & \textbf{0.7931} & \textbf{0.7931} & \underline{\textbf{0.8103}} & \textbf{0.7586} & \textbf{0.7931} & \textbf{0.7759} & \textbf{0.7759} & \textbf{0.6034} & \textbf{0.7931} & \textbf{0.7759}\\
11 & \underline{\textbf{0.8276}} & \textbf{0.7931} & \textbf{0.8103} & \underline{\textbf{0.8276}} & \textbf{0.7586} & \underline{\textbf{0.8276}} & \textbf{0.7931} & \textbf{0.7759} & \textbf{0.6034} & \textbf{0.7931} & \textbf{0.7759}\\
12 & \underline{\textbf{0.7931}} & \textbf{0.7586} & \underline{\textbf{0.7931}} & \underline{\textbf{0.7931}} & \textbf{0.7241} & \underline{\textbf{0.7931}} & \textbf{0.7586} & \textbf{0.7586} & \textbf{0.5862} & \textbf{0.7759} & \underline{\textbf{0.7931}}\\\hline
\end{tabularx}
\label{tab:Tab02BVARHS}
\caption*{\scriptsize\textit{Notes}: see notes to Table \ref{tab:Tab02AppendixBVARS}}
\end{table}

\begin{table}[htbp]
\centering
\caption{Relative RMSFE and Success Ratio of Bayesian Vector Autoregressive models using nominal price data}
\newcolumntype{C}[1]{>{\scriptsize\hsize=#1\hsize\centering\arraybackslash}X}
\begin{tabularx}{\textwidth}{C{.1}C{1.1}C{1.1}C{1.1}|C{1.1}C{1.1}|C{1.1}C{1.1}|C{1.1}C{1.1}}
\multicolumn{10}{l}{\scriptsize (a) Relative RMSFE}\\\hline
 & \multicolumn{3}{c|}{} & \multicolumn{2}{c}{\scriptsize 1 Factor} & \multicolumn{2}{c}{\scriptsize 2 Factors} & \multicolumn{2}{c}{\scriptsize 3 Factors} \\\cline{5-10}
h & BVAR(1) & BVAR(3) & BVAR(12) & BVAR(1) & BVAR(3) & BVAR(1) & BVAR(3) & BVAR(1) & BVAR(3) \\\hline
1 & 1.0566 & 1.0736 & 1.0726 & 1.0631 & 1.0710 & 1.0437 & 1.0847 & 1.0636 & 1.1241\\
2 & 1.0250 & 1.0329 & 1.0416 & 1.0261 & 1.0404 & 1.0429 & 1.1227 & 1.0806 & 1.1683\\
3 & 1.0197 & 1.0146 & 1.0242 & 1.0287 & 1.0431 & 1.0243 & 1.0833 & 1.0377 & 1.0897\\
4 & \textbf{0.9887} & \underline{\textbf{0.9829}} & 1.0036 & \textbf{0.9831} & \textbf{0.9886} & \textbf{0.9985} & 1.0071 & 1.0100 & 1.0166\\
5 & \textbf{0.9971} & \textbf{0.9938} & 1.0217 & \underline{\textbf{0.9912}} & \textbf{0.9947} & 1.0024 & 1.0150 & 1.0169 & 1.0255\\
6 & \textbf{0.9592} & \textbf{0.9581} & \textbf{0.9918} & \underline{\textbf{0.9529}} & \textbf{0.9647} & \textbf{0.9649} & \textbf{0.9815} & \textbf{0.9788} & \textbf{0.9873}\\
7 & \textbf{0.9447} & \textbf{0.9454} & \textbf{0.9808} & \underline{\textbf{0.9373}} & \textbf{0.9511} & \textbf{0.9514} & \textbf{0.9609} & \textbf{0.9630} & \textbf{0.9663}\\
8 & \textbf{0.9019} & \textbf{0.9056} & \textbf{0.9347} & \underline{\textbf{0.8914}} & \textbf{0.9071} & \textbf{0.9078} & \textbf{0.9334} & \textbf{0.9239} & \textbf{0.9387}\\
9 & \textbf{0.9147} & \textbf{0.9172} & \textbf{0.9493} & \underline{\textbf{0.9117}} & \textbf{0.9316} & \textbf{0.9141} & \textbf{0.9762} & \textbf{0.9203} & \textbf{0.9698}\\
10 & \underline{\textbf{0.8868}} & \textbf{0.8944} & \textbf{0.9227} & \textbf{0.8895} & \textbf{0.9144} & \textbf{0.8919} & \textbf{0.9686} & \textbf{0.8939} & \textbf{0.9621}\\
11 & \underline{\textbf{0.8836}} & \textbf{0.8945} & \textbf{0.9287} & \textbf{0.8872} & \textbf{0.9134} & \textbf{0.8876} & \textbf{0.9658} & \textbf{0.8896} & \textbf{0.9630}\\
12 & \underline{\textbf{0.8774}} & \textbf{0.8917} & \textbf{0.9352} & \textbf{0.8848} & \textbf{0.9131} & \textbf{0.8859} & \textbf{0.9707} & \textbf{0.8919} & \textbf{0.9694}\\\hline
\multicolumn{10}{l}{\scriptsize (b) Success Ratio}\\\hline
 & \multicolumn{3}{c|}{} & \multicolumn{2}{c}{\scriptsize 1 Factor} & \multicolumn{2}{c}{\scriptsize 2 Factors} & \multicolumn{2}{c}{\scriptsize 3 Factors} \\\cline{5-10}
h & BVAR(1) & BVAR(3) & BVAR(12) & BVAR(1) & BVAR(3) & BVAR(1) & BVAR(3) & BVAR(1) & BVAR(3) \\\hline
1 & \textbf{0.5345} & \textbf{0.5517} & \textbf{0.5517} & \textbf{0.5517} & \textbf{0.5517} & \textbf{0.5517} & \underline{\textbf{0.6034}} & \textbf{0.5517} & \textbf{0.5690}\\
2 & \textbf{0.5345} & \textbf{0.5517} & \underline{\textbf{0.5690}} & \textbf{0.5345} & \textbf{0.5517} & \textbf{0.5517} & \underline{\textbf{0.5690}} & \textbf{0.5345} & \textbf{0.5172}\\
3 & \textbf{0.5862} & \textbf{0.6034} & \underline{\textbf{0.6207}} & \textbf{0.6034} & \textbf{0.5690} & \underline{\textbf{0.6207}} & \underline{\textbf{0.6207}} & \textbf{0.5862} & \textbf{0.5862}\\
4 & \underline{\textbf{0.6897}} & \textbf{0.6552} & \textbf{0.6552} & \textbf{0.6552} & \textbf{0.6379} & \textbf{0.6724} & \textbf{0.6724} & \textbf{0.6379} & \textbf{0.6552}\\
5 & \underline{\textbf{0.6379}} & \underline{\textbf{0.6379}} & \textbf{0.6207} & \underline{\textbf{0.6379}} & \textbf{0.6034} & \underline{\textbf{0.6379}} & \textbf{0.6207} & \underline{\textbf{0.6379}} & \underline{\textbf{0.6379}}\\
6 & \underline{\textbf{0.7414}} & \textbf{0.7241} & \textbf{0.7241} & \underline{\textbf{0.7414}} & \textbf{0.7069} & \underline{\textbf{0.7414}} & \textbf{0.6897} & \underline{\textbf{0.7414}} & \textbf{0.7069}\\
7 & \underline{\textbf{0.7759}} & \textbf{0.7586} & \underline{\textbf{0.7759}} & \underline{\textbf{0.7759}} & \textbf{0.7414} & \underline{\textbf{0.7759}} & \textbf{0.7414} & \underline{\textbf{0.7759}} & \textbf{0.7586}\\
8 & \underline{\textbf{0.7931}} & \textbf{0.7759} & \underline{\textbf{0.7931}} & \underline{\textbf{0.7931}} & \textbf{0.7586} & \underline{\textbf{0.7931}} & \textbf{0.7586} & \underline{\textbf{0.7931}} & \textbf{0.7759}\\
9 & \underline{\textbf{0.8103}} & \textbf{0.7931} & \underline{\textbf{0.8103}} & \underline{\textbf{0.8103}} & \textbf{0.7759} & \underline{\textbf{0.8103}} & \textbf{0.7759} & \underline{\textbf{0.8103}} & \textbf{0.7931}\\
10 & \underline{\textbf{0.8276}} & \textbf{0.7931} & \underline{\textbf{0.8276}} & \underline{\textbf{0.8276}} & \textbf{0.7931} & \underline{\textbf{0.8276}} & \textbf{0.7759} & \underline{\textbf{0.8276}} & \textbf{0.7759}\\
11 & \underline{\textbf{0.8966}} & \textbf{0.8621} & \underline{\textbf{0.8966}} & \underline{\textbf{0.8966}} & \textbf{0.8621} & \underline{\textbf{0.8966}} & \textbf{0.8448} & \underline{\textbf{0.8966}} & \textbf{0.8448}\\
12 & \textbf{0.8448} & \textbf{0.8276} & \underline{\textbf{0.8621}} & \underline{\textbf{0.8621}} & \textbf{0.8276} & \textbf{0.8448} & \textbf{0.8103} & \textbf{0.8448} & \textbf{0.8103}\\\hline
\end{tabularx}
\label{tab:Tab02BVAR_nominal}
\caption*{\scriptsize\textit{Notes}: see notes to Table \ref{tab:Tab02AppendixBVARS}}
\end{table}

\FloatBarrier

\subsection{Bayesian VAR models with stochastic volatility}\label{sec:bvarssv}
In this section, we compare the performance of the baseline BVAR models with that of otherwise identical specifications augmented with SV. Full details about specification and estimation are provided in Section \ref{sec:bvarsvestsv}. 

Table \ref{tab:Tab02BVARSV} reports relative RMSFE and success ratios for the real carbon price. Panel (a) shows that allowing for SV does not improve point-forecast accuracy. Across almost all horizons, the homoskedastic specifications deliver lower RMSFEs than their SV counterparts. This difference is especially visible at medium and long horizons, where the homoskedastic BVAR and factor-augmented BVARs often outperform the benchmark, while the corresponding SV models remain close to, or above, one.

Panel (b) of Table \ref{tab:Tab02BVARSV} reports success ratios for sign forecasts. In this case, the evidence is more mixed. At short horizons, SV improves directional accuracy: the BVAR-SV model without factors attains the highest success ratio at horizons one to four, and all SV specifications perform well between horizons three and five. From horizon six onward, however, differences across models become negligible, as most specifications deliver identical or nearly identical success ratios.

To evaluate density performance, we rely on qCRPS \citep{supp:gneiting2011comparing}, where the model with the lowest score is ranked as the most accurate. Note that, contrary to the previous tables, these numbers are not ratios to a benchmark but are expressed in the same scale as real prices. In Panel (a) of Table \ref{tab:Tab03Dens}, we focus on the center of the distribution, while the accuracy in forecasting the right and left tails of the distribution, which are of interest to assess the probability of extreme price movements, is evaluated in Panels (b) and (c), respectively.

With very few exceptions, homoskedastic factor-augmented BVAR models -- especially the specification with one factor -- are more accurate at forecasting the center and both tails of the predictive distribution than the alternative SV specifications. In Panel (a), SV delivers the lowest qCRPS only at the one-month-ahead horizon, while from horizon two onward the best model is always homoskedastic. A similar pattern emerges for the right tail in Panel (b): although the BVAR-SV specification performs best at horizon two, the lowest qCRPS is otherwise obtained by homoskedastic models, most frequently the one-factor BVAR. The evidence for the left tail in Panel (c) is also consistent with this conclusion. Apart from horizons one and six, where the BVAR-SV without factors records the lowest score, the best-performing models are again homoskedastic.

\begin{table}[ht]
\centering
\caption{Relative RMSFE and Success Ratio of Bayesian Vector Autoregressive models with SV}
\newcolumntype{C}[1]{>{\scriptsize\hsize=#1\hsize\centering\arraybackslash}X}
\renewcommand{\arraystretch}{0.8}
\begin{tabularx}{\textwidth}{C{.4}C{1.1}|C{1.1}|C{1.1}|C{1.1}|C{1.1}|C{1.1}}
\multicolumn{7}{l}{\scriptsize (a) Relative RMSFE}\\\hline
 & & 1 Factor & 2 Factors & & 1 Factor & 2 Factors  \\\cline{3-4}\cline{6-7}
h & BVAR(1) & BVAR(1) & BVAR(1) & BVAR-SV(1) & BVAR-SV(1) & BVAR-SV(1) \\\hline
1 & 1.0505 & 1.0604 & 1.0500 & 1.0421 & 1.0498 & 1.0453\\
2 & 1.0337 & 1.0407 & 1.0647 & 1.0378 & 1.0454 & 1.0554\\
3 & 1.0366 & 1.0499 & 1.0468 & 1.0468 & 1.0550 & 1.0546\\
4 & 1.0178 & 1.0130 & 1.0301 & 1.0367 & 1.0432 & 1.0454\\
5 & 1.0259 & 1.0187 & 1.0320 & 1.0517 & 1.0539 & 1.0582\\
6 & \textbf{0.9997} & \underline{\textbf{0.9921}} & 1.0036 & 1.0288 & 1.0317 & 1.0371\\
7 & \textbf{0.9838} & \underline{\textbf{0.9806}} & \textbf{0.9940} & 1.0246 & 1.0263 & 1.0328\\
8 & \textbf{0.9533} & \underline{\textbf{0.9398}} & \textbf{0.9642} & \textbf{0.9998} & \textbf{0.9983} & 1.0020\\
9 & \textbf{0.9667} & \underline{\textbf{0.9610}} & \textbf{0.9695} & 1.0115 & 1.0136 & 1.0128\\
10 & \textbf{0.9455} & \underline{\textbf{0.9448}} & \textbf{0.9539} & \textbf{0.9971} & 1.0002 & \textbf{0.9981}\\
11 & \underline{\textbf{0.9412}} & \textbf{0.9433} & \textbf{0.9509} & 1.0020 & 1.0027 & 1.0010\\
12 & \underline{\textbf{0.9372}} & \textbf{0.9416} & \textbf{0.9492} & 1.0069 & 1.0068 & 1.0074\\\hline
\multicolumn{7}{l}{\scriptsize (b) Success Ratio}\\\hline
 & & 1 Factor & 2 Factors & & 1 Factor & 2 Factors  \\\cline{3-4}\cline{6-7}
h & BVAR(1) & BVAR(1) & BVAR(1) & BVAR-SV(1) & BVAR-SV(1) & BVAR-SV(1) \\\hline
1 & \textbf{0.5517} & \textbf{0.5517} & \textbf{0.5517} & \underline{\textbf{0.5690}} & \textbf{0.5517} & \textbf{0.5172}\\
2 & \textbf{0.5517} & \textbf{0.5517} & \textbf{0.5517} & \underline{\textbf{0.6034}} & \textbf{0.5862} & \underline{\textbf{0.6034}}\\
3 & \textbf{0.5690} & \textbf{0.5517} & \textbf{0.5690} & \underline{\textbf{0.6207}} & \underline{\textbf{0.6207}} & \underline{\textbf{0.6207}}\\
4 & \textbf{0.6034} & \textbf{0.6207} & \textbf{0.6034} & \underline{\textbf{0.6379}} & \underline{\textbf{0.6379}} & \underline{\textbf{0.6379}}\\
5 & \underline{\textbf{0.6379}} & \underline{\textbf{0.6379}} & \textbf{0.6207} & \underline{\textbf{0.6379}} & \underline{\textbf{0.6379}} & \underline{\textbf{0.6379}}\\
6 & \underline{\textbf{0.6897}} & \underline{\textbf{0.6897}} & \underline{\textbf{0.6897}} & \underline{\textbf{0.6897}} & \underline{\textbf{0.6897}} & \underline{\textbf{0.6897}}\\
7 & \underline{\textbf{0.7069}} & \underline{\textbf{0.7069}} & \underline{\textbf{0.7069}} & \underline{\textbf{0.7069}} & \underline{\textbf{0.7069}} & \underline{\textbf{0.7069}}\\
8 & \underline{\textbf{0.7759}} & \underline{\textbf{0.7759}} & \underline{\textbf{0.7759}} & \underline{\textbf{0.7759}} & \underline{\textbf{0.7759}} & \underline{\textbf{0.7759}}\\
9 & \underline{\textbf{0.7931}} & \underline{\textbf{0.7931}} & \underline{\textbf{0.7931}} & \underline{\textbf{0.7931}} & \underline{\textbf{0.7931}} & \underline{\textbf{0.7931}}\\
10 & \underline{\textbf{0.8103}} & \underline{\textbf{0.8103}} & \underline{\textbf{0.8103}} & \underline{\textbf{0.8103}} & \underline{\textbf{0.8103}} & \underline{\textbf{0.8103}}\\
11 & \underline{\textbf{0.8276}} & \underline{\textbf{0.8276}} & \underline{\textbf{0.8276}} & \underline{\textbf{0.8276}} & \underline{\textbf{0.8276}} & \underline{\textbf{0.8276}}\\
12 & \textbf{0.7931} & \textbf{0.7931} & \textbf{0.7931} & \underline{\textbf{0.8103}} & \underline{\textbf{0.8103}} & \underline{\textbf{0.8103}}\\\hline
\end{tabularx}
\label{tab:Tab02BVARSV}
\caption*{\scriptsize\textit{Notes}: see notes to Table \ref{tab:Tab02AppendixBVARS}.}
\end{table}

\begin{table}[ht]
\centering
\caption{Quantile-weighted Continuous Ranked Probability Score (qCRPS)}
\newcolumntype{C}[1]{>{\scriptsize\hsize=#1\hsize\centering\arraybackslash}X}
\renewcommand{\arraystretch}{0.8}
\begin{tabularx}{\textwidth}{C{.4}C{1.1}|C{1.1}|C{1.1}|C{1.1}|C{1.1}|C{1.1}}
\multicolumn{7}{l}{\scriptsize (a) Quantile-weighted CRPS: center}\\\hline
 & & 1 Factor & 2 Factors & & 1 Factor & 2 Factors  \\\cline{3-4}\cline{6-7}
h & BVAR(1) & BVAR(1) & BVAR(1) &  BVAR-SV(1) & BVAR-SV(1) & BVAR-SV(1) \\\hline
1 & 0.5025 & 0.4973 & 0.4919 & \underline{\textbf{0.4900}} & 0.4936 & 0.4925\\
2 & \underline{\textbf{0.7263}} & 0.7276 & 0.7322 & 0.7277 & 0.7315 & 0.7373\\
3 & 0.9037 & 0.9085 & \underline{\textbf{0.9028}} & 0.9173 & 0.9226 & 0.9243\\
4 & 1.0083 & \underline{\textbf{1.0029}} & 1.0175 & 1.0273 & 1.0323 & 1.0407\\
5 & 1.1723 & \underline{\textbf{1.1657}} & 1.1855 & 1.1918 & 1.1922 & 1.2088\\
6 & 1.2594 & \underline{\textbf{1.2506}} & 1.2662 & 1.2703 & 1.2762 & 1.2880\\
7 & 1.3617 & \underline{\textbf{1.3472}} & 1.3676 & 1.3729 & 1.3740 & 1.3887\\
8 & 1.4506 & \underline{\textbf{1.4381}} & 1.4597 & 1.4781 & 1.4766 & 1.4922\\
9 & 1.6191 & \underline{\textbf{1.6149}} & 1.6221 & 1.6503 & 1.6524 & 1.6648\\
10 & \underline{\textbf{1.7170}} & 1.7231 & 1.7290 & 1.7748 & 1.7791 & 1.7920\\
11 & \underline{\textbf{1.8297}} & 1.8391 & 1.8378 & 1.9037 & 1.9059 & 1.9157\\
12 & \underline{\textbf{1.9409}} & 1.9645 & 1.9573 & 2.0465 & 2.0520 & 2.0628\\\hline
\multicolumn{7}{l}{\scriptsize (b) Quantile-weighted CRPS: right tail}\\\hline
 & & 1 Factor & 2 Factors & & 1 Factor & 2 Factors  \\\cline{3-4}\cline{6-7}
h & BVAR(1) & BVAR(1) & BVAR(1) &  BVAR-SV(1) & BVAR-SV(1) & BVAR-SV(1) \\\hline
1 & 0.7592 & 0.7542 & \underline{\textbf{0.7410}} & 0.7457 & 0.7500 & 0.7441\\
2 & 1.1810 & 1.1756 & 1.1829 & \underline{\textbf{1.1724}} & 1.1815 & 1.1835\\
3 & 1.4730 & \underline{\textbf{1.4646}} & 1.4662 & 1.4938 & 1.4976 & 1.5065\\
4 & 1.7025 & \underline{\textbf{1.6841}} & 1.7114 & 1.7495 & 1.7531 & 1.7677\\
5 & 1.9380 & \underline{\textbf{1.9209}} & 1.9553 & 1.9766 & 1.9884 & 2.0026\\
6 & 2.1408 & \underline{\textbf{2.1228}} & 2.1487 & 2.1814 & 2.1988 & 2.2061\\
7 & 2.3171 & \underline{\textbf{2.2929}} & 2.3179 & 2.3617 & 2.3705 & 2.3839\\
8 & 2.4769 & \underline{\textbf{2.4450}} & 2.4698 & 2.5403 & 2.5411 & 2.5594\\
9 & 2.7457 & \underline{\textbf{2.7269}} & 2.7309 & 2.8172 & 2.8234 & 2.8425\\
10 & 2.9955 & \underline{\textbf{2.9837}} & 2.9901 & 3.0953 & 3.1089 & 3.1294\\
11 & 3.2650 & 3.2576 & \underline{\textbf{3.2491}} & 3.3890 & 3.3967 & 3.4118\\
12 & 3.5462 & 3.5620 & \underline{\textbf{3.5432}} & 3.7103 & 3.7282 & 3.7499\\\hline
\multicolumn{7}{l}{\scriptsize (c) Quantile-weighted CRPS: left tail}\\\hline
 & & 1 Factor & 2 Factors & & 1 Factor & 2 Factors  \\\cline{3-4}\cline{6-7}
h & BVAR(1) & BVAR(1) & BVAR(1) &  BVAR-SV(1) & BVAR-SV(1) & BVAR-SV(1) \\\hline
1 & 0.8016 & 0.7983 & 0.7968 & \underline{\textbf{0.7965}} & 0.8024 & 0.8033\\
2 & \underline{\textbf{1.0537}} & 1.0667 & 1.0734 & 1.0876 & 1.0939 & 1.1067\\
3 & \underline{\textbf{1.2809}} & 1.3010 & 1.2883 & 1.3118 & 1.3269 & 1.3230\\
4 & 1.4099 & \underline{\textbf{1.4091}} & 1.4290 & 1.4249 & 1.4380 & 1.4462\\
5 & \underline{\textbf{1.6347}} & 1.6364 & 1.6588 & 1.6605 & 1.6573 & 1.6822\\
6 & 1.7339 & 1.7304 & 1.7511 & \underline{\textbf{1.7287}} & 1.7342 & 1.7498\\
7 & 1.9166 & \underline{\textbf{1.9048}} & 1.9369 & 1.9129 & 1.9189 & 1.9361\\
8 & 2.0647 & \underline{\textbf{2.0544}} & 2.0958 & 2.0864 & 2.0876 & 2.1011\\
9 & 2.2823 & \underline{\textbf{2.2821}} & 2.2989 & 2.3107 & 2.3174 & 2.3185\\
10 & \underline{\textbf{2.3575}} & 2.3712 & 2.3809 & 2.3960 & 2.4049 & 2.4046\\
11 & \underline{\textbf{2.4624}} & 2.4781 & 2.4871 & 2.5083 & 2.5158 & 2.5151\\
12 & \underline{\textbf{2.5460}} & 2.5767 & 2.5805 & 2.6116 & 2.6200 & 2.6258\\\hline
\end{tabularx}
\label{tab:Tab03Dens}
\caption*{\scriptsize\textit{Notes}: Panel (\textit{a}), (\textit{b}) and (\textit{c}) show quantile-weighted Continuous Ranked Probability Scores (CRPS) with weighting that emphasizes the center and the right or the left tail of the distribution, respectively. The best forecasts, associated with the lowest scores, are underlined.}
\end{table}

\FloatBarrier

\subsection{Instabilities in forecasting performance: the fluctuation test}
\label{sec:appFlcutuationTest}

There is widespread evidence that the relative forecasting performance of competing models may vary over time because of parameter instability, shocks with time-varying volatility, and changes in the variance of predictors \citep{supp:giacomini2010forecast,supp:RossiJEL}. In this case, full-sample averages of forecast losses may conceal episodes in which one model temporarily outperforms another, or periods in which its relative performance deteriorates.

To assess whether this issue is relevant in our application, we implement the fluctuation test statistic, $\mathcal{F}_t^{OOS}$, of \citet{supp:giacomini2010forecast}. The statistic is computed as a centered moving average of the loss differential. For point forecasts, the loss differential is defined as the standardized difference between the squared forecast error of a benchmark model and that of a factor-augmented BVAR(1) model. Positive values of the statistic indicate that the BFAVAR model has a lower MSFE than the benchmark. The tests are one-sided: the null hypothesis is equal MSFE, while the alternative is that the BFAVAR model is more accurate. The dashed line in the figures reports the 5\% critical value, $CV_{0.05}$; the null is rejected when $\max \mathcal{F}_t^{OOS}>CV_{0.05}$.\footnote{The test has a nonstandard distribution. The critical values provided by \citet{supp:giacomini2010forecast} depend on the ratio between the window length and the number of observations in the evaluation sample. We consider centered rolling windows of 19 and 25 months.}

We focus on three forecast horizons: one month, one quarter, and one year ahead. To provide a demanding comparison, we evaluate the BFAVAR(1) models not only against the RW, but also against the BAR(1) and BVAR(1), which the main text shows to be more competitive benchmarks.

\begin{figure}[h!]
    \centering
    \caption{Fluctuation test statistic from December 2019 to September 2022.}
   \includegraphics[width=0.95\linewidth]{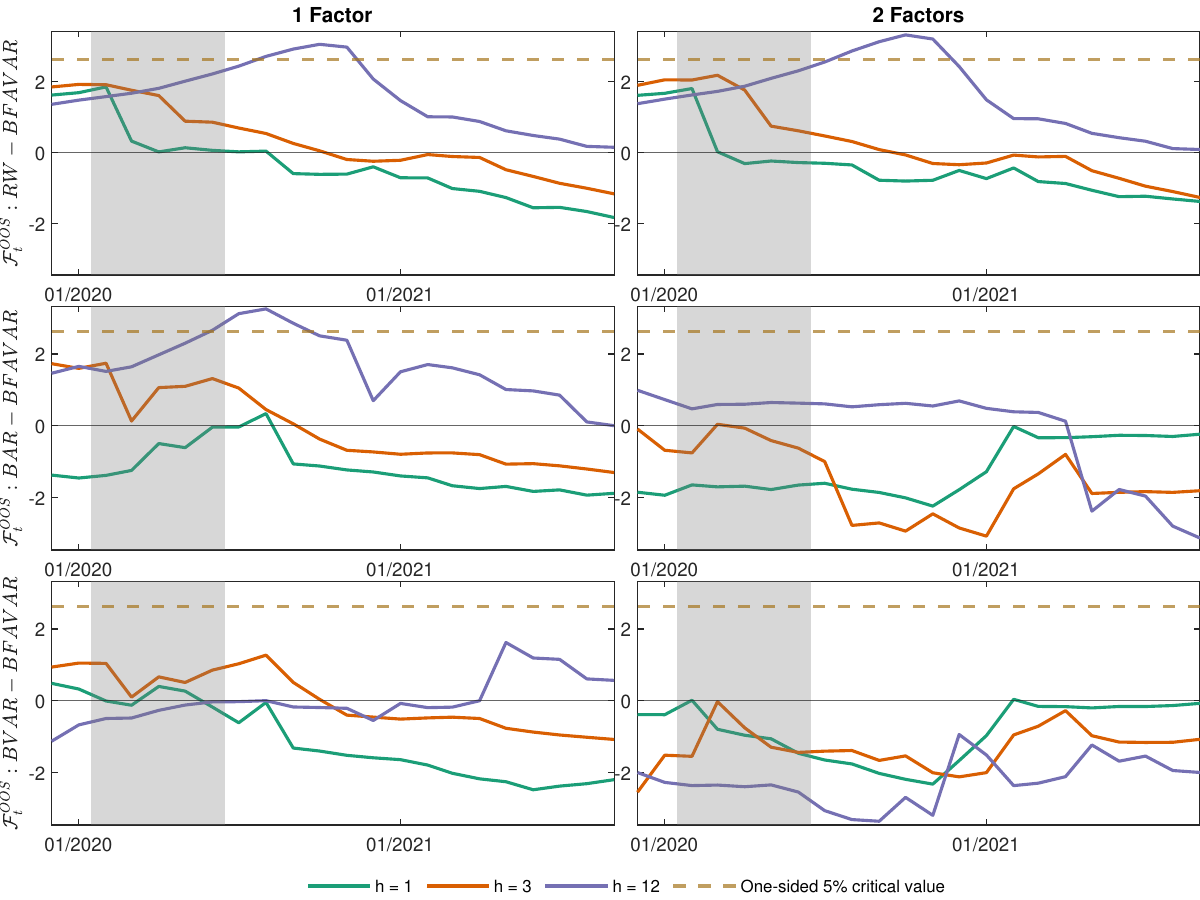}
    \label{fig:FigGW24}
        \caption*{\scriptsize{\textit{Notes:} fluctuation test statistic of \citet{supp:giacomini2010forecast} obtained with a 25-month centered rolling window.}}
\end{figure}

Figure \ref{fig:FigGW24} reports fluctuation-test statistics for BFAVAR(1) specifications with one and two factors, using a 25-month centered rolling window. The relative performance of the factor-augmented models varies over time, the one-factor specification is generally more competitive than the two-factor specification, and the evidence in favor of factor augmentation is stronger at medium and long horizons than at the one-month horizon. The longer window smooths some of the short-run fluctuations in the test statistic, but it does not alter the conclusion that relative forecast accuracy deteriorates toward the end of 2022.

We also examine instability in density forecast accuracy. For this purpose, we compute fluctuation-test statistics using the qCRPS instead of squared forecast errors. Specifically, the statistic is obtained as the standardized difference between the qCRPS of the benchmark model and the qCRPS of the BFAVAR(1) model. Positive values therefore indicate that the BFAVAR model delivers a lower qCRPS than the benchmark. We consider losses that emphasize the center of the predictive distribution, the left tail, and the right tail.

\begin{figure}[h!]
    \centering
    \caption{Fluctuation test statistic, qCRPS, with center weights from September 2019 to December 2022 calculated with a centered rolling window of size 19 months.}
   \includegraphics[width=0.95\linewidth]{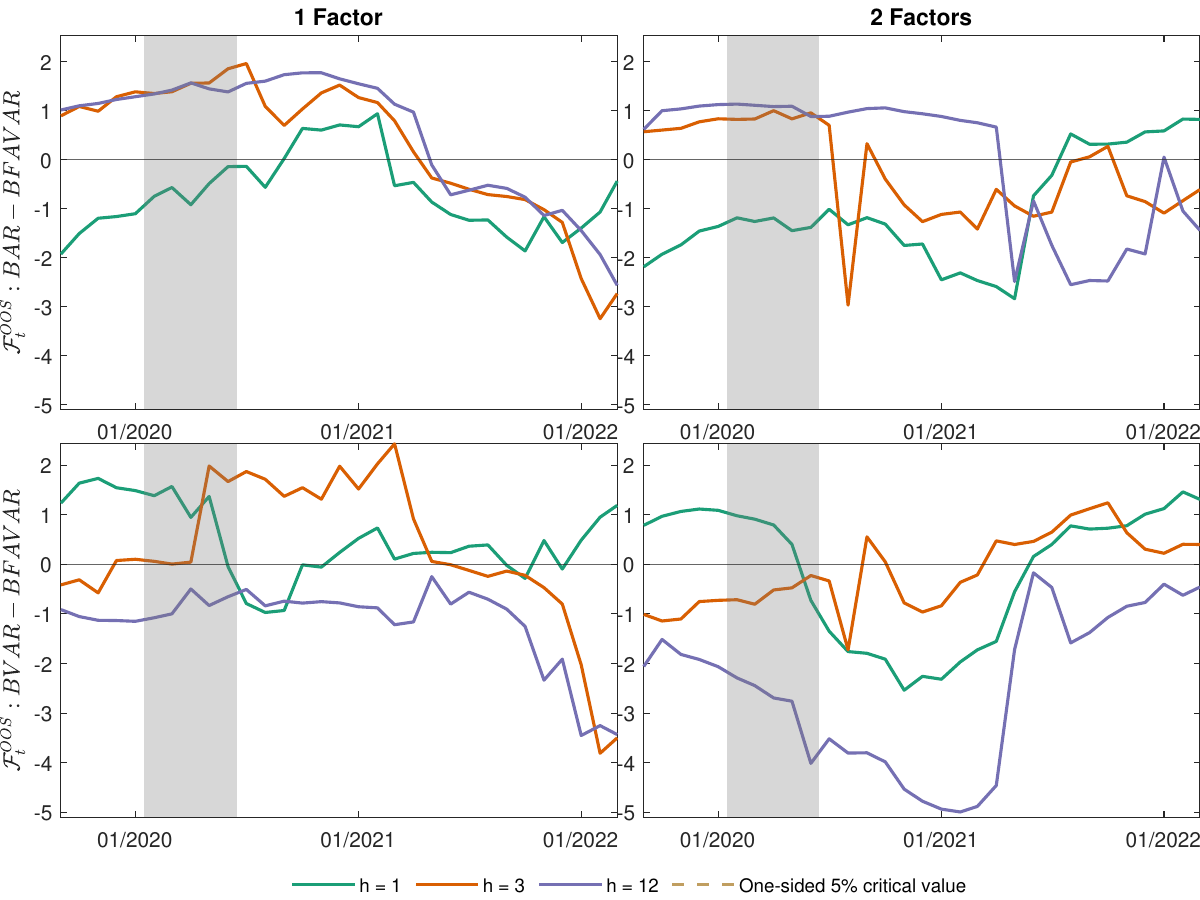}
    \label{fig:FigGW18c}
\end{figure}
\begin{figure}[H]
    \centering
    \caption{Fluctuation test statistic, qCRPS, with center weights from September 2019 to December 2022 calculated with a centered rolling window of size 25 months.}
   \includegraphics[width=0.95\linewidth]{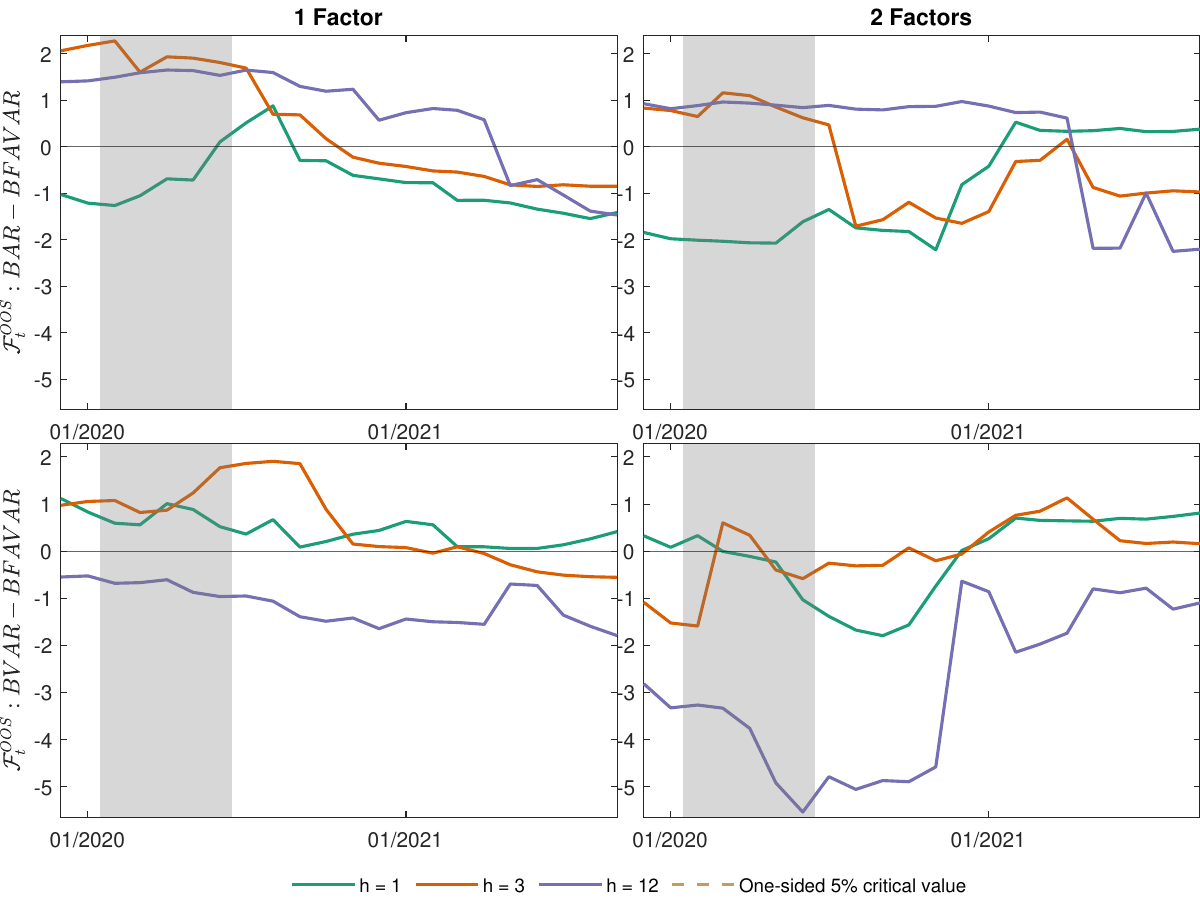}
    \label{fig:FigGW24c}
\end{figure}
\begin{figure}[h!]
    \centering
    \caption{Fluctuation test statistic, qCRPS, with left tails weights from September 2019 to December 2022 calculated with a centered rolling window of size 19 months.}
   \includegraphics[width=0.95\linewidth]{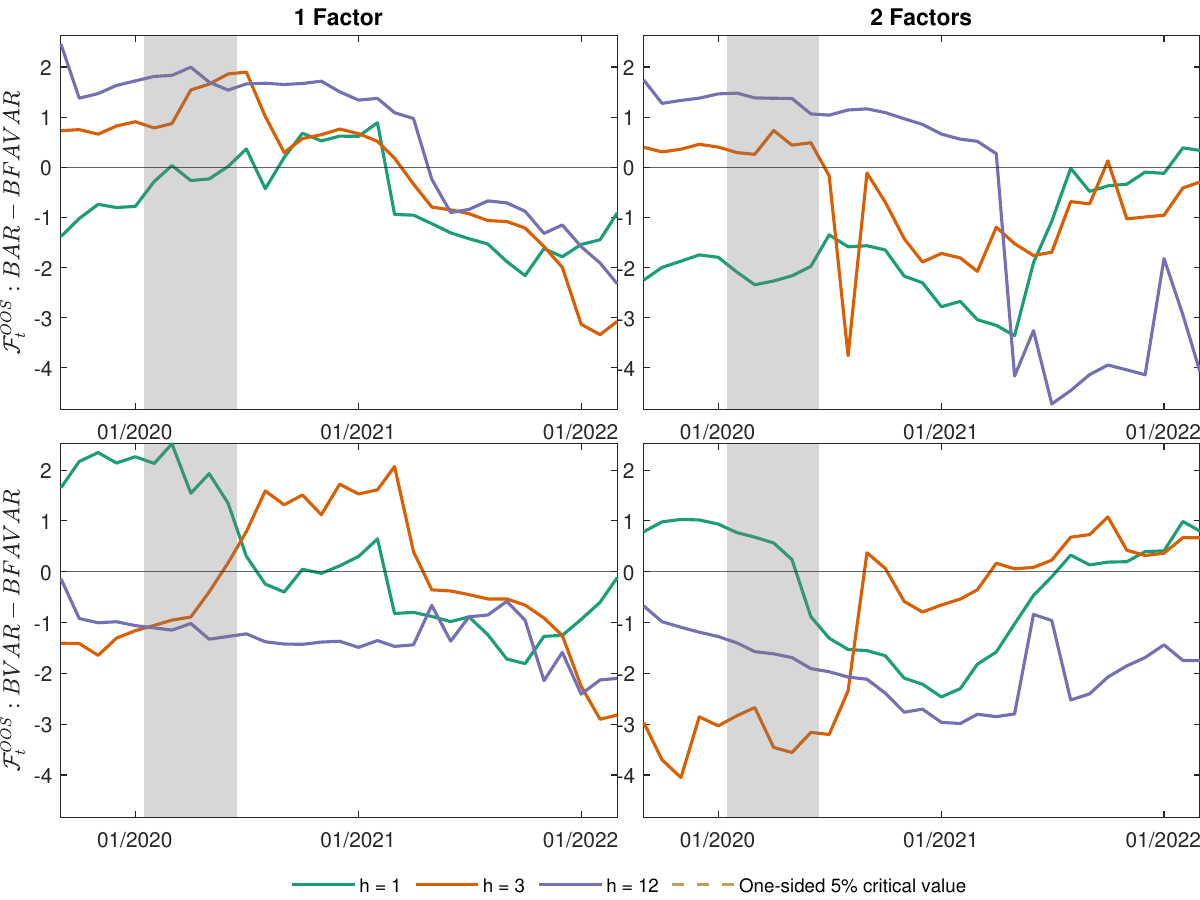}
    \label{fig:FigGW18LT}
\end{figure}
\begin{figure}[H]
    \centering
    \caption{Fluctuation test statistic, qCRPS, with left tails weights from September 2019 to December 2022 calculated with a centered rolling window of size 25 months.}
   \includegraphics[width=0.95\linewidth]{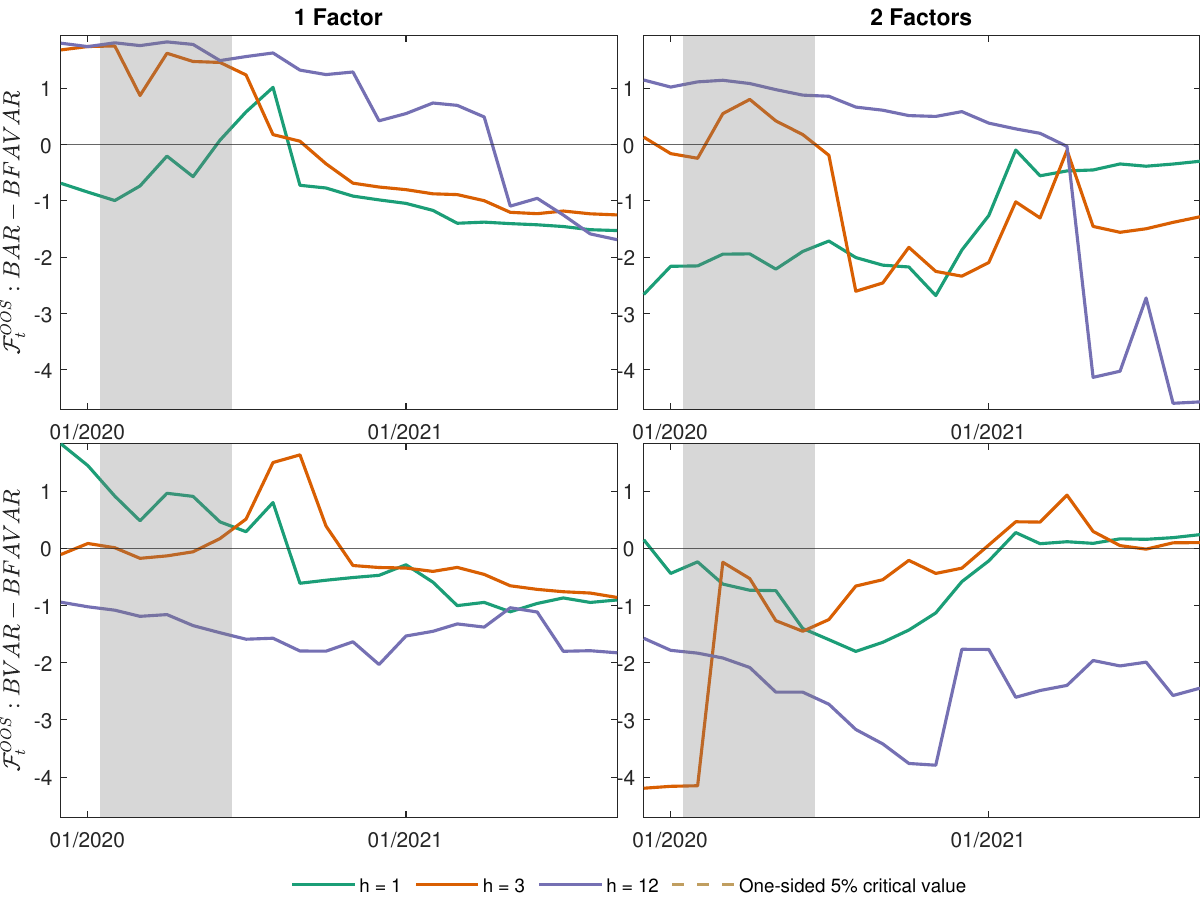}
    \label{fig:FigGW24LT}
\end{figure}
\begin{figure}[H]
    \centering
    \caption{Fluctuation test statistic for the right tail for BFAVAR with 1 factor (left) and 2 factors (right). September 2019 to December 2022.}
   \includegraphics[width=0.9\linewidth]{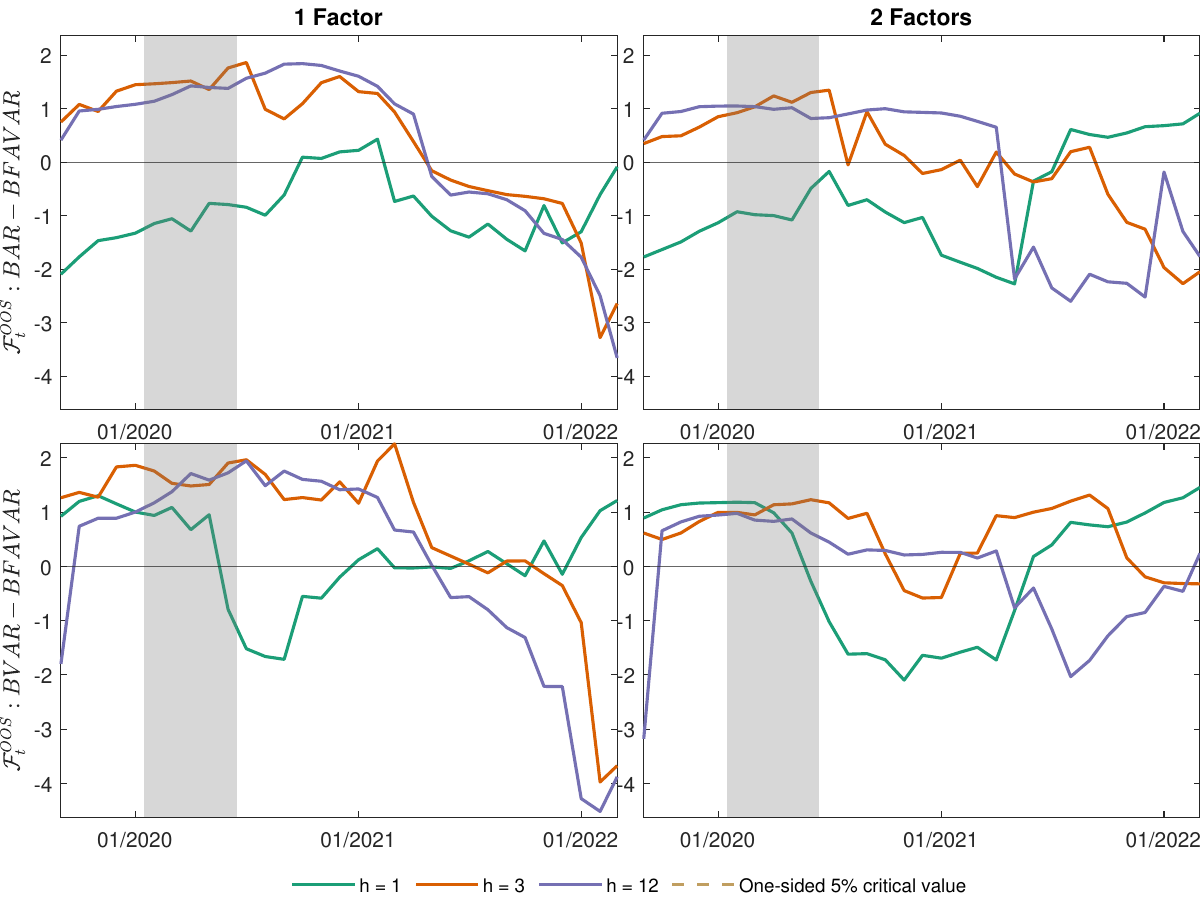}
    \label{fig:FigGW18RT}
\end{figure}
\begin{figure}[H]
    \centering
    \caption{Fluctuation test statistic, qCRPS, with right tails weights from September 2019 to December 2022 calculated with a centered rolling window of size 25 months.}
   \includegraphics[width=1\linewidth]{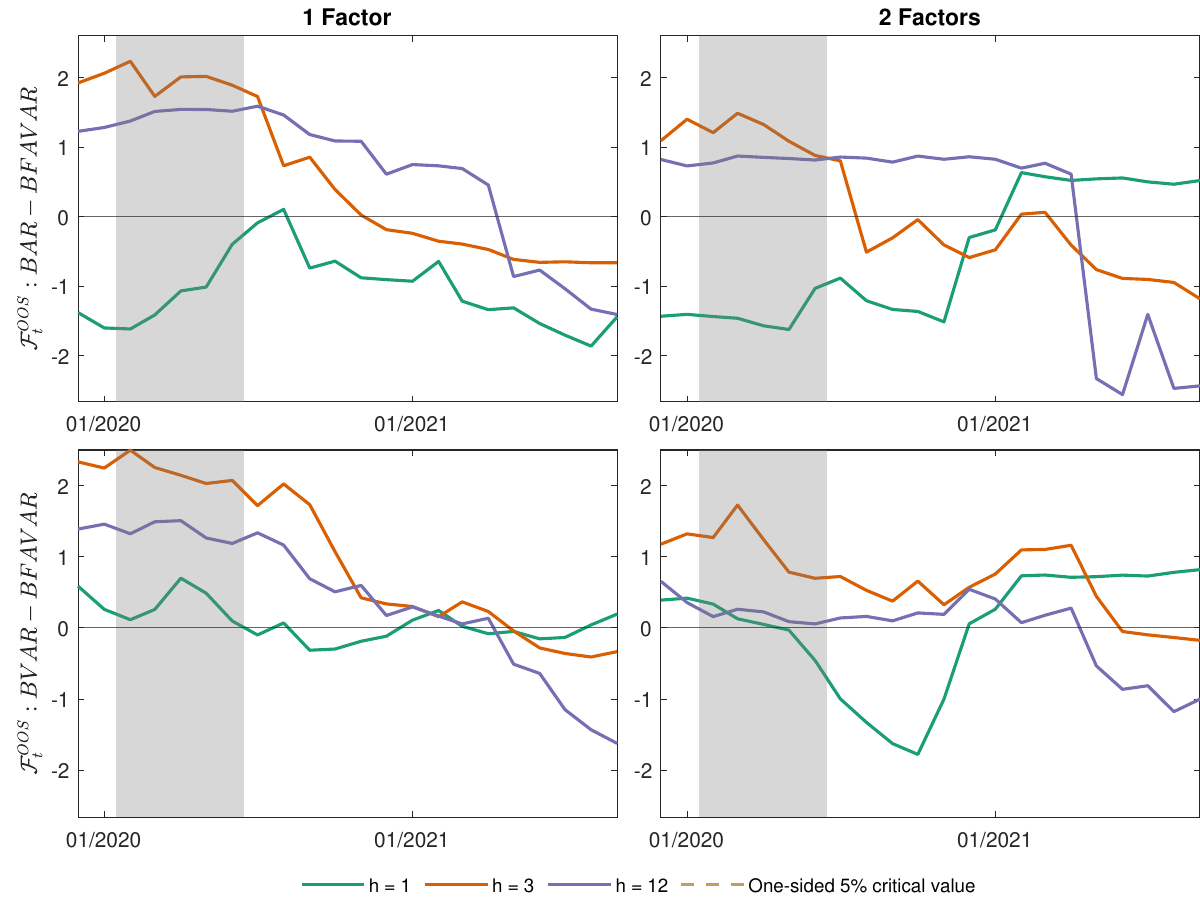}
    \label{fig:FigGW24RT}
\end{figure}

Figures \ref{fig:FigGW18c} and \ref{fig:FigGW24c} report the fluctuation-test statistics for the center of the predictive distribution, using 19- and 25-month windows, respectively. Figures \ref{fig:FigGW18LT} and \ref{fig:FigGW24LT} repeat the exercise for the left tail, while Figures \ref{fig:FigGW18RT} and \ref{fig:FigGW24RT} focus on the right tail. The density-forecast results reinforce the evidence from point forecasts: relative performance is unstable over time, and the one-factor BFAVAR model delivers gains during some parts of the evaluation sample. For right-tail forecast losses, the one-factor BFAVAR model often improves upon the BAR(1) and BVAR(1), especially at horizons $h=1$ and $h=3$, while evidence is weaker at $h=12$. Results based on the 25-month window are smoother but qualitatively similar.

Overall, the fluctuation-test evidence suggests that the factor-augmented BVAR models contain useful predictive information, especially beyond very short horizons and for some regions of the predictive distribution. At the same time, their relative performance is not stable over time and deteriorates around late 2022. This motivates the forecast-combination exercise in the main text, where we ask whether simple pooling rules can mitigate this instability in practice.

\FloatBarrier

\subsection{Forecast pooling}\label{sec:pooling}
The fluctuation-test evidence reported in the Supplement shows that the relative performance of the factor-augmented BVAR models varies over time. In particular, the one-factor BFAVAR(1) performs well during parts of the evaluation sample, especially at medium and long horizons, but its relative accuracy deteriorates toward the end of 2022. We therefore examine whether such instability can be mitigated in practice by combining forecasts across parsimonious specifications.

Forecast combinations have a long tradition in econometrics. The classic contribution of \citet{supp:bates1969combination} shows that combining forecasts may reduce mean-squared forecast errors relative to individual forecasts. We focus on simple combination rules rather than estimating time-varying weights. This choice is motivated by the short evaluation sample. It is also consistent with evidence from energy-commodity forecasting showing that simple pooling methods often outperform individual models \citep{supp:aag2025forecasting,supp:baumeister2015forecasting}. We consider two simple combination rules. The first is the simple average, computed as the equal-weight average of the forecasts in a given model pool. The second is the median forecast, computed as the median across the forecasts in the same pool. The median is a robust combination rule because it is less sensitive than the mean to occasional outlying model forecasts \citep{supp:stock1999comparison,supp:StockWatson2004}.

We apply the two combination rules to three pools of parsimonious lag-one models.\footnote{We exclude specifications that the previous results show to be less competitive, such as higher-lag models, large BVARs including all predictors individually, and stochastic-volatility models for the real price of carbon.} The first pool (\textit{Pool 1}) combines forecasts from the AR(1), the BVAR(1), and the one-factor BFAVAR(1). This pool averages across three forecasting approaches: univariate price persistence, small multivariate EU ETS dynamics, and macro-financial factor information. The second pool (\textit{Pool 2}) combines the factor-augmented BVAR(1) models with one, two, and three factors. This pool addresses uncertainty about the number of factors needed to summarize the predictor panel. The third pool (\textit{Pool 3}) combines all five parsimonious lag-one specifications: the AR(1), the BVAR(1), and the BFAVAR(1) models with one, two, and three factors. This is the broadest pooling specification we consider.

For each pool, we compute both the mean and the median forecast. This gives six pooled forecasts: Mean-Pool 1, Median-Pool 1, Mean-Pool 2, Median-Pool 2, Mean-Pool 3, and Median-Pool 3. We evaluate them using the RMSFE relative to the RW benchmark.

\begin{table}[ht]
\centering
\caption{Relative RMSFE of individual models and forecast pools}
\newcolumntype{C}[1]{>{\scriptsize\hsize=#1\hsize\centering\arraybackslash}X}
\renewcommand{\arraystretch}{0.8}
\begin{tabularx}{\textwidth}{C{.45}C{1}C{1}C{1}C{1}C{1}}
\hline
\multicolumn{6}{l}{\scriptsize (a) Individual models}\\\hline
$h$ & AR(1) & BVAR(1) & BFAVAR(1)-1F & BFAVAR(1)-2F & BFAVAR(1)-3F \\\hline
1 & \underline{1.0437} & 1.0505 & 1.0604 & 1.0500 & 1.0577\\
2 & \underline{1.0324} & 1.0337 & 1.0407 & 1.0647 & 1.0816\\
3 & \underline{1.0355} & 1.0366 & 1.0499 & 1.0468 & 1.0725\\
4 & 1.0127 & 1.0178 & 1.0130 & 1.0301 & 1.0481\\
5 & 1.0193 & 1.0259 & 1.0187 & 1.0320 & 1.0492\\
6 & \underline{\textbf{0.9870}} & \textbf{0.9997} & \textbf{0.9921} & 1.0036 & 1.0184\\
7 & \textbf{0.9813} & \textbf{0.9838} & \textbf{0.9806} & \textbf{0.9940} & 1.0093\\
8 & \textbf{0.9566} & \textbf{0.9533} & \underline{\textbf{0.9398}} & \textbf{0.9642} & \textbf{0.9783}\\
9 & \textbf{0.9695} & \textbf{0.9667} & \underline{\textbf{0.9610}} & \textbf{0.9695} & \textbf{0.9843}\\
10 & \textbf{0.9514} & \textbf{0.9455} & \textbf{0.9448} & \textbf{0.9539} & \textbf{0.9637}\\
11 & \textbf{0.9499} & \underline{\textbf{0.9412}} & \textbf{0.9433} & \textbf{0.9509} & \textbf{0.9574}\\
12 & \textbf{0.9495} & \underline{\textbf{0.9372}} & \textbf{0.9416} & \textbf{0.9492} & \textbf{0.9558}\\\hline
\multicolumn{6}{l}{\scriptsize (b) Mean forecast pools}\\\hline
$h$ & Mean-Pool 1 & Mean-Pool 2 & Mean-Pool 3 & & \\\hline
1 & 1.0500 & 1.0538 & 1.0498 & & \\
2 & 1.0335 & 1.0603 & 1.0479 & & \\
3 & 1.0381 & 1.0546 & 1.0454 & & \\
4 & \underline{1.0116} & 1.0289 & 1.0215 & & \\
5 & 1.0183 & 1.0320 & 1.0264 & & \\
6 & \textbf{0.9898} & 1.0034 & \textbf{0.9975} & & \\
7 & \underline{\textbf{0.9789}} & \textbf{0.9936} & \textbf{0.9873} & & \\
8 & \textbf{0.9470} & \textbf{0.9598} & \textbf{0.9561} & & \\
9 & \textbf{0.9631} & \textbf{0.9709} & \textbf{0.9682} & & \\
10 & \underline{\textbf{0.9445}} & \textbf{0.9534} & \textbf{0.9499} & & \\
11 & \textbf{0.9422} & \textbf{0.9499} & \textbf{0.9467} & & \\
12 & \textbf{0.9402} & \textbf{0.9483} & \textbf{0.9449} & & \\\hline
\multicolumn{6}{l}{\scriptsize (c) Median forecast pools}\\\hline
$h$ & Median-Pool 1 & Median-Pool 2 & Median-Pool 3 & & \\\hline
1 & 1.0481 & 1.0506 & 1.0585 & & \\
2 & 1.0344 & 1.0667 & 1.0474 & & \\
3 & 1.0375 & 1.0560 & 1.0460 & & \\
4 & 1.0120 & 1.0317 & 1.0189 & & \\
5 & \underline{1.0181} & 1.0335 & 1.0261 & & \\
6 & \textbf{0.9880} & 1.0063 & \textbf{0.9964} & & \\
7 & \textbf{0.9799} & \textbf{0.9972} & \textbf{0.9874} & & \\
8 & \textbf{0.9509} & \textbf{0.9675} & \textbf{0.9554} & & \\
9 & \textbf{0.9660} & \textbf{0.9764} & \textbf{0.9675} & & \\
10 & \textbf{0.9476} & \textbf{0.9571} & \textbf{0.9480} & & \\
11 & \textbf{0.9445} & \textbf{0.9525} & \textbf{0.9466} & & \\
12 & \textbf{0.9441} & \textbf{0.9498} & \textbf{0.9446} & & \\\hline
\end{tabularx}
\label{tab:TabPoolingRMSFE}
\caption*{\scriptsize\textit{Notes}: Relative RMSFE with respect to the random walk. Bold entries denote RMSFE ratios below one. Underlined entries denote the best-performing specification at each horizon among all individual and pooled forecasts. $^{\ast}$ and $^{\ast\ast}$ indicate rejection of equal predictive accuracy against the random walk at the 10\% and 5\% levels, respectively, using the one-sided fixed-$b$ Diebold--Mariano test.}
\end{table}

Table \ref{tab:TabPoolingRMSFE} reports the relative RMSFE of the individual models and pooled forecasts. The results confirm that pooling is not useful at very short horizons. For $h=1,\ldots,5$, all individual and pooled forecasts have relative RMSFE above one, indicating that the random walk remains difficult to beat. This is consistent with the previous evidence that multivariate information provides limited gains at very short horizons.

Forecast gains emerge from medium horizons onward. From $h=6$, most pooled forecasts deliver relative RMSFE below one. The strongest results are obtained for \textit{Pool 1}, which combines the AR(1), the BVAR(1), and the one-factor BFAVAR(1). In particular, Mean-Pool 1 is the best-performing specification at horizons $h=7$ and $h=10$, and remains very close to the best individual model at longer horizons. At $h=12$, Mean-Pool 1 has a relative RMSFE of 0.9402, close to both the BVAR(1) and the one-factor BFAVAR(1), whose relative RMSFEs are 0.9372 and 0.9416, respectively.

The comparison across pools is informative. \textit{Pool 2}, which averages across BFAVAR(1) models with one, two, and three factors, does not improve upon the one-factor BFAVAR. \textit{Pool 3}, which combines all five parsimonious lag-one specifications, performs well at medium and long horizons, but generally does not improve upon \textit{Pool 1}. Thus, expanding the pool beyond the AR(1), BVAR(1), and one-factor BFAVAR(1) tends to worsen rather than strengthen forecast accuracy.

The median combinations are competitive but do not systematically dominate the simple averages. Overall, the table shows that forecast averaging delivers competitive forecasts at medium and long horizons. Importantly, the best pooling results are obtained when the one-factor BFAVAR is combined with the parsimonious AR(1) and BVAR(1) benchmarks.

\begin{figure}[!htbp]
\centering
\includegraphics[width=1\linewidth]{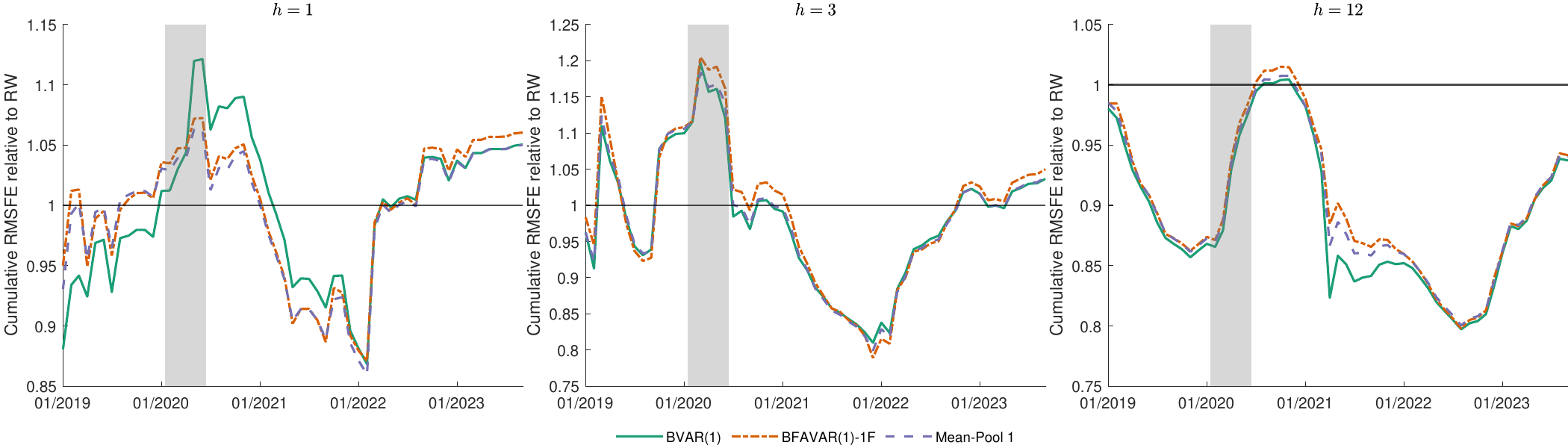}
\caption{Cumulative RMSFE ratios of individual and pooled forecasts relative to the random walk.}
\caption*{\scriptsize\textit{Notes}: The figure reports cumulative RMSFE ratios for the BVAR(1), the one-factor BFAVAR(1), and Mean-Pool 1, defined as the simple average of the AR(1), BVAR(1), and one-factor BFAVAR(1) forecasts. Values below one indicate that a model has accumulated smaller forecast errors than the random walk up to that date.}
\label{fig:FigPooling}
\end{figure}

Figure \ref{fig:FigPooling} complements the full-sample results by plotting cumulative RMSFE ratios relative to the RW for the BVAR(1), the one-factor BFAVAR(1), and Mean-Pool 1. The figure confirms that relative forecast accuracy changes substantially over the evaluation sample. The one-factor BFAVAR(1) performs well before and after the COVID-19 recession, but its cumulative performance deteriorates around 2022. Mean-Pool 1 does not fully offset this deterioration, but it remains competitive and broadly tracks the best-performing individual specifications. Thus, pooling does not mitigate the instability documented by the fluctuation tests, while confirming that the useful information in the factor-augmented specification is concentrated in the parsimonious one-factor BFAVAR.
\FloatBarrier

\subsection{The role of verified emissions in point and density forecasting}\label{sec:supp-emissions}
Table \ref{tab:Tab03RMSFENoEmiss} reports relative RMSFE and Success Ratios for the BVAR(1) and BFAVAR(1) specifications with one and two factors, estimated with and without interpolated verified emissions. Differences between the two sets of models are small across all horizons, with no systematic pattern favoring either specification. If anything, models without verified emissions record marginally lower RMSFE ratios at some short horizons, while results become virtually indistinguishable at longer horizons. Success Ratios are also very similar across specifications. Overall, the results indicate that interpolated verified emissions do not materially affect point forecast accuracy, alleviating concerns that the interpolation process induces a look-ahead bias in the main results.

Table \ref{tab:Tab03DensNoEmiss} reports quantile-weighted CRPS scores for the center, right tail, and left tail of the predictive distribution for the same set of models. Differences in qCRPS between models estimated with and without interpolated verified emissions are small in absolute terms at all horizons and across all panels, with no systematic pattern favoring either specification. If anything, models without verified emissions record marginally lower scores at several horizons, most visibly for the center and the right tail of the distribution at medium horizons. 

Overall, the point and density forecast results indicate that interpolated verified emissions do not materially affect forecast accuracy, alleviating concerns that the interpolation process induces a look-ahead bias or otherwise drives the main findings.

\begin{table}[ht]
\centering
\caption{Relative RMSFE and Success Ratio: the role of verified emissions}
\newcolumntype{C}[1]{>{\scriptsize\hsize=#1\hsize\centering\arraybackslash}X}
\begin{tabularx}{\textwidth}{C{.4}C{1.1}|C{1.1}|C{1.1}|C{1.1}|C{1.1}|C{1.1}}
\hline
\multicolumn{7}{l}{\scriptsize (a) Relative RMSFE}\\\hline
 & \multicolumn{3}{c}{\scriptsize With Verified Emissions} & \multicolumn{3}{c}{\scriptsize Without Verified Emissions} \\\cline{2-7}
 & & 1 Factor & 2 Factors & & 1 Factor & 2 Factors \\\cline{3-4}\cline{6-7}
h & BVAR(1) & BVAR(1) & BVAR(1) & BVAR(1) & BVAR(1) & BVAR(1) \\\hline
1 & 1.0505 & 1.0604 & 1.0500 & 1.0498 & 1.0596 & \underline{1.0486} \\
2 & 1.0337 & 1.0407 & 1.0647 & \underline{1.0296} & 1.0371 & 1.0597 \\
3 & 1.0366 & 1.0499 & 1.0468 & \underline{1.0260} & 1.0521 & 1.0433 \\
6 & \textbf{0.9997} & \textbf{0.9921} & 1.0036 & \underline{\textbf{0.9900}} & \textbf{0.9918} & 1.0098 \\
9 & \textbf{0.9667} & \underline{\textbf{0.9610}} & \textbf{0.9695} & \textbf{0.9612} & \textbf{0.9699} & \textbf{0.9736} \\
12 & \underline{\textbf{0.9372}} & \textbf{0.9416} & \textbf{0.9492} & \textbf{0.9389} & \textbf{0.9545} & \textbf{0.9535} \\\hline
\multicolumn{7}{l}{\scriptsize (b) Success Ratio}\\\hline
 & \multicolumn{3}{c}{\scriptsize With Verified Emissions} & \multicolumn{3}{c}{\scriptsize Without Verified Emissions} \\\cline{2-7}
 & & 1 Factor & 2 Factors & & 1 Factor & 2 Factors \\\cline{3-4}\cline{6-7}
h & BVAR(1) & BVAR(1) & BVAR(1) & BVAR(1) & BVAR(1) & BVAR(1) \\\hline
1 & \textbf{0.5517} & \textbf{0.5517} & \textbf{0.5517} & \underline{\textbf{0.5690}} & \underline{\textbf{0.5690}} & \textbf{0.5172} \\
2 & \underline{\textbf{0.5517}} & \underline{\textbf{0.5517}} & \underline{\textbf{0.5517}} & \underline{\textbf{0.5517}} & \underline{\textbf{0.5517}} & \underline{\textbf{0.5517}} \\
3 & \underline{\textbf{0.5690}} & \textbf{0.5517} & \underline{\textbf{0.5690}} & \textbf{0.5517} & \underline{\textbf{0.5690}} & \underline{\textbf{0.5690}} \\
6 & \underline{\textbf{0.6897}} & \underline{\textbf{0.6897}} & \underline{\textbf{0.6897}} & \underline{\textbf{0.6897}} & \underline{\textbf{0.6897}} & \underline{\textbf{0.6897}} \\
9 & \underline{\textbf{0.7931}} & \underline{\textbf{0.7931}} & \underline{\textbf{0.7931}} & \underline{\textbf{0.7931}} & \underline{\textbf{0.7931}} & \underline{\textbf{0.7931}} \\
12 & \underline{\textbf{0.7931}} & \underline{\textbf{0.7931}} & \underline{\textbf{0.7931}} & \underline{\textbf{0.7931}} & \underline{\textbf{0.7931}} & \underline{\textbf{0.7931}} \\\hline
\end{tabularx}
\label{tab:Tab03RMSFENoEmiss}
\caption*{\scriptsize\textit{Notes}: Panel (a) reports relative RMSFE. Values below one are in bold and the lowest score at each horizon is underlined. $^\ast$ ($^{\ast\ast}$) denotes rejection of the one-sided Diebold--Mariano test at the 10\% (5\%) significance level. Panel (b) reports Success Ratios. Values above 0.5 are in bold and the highest value at each horizon is underlined. $^\ast$ denotes a $p$-value below 0.1 for the Pesaran--Timmermann test.}
\end{table}

\begin{table}[ht]
\centering
\caption{Quantile-weighted Continuous Ranked Probability Score (CRPS): the role of verified emissions}
\newcolumntype{C}[1]{>{\scriptsize\hsize=#1\hsize\centering\arraybackslash}X}
\renewcommand{\arraystretch}{0.8}
\begin{tabularx}{\textwidth}{C{.4}C{1.1}|C{1.1}|C{1.1}|C{1.1}|C{1.1}|C{1.1}}
\multicolumn{7}{l}{\scriptsize (a) Quantile-weighted CRPS: center}\\\hline
 & \multicolumn{3}{c}{\scriptsize With Verified Emissions} & \multicolumn{3}{c}{\scriptsize Without Verified Emissions}  \\\cline{2-7}
 & & 1 Factor & 2 Factors & & 1 Factor & 2 Factors  \\\cline{3-4}\cline{6-7}
h & BVAR(1) & BVAR(1) & BVAR(1) &  BVAR(1) & BVAR(1) & BVAR(1) \\\hline
1 & 0.5025 & 0.4973 & 0.4919 & 0.4966 & 0.4975 & \underline{\textbf{0.4907}}\\
2 & 0.7263 & 0.7276 & 0.7322 & \underline{\textbf{0.7205}} & 0.7249 & 0.7339\\
3 & 0.9037 & 0.9085 & 0.9028 & \underline{\textbf{0.8918}} & 0.9117 & 0.9026\\
4 & 1.0083 & 1.0029 & 1.0175 & \underline{\textbf{0.9964}} & 1.0054 & 1.0152\\
5 & 1.1723 & 1.1657 & 1.1855 & \underline{\textbf{1.1639}} & 1.1672 & 1.1785\\
6 & 1.2594 & 1.2506 & 1.2662 & 1.2460 & \underline{\textbf{1.2453}} & 1.2613\\
7 & 1.3617 & 1.3472 & 1.3676 & 1.3461 & \underline{\textbf{1.3436}} & 1.3643\\
8 & 1.4506 & 1.4381 & 1.4597 & 1.4380 & \underline{\textbf{1.4376}} & 1.4631\\
9 & 1.6191 & 1.6149 & 1.6221 & \underline{\textbf{1.6057}} & 1.6146 & 1.6210\\
10 & 1.7170 & 1.7231 & 1.7290 & \underline{\textbf{1.7064}} & 1.7286 & 1.7295\\
11 & 1.8297 & 1.8391 & 1.8378 & \underline{\textbf{1.8219}} & 1.8483 & 1.8456\\
12 & 1.9409 & 1.9645 & 1.9573 & \underline{\textbf{1.9378}} & 1.9733 & 1.9694\\\hline
\multicolumn{7}{l}{\scriptsize (b) Quantile-weighted CRPS: right tail}\\\hline
 & \multicolumn{3}{c}{\scriptsize With Verified Emissions} & \multicolumn{3}{c}{\scriptsize Without Verified Emissions}  \\\cline{2-7}
 & & 1 Factor & 2 Factors & & 1 Factor & 2 Factors  \\\cline{3-4}\cline{6-7}
h & BVAR(1) & BVAR(1) & BVAR(1) &  BVAR(1) & BVAR(1) & BVAR(1) \\\hline
1 & 0.7592 & 0.7542 & 0.7410 & 0.7488 & 0.7546 & \underline{\textbf{0.7396}}\\
2 & 1.1810 & 1.1756 & 1.1829 & 1.1659 & \underline{\textbf{1.1657}} & 1.1842\\
3 & 1.4730 & 1.4646 & 1.4662 & \underline{\textbf{1.4460}} & 1.4604 & 1.4630\\
4 & 1.7025 & 1.6841 & 1.7114 & \underline{\textbf{1.6779}} & 1.6786 & 1.7063\\
5 & 1.9380 & 1.9209 & 1.9553 & 1.9166 & \underline{\textbf{1.9139}} & 1.9397\\
6 & 2.1408 & 2.1228 & 2.1487 & 2.1155 & \underline{\textbf{2.1092}} & 2.1420\\
7 & 2.3171 & 2.2929 & 2.3179 & 2.2857 & \underline{\textbf{2.2691}} & 2.3114\\
8 & 2.4769 & 2.4450 & 2.4698 & 2.4372 & \underline{\textbf{2.4224}} & 2.4669\\
9 & 2.7457 & 2.7269 & 2.7309 & \underline{\textbf{2.7033}} & 2.7110 & 2.7243\\
10 & 2.9955 & 2.9837 & 2.9901 & \underline{\textbf{2.9542}} & 2.9823 & 2.9817\\
11 & 3.2650 & 3.2576 & 3.2491 & \underline{\textbf{3.2257}} & 3.2669 & 3.2574\\
12 & 3.5462 & 3.5620 & 3.5432 & \underline{\textbf{3.5136}} & 3.5691 & 3.5619\\\hline
\multicolumn{7}{l}{\scriptsize (c) Quantile-weighted CRPS: left tail}\\\hline
 & \multicolumn{3}{c}{\scriptsize With Verified Emissions} & \multicolumn{3}{c}{\scriptsize Without Verified Emissions}  \\\cline{2-7}
 & & 1 Factor & 2 Factors & & 1 Factor & 2 Factors  \\\cline{3-4}\cline{6-7}
h & BVAR(1) & BVAR(1) & BVAR(1) &  BVAR(1) & BVAR(1) & BVAR(1) \\\hline
1 & 0.8016 & 0.7983 & 0.7968 & \underline{\textbf{0.7967}} & 0.7976 & 0.7979\\
2 & \underline{\textbf{1.0537}} & 1.0667 & 1.0734 & 1.0537 & 1.0670 & 1.0781\\
3 & 1.2809 & 1.3010 & 1.2883 & \underline{\textbf{1.2729}} & 1.3080 & 1.2873\\
4 & 1.4099 & 1.4091 & 1.4290 & \underline{\textbf{1.4007}} & 1.4166 & 1.4280\\
5 & 1.6347 & 1.6364 & 1.6588 & \underline{\textbf{1.6337}} & 1.6388 & 1.6549\\
6 & 1.7339 & 1.7304 & 1.7511 & \underline{\textbf{1.7252}} & 1.7252 & 1.7459\\
7 & 1.9166 & \underline{\textbf{1.9048}} & 1.9369 & 1.9058 & 1.9087 & 1.9339\\
8 & 2.0647 & \underline{\textbf{2.0544}} & 2.0958 & 2.0620 & 2.0638 & 2.1017\\
9 & 2.2823 & 2.2821 & 2.2989 & \underline{\textbf{2.2778}} & 2.2912 & 2.2967\\
10 & 2.3575 & 2.3712 & 2.3809 & \underline{\textbf{2.3550}} & 2.3817 & 2.3774\\
11 & \underline{\textbf{2.4624}} & 2.4781 & 2.4871 & 2.4643 & 2.4942 & 2.4896\\
12 & \underline{\textbf{2.5460}} & 2.5767 & 2.5805 & 2.5547 & 2.5881 & 2.5876\\\hline
\end{tabularx}
\label{tab:Tab03DensNoEmiss}
\caption*{\scriptsize\textit{Notes}: Panel (\textit{a}) (\textit{b}), and (\textit{c}) show quantile-weighted Continuous Ranked Probability Scores (CRPS) with weighting that emphasizes the center and the right or the left tail of the distribution, respectively. The best forecasts, associated with the lowest scores, are underlined.}
\end{table}

\FloatBarrier

\subsection{Expert forecasts of the nominal price of carbon}
Our qualitative comparison of the single-factor BVAR(1) point forecasts against those issued by the Carbon Team at the London Stock Exchange Group (LSEG; formerly Refinitiv), and its survey forecasts, involves two preliminary steps. First, LSEG provides forecasts expressed in current euros; therefore, in this case, we rely on nominal price forecasts.\footnote{We have considered an alternative approach, where we transform our real price forecasts into nominal terms by producing inflation forecasts from an Unobserved Component SV model and using them to get nominal price forecasts. Moreover, given the short time span of our evaluation sample, using different inflation forecasts, such as RW forecasts, does not alter the results. Results are available from the authors upon request.} The second challenge of working with these data is that they are ``fixed event forecasts'', while our models produce ``fixed horizon forecasts''.

The characteristic of ``fixed event forecasts'' is that the forecast horizon changes as the forecast origin moves forward. 
At each forecast origin, the LSEG team produces forecasts for the current year, $f^{FE}_{t+k|t}$, and for the next year, $f^{FE}_{t+k+12|t}$, where $k=1,...,12$ represents the number of months until the end of the year (e.g. $k=12$ in January and $k=1$ in December). 
To approximate one-year-ahead fixed horizon forecasts, $f_{t+12|t}^{FH}$, using LSEG's fixed event forecasts, we follow \citet{supp:dovern2012disagreement}:

\begin{equation}
    f_{t+12|t}^{FH} = \frac{k}{12} f^{FE}_{t+k|t} +\frac{12-k}{12} f^{FE}_{t+k+12|t},
\end{equation}
where weights are proportional to the degree of overlap of the two fixed event forecasts.\footnote{A fixed horizon forecast issued in January, $k=12$, would therefore be equal to $f_{t+12|t}^{FH} = f^{FE}_{t+12|t}$.} 

\bigskip

\begin{figure}[!ht]
    \centering
        \caption{\small{Nominal EU ETS price and nominal price forecasts.}}
\includegraphics[width=14cm]{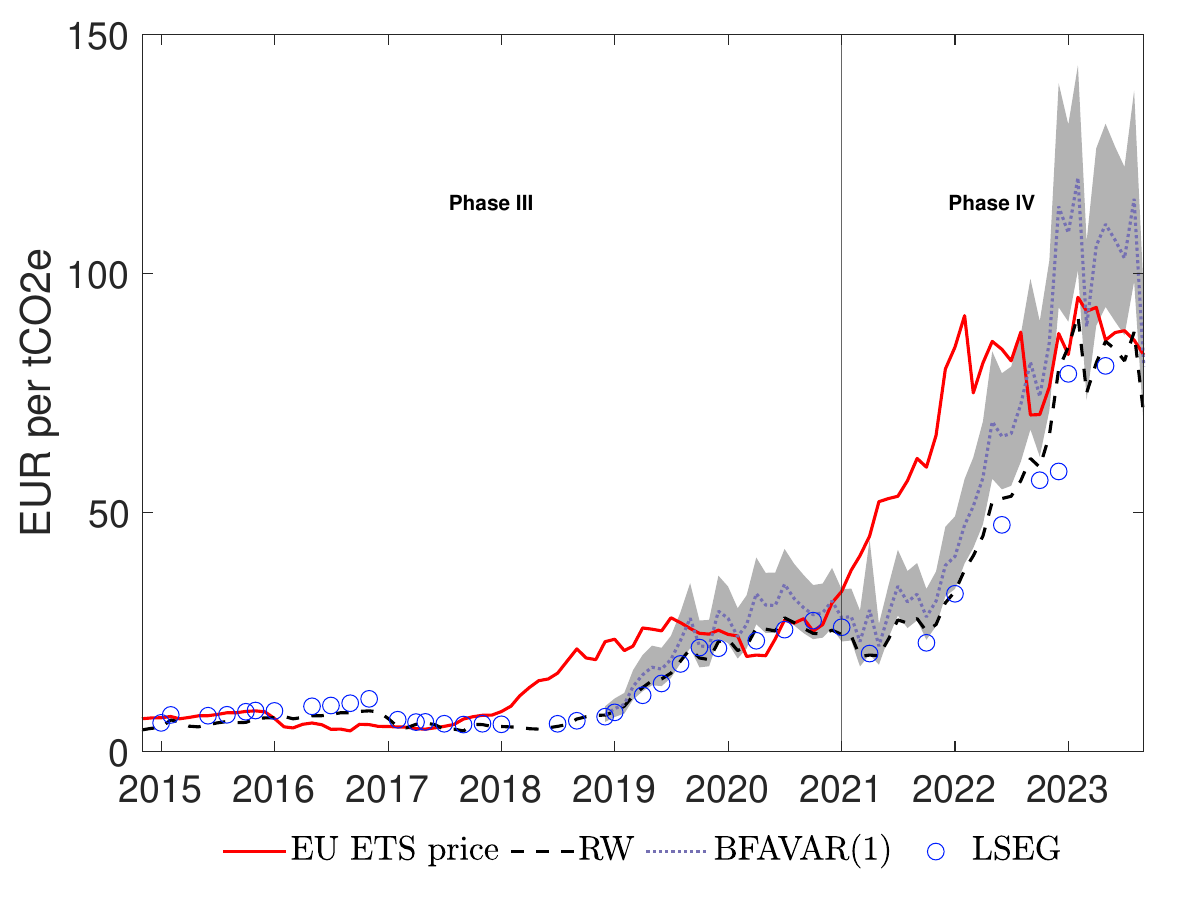}
\label{fig:FigLSEG_nominal}
    \caption*{\scriptsize{\textit{Notes}: Nominal EU ETS price (red line), LSEG (blue dots), RW (dashed black line), BFAVAR(1) one-year-ahead forecasts using nominal price data (dotted blue line). The shaded gray area represents the $68\%$ credible intervals from the BFAVAR(1) model. The shaded vertical area represents the COVID recession, as determined by the CEPR-EABCN Euro Area Business Cycle Dating Committee (\url{https://eabcn.org}).}}
\end{figure}

\noindent\textit{LSEG Carbon Team's forecasts.} Using the procedure outlined above, we obtain a set of 39 one-year-ahead fixed horizon forecasts irregularly spaced over the period January 2015--May 2023; only 19 of these forecasts overlap with those in our evaluation period, spanning from December 2018 to September 2023. Figure~\ref{fig:FigLSEG_nominal} reports the nominal EU ETS price together with LSEG, random-walk, and BFAVAR(1) one-year-ahead forecasts. 

\FloatBarrier

\bigskip

\noindent\textit{Survey forecasts.} Carbon Market Surveys were run by LSEG each year from 2020 to 2022 between February and April (in 2020, the survey was closed in March). Participants who answered the section of the survey concerning EU ETS price expectations numbered 60 in 2020, 119 in 2021, and 88 in 2022. 
Since survey forecasts are also of the fixed event type, we approximate fixed horizon one-year-ahead prediction densities derived from surveys and compare them with prediction densities from the BFAVAR model and the realized nominal price at the target date.

\FloatBarrier

\subsection{Verified Emission Forecasts}\label{sec:emissioni}
Table \ref{tab:Tab06EmissRMSE} reports relative RMSFEs for verified emissions forecasts across all horizons from one to twelve months ahead. No specification beats the RW at $h=1$ and $h=2$; from $h=3$ onward, all SV models deliver relative RMSFEs below one, with the one-factor BFAVAR(1)-SV performing best at horizons three through seven and the baseline BVAR(1)-SV dominating from $h=8$ 
onward. The AR(1) is uniformly worse than the RW at all horizons, providing evidence against its use as a univariate benchmark. At $h=12$, SV models reduce RMSFE by between 6.5\% and 7.1\% relative to the RW.

\begin{table}[ht]
\centering
\caption{Relative RMSFE: verified emissions}
\newcolumntype{C}[1]{>{\scriptsize\hsize=#1\hsize\centering\arraybackslash}X}
\renewcommand{\arraystretch}{0.8}
\begin{tabularx}{\textwidth}{C{.2}|C{.9}|C{.9}|C{.9}|C{.9}|C{1.3}|C{1.3}|C{1.3}|C{1.3}}\hline
 &  & &  1 Factor &  2 Factors   & & &  1 Factor &  2 Factors  \\\cline{4-5}\cline{8-9}
h & BAR(1) & BVAR(1) & BVAR(1) & BVAR(1) & BAR(1)-SV & BVAR(1)-SV & BVAR(1)-SV & BVAR(1)-SV \\\hline
1 & 1.1050 & 1.2382 & 1.1805 & 1.2370 & 1.0064 & 1.0206 & 1.0320 & 1.0134\\
2 & 1.0567 & 1.2274 & 1.1630 & 1.2034 & 1.0020 & 1.0003 & 1.0017 & 1.0002\\
3 & 1.0470 & 1.2477 & 1.1716 & 1.2162 & \textbf{0.9999} & \textbf{0.9882} & \underline{\textbf{0.9867}} & \textbf{0.9915}\\
4 & 1.0434 & 1.2634 & 1.1719 & 1.2247 & \textbf{0.9988} & \textbf{0.9782} & \underline{\textbf{0.9773}} & \textbf{0.9815}\\
5 & 1.0331 & 1.2419 & 1.1515 & 1.2012 & \textbf{0.9972} & \textbf{0.9663} & \underline{\textbf{0.9646}} & \textbf{0.9676}\\
6 & 1.0296 & 1.2332 & 1.1379 & 1.1881 & \textbf{0.9967} & \textbf{0.9593} & \underline{\textbf{0.9576}} & \textbf{0.9584}\\
7 & 1.0230 & 1.2055 & 1.1141 & 1.1602 & \textbf{0.9956} & \textbf{0.9517} & \underline{\textbf{0.9513}} & \textbf{0.9518}\\
8 & 1.0210 & 1.1911 & 1.1009 & 1.1466 & \textbf{0.9960} & \underline{\textbf{0.9476}} & \textbf{0.9482} & \textbf{0.9484}\\
9 & 1.0179 & 1.1705 & 1.0836 & 1.1264 & \textbf{0.9964} & \underline{\textbf{0.9411}} & \textbf{0.9454} & \textbf{0.9441}\\
10 & 1.0139 & 1.1476 & 1.0686 & 1.1070 & \textbf{0.9960} & \underline{\textbf{0.9371}} & \textbf{0.9417} & \textbf{0.9411}\\
11 & 1.0139 & 1.1417 & 1.0621 & 1.1009 & \textbf{0.9962} & \underline{\textbf{0.9324}} & \textbf{0.9381} & \textbf{0.9370}\\
12 & 1.0124 & 1.1280 & 1.0502 & 1.0878 & \textbf{0.9962} & \underline{\textbf{0.9295}} & \textbf{0.9347} & \textbf{0.9330}\\\hline
\end{tabularx}
\label{tab:Tab06EmissRMSE}
\caption*{\scriptsize\textit{Notes}: see notes to Table \ref{tab:Tab01UnivApp}}
\end{table}

Table \ref{tab:Tab05EmissDens} reports quantile-weighted CRPS scores for the center and tails of the predictive distribution of verified emissions. For the center, SV models dominate throughout, with the one-factor BFAVAR(1)-SV recording the lowest score at most horizons up to $h=10$ and the BVAR(1)-SV prevailing at longer horizons. For the tails, the BAR(1)-SV model records the lowest qCRPS from $h=7$ onward, while the one-factor BFAVAR(1)-SV performs best at shorter horizons. Overall, incorporating stochastic volatility delivers consistent gains for verified emissions forecasting across all evaluation dimensions, in contrast to the results for the real price of carbon.

\begin{table}[ht]
\centering
\caption{Quantile-weighted Continuous Ranked Probability Score (qCRPS): verified emissions}
\newcolumntype{C}[1]{>{\scriptsize\hsize=#1\hsize\centering\arraybackslash}X}
\renewcommand{\arraystretch}{0.8}
\begin{tabularx}{\textwidth}{C{.2}|C{.9}|C{.9}|C{.9}|C{.9}|C{1.3}|C{1.3}|C{1.3}|C{1.3}}\hline
\multicolumn{9}{l}{\scriptsize (a) Quantile-weighted CRPS: center}\\\hline
 &  & &  1 Factor &  2 Factors   & & &  1 Factor &  2 Factors  \\\cline{4-5}\cline{8-9}
h & BAR(1) & BVAR(1) & BVAR(1) & BVAR(1) &  BAR(1)-SV & BVAR-SV(1) & BVAR-SV(1) & BVAR-SV(1) \\\hline
1 & 157.0242 & 157.0021 & 156.9899 & 156.9887 & 156.9946 & 156.9781 & \underline{\textbf{156.9780}} & 156.9921\\
2 & 156.8855 & 156.8483 & 156.8334 & 156.8374 & 156.8151 & 156.7982 & \underline{\textbf{156.7928}} & 156.8204\\
3 & 156.7733 & 156.7232 & 156.7100 & 156.7176 & 156.6598 & 156.6397 & \underline{\textbf{156.6362}} & 156.6704\\
4 & 156.6913 & 156.6291 & 156.6185 & 156.6298 & 156.5277 & 156.5124 & \underline{\textbf{156.5010}} & 156.5420\\
5 & 156.6258 & 156.5507 & 156.5448 & 156.5562 & 156.4141 & 156.3982 & \underline{\textbf{156.3833}} & 156.4253\\
6 & 156.6003 & 156.5167 & 156.5132 & 156.5291 & 156.3403 & 156.3271 & \underline{\textbf{156.3116}} & 156.3555\\
7 & 156.5850 & 156.4944 & 156.4932 & 156.5165 & 156.2708 & 156.2593 & \underline{\textbf{156.2520}} & 156.2946\\
8 & 156.6132 & 156.5156 & 156.5159 & 156.5499 & 156.2490 & \underline{\textbf{156.2374}} & 156.2385 & 156.2747\\
9 & 156.6089 & 156.5020 & 156.5019 & 156.5593 & 156.1928 & 156.1849 & \underline{\textbf{156.1826}} & 156.2263\\
10 & 156.6533 & 156.5480 & 156.5508 & 156.6002 & 156.1893 & 156.1863 & \underline{\textbf{156.1810}} & 156.2230\\
11 & 156.6695 & 156.5500 & 156.5593 & 156.6120 & 156.1585 & \underline{\textbf{156.1477}} & 156.1487 & 156.1910\\
12 & 156.6652 & 156.5383 & 156.5498 & 156.6039 & 156.1081 & \underline{\textbf{156.0949}} & 156.0967 & 156.1433\\\hline
\multicolumn{9}{l}{\scriptsize (b) Quantile-weighted CRPS: tails}\\\hline
 &  & &  1 Factor &  2 Factors   & & &  1 Factor &  2 Factors  \\\cline{4-5}\cline{8-9}
h & BAR(1) & BVAR(1) & BVAR(1) & BVAR(1) &  BAR(1)-SV & BVAR-SV(1) & BVAR-SV(1) & BVAR-SV(1) \\\hline
1 & 268.3299 & 268.2731 & 268.2503 & 268.2322 & 268.2331 & \underline{\textbf{268.1990}} & 268.2043 & 268.2284\\
2 & 267.5153 & 267.4249 & 267.3934 & 267.3871 & 267.3189 & 267.3001 & \underline{\textbf{267.2775}} & 267.3422\\
3 & 266.8291 & 266.7003 & 266.6655 & 266.6619 & 266.5056 & 266.5011 & \underline{\textbf{266.4760}} & 266.5737\\
4 & 266.2506 & 266.0752 & 266.0567 & 266.0522 & 265.7708 & 265.8089 & \underline{\textbf{265.7487}} & 265.9089\\
5 & 265.7319 & 265.5114 & 265.5022 & 265.5110 & 265.1168 & 265.1746 & \underline{\textbf{265.0863}} & 265.2611\\
6 & 265.3104 & 265.0560 & 265.0518 & 265.0623 & 264.5733 & 264.6309 & \underline{\textbf{264.5444}} & 264.7302\\
7 & 264.9266 & 264.6453 & 264.6548 & 264.6688 & \underline{\textbf{264.0273}} & 264.1117 & 264.0343 & 264.2273\\
8 & 264.6624 & 264.3254 & 264.3593 & 264.3907 & \underline{\textbf{263.6092}} & 263.7132 & 263.6682 & 263.8356\\
9 & 264.3227 & 263.9766 & 264.0001 & 264.0878 & \underline{\textbf{263.1151}} & 263.2528 & 263.1862 & 263.3792\\
10 & 264.1279 & 263.7373 & 263.7826 & 263.8496 & \underline{\textbf{262.7595}} & 262.9314 & 262.8339 & 263.0278\\
11 & 263.8515 & 263.4195 & 263.4950 & 263.5534 & \underline{\textbf{262.3493}} & 262.5213 & 262.4200 & 262.6357\\
12 & 263.5330 & 263.0531 & 263.1335 & 263.2389 & \underline{\textbf{261.8955}} & 262.0665 & 261.9731 & 262.2218\\\hline
\end{tabularx}
\label{tab:Tab05EmissDens}
\caption*{\scriptsize\textit{Notes}: Panel (\textit{a}) and (\textit{b}) show quantile-weighted Continuous Ranked Probability Scores (CRPS) with weighting that emphasizes the center and the tails of the distribution, respectively. The best forecasts, associated with the lowest scores, are underlined.}
\end{table}

\FloatBarrier

\section{Data Sources}
\label{sec:data_sources}
\begingroup
\setstretch{1}
\small
\setlength{\tabcolsep}{5pt}
\renewcommand{\arraystretch}{1.04}

\begin{minipage}{\textwidth}
\begin{tabularx}{\textwidth}{@{}>{\raggedright\arraybackslash}X >{\raggedright\arraybackslash}p{0.36\textwidth}@{}}\hline
\toprule
\textbf{Variable} & \textbf{Source}\\ \hline
\midrule

\multicolumn{2}{@{}l}{\textbf{Carbon price}}\\[-0.15em]
EUA one-month-ahead futures price & London Stock Exchange Group (LSEG) \\
\\[-0.35em]

\multicolumn{2}{@{}l}{\textbf{Verified emissions}}\\[-0.15em]
Verified emissions, EU-19 & European Environment Agency \\
\\[-0.35em]

\multicolumn{2}{@{}l}{\textbf{Economic activity}}\\[-0.15em]
Aggregate industrial production, EU-19 & Eurostat \\
Sectoral industrial production indices:\newline
\hspace*{1em}-- Electricity, gas, steam, and air conditioning supply\newline
\hspace*{1em}-- Basic metals\newline
\hspace*{1em}-- Manufacture of paper and its products\newline
\hspace*{1em}-- Coke and refined petroleum products\newline
\hspace*{1em}-- Chemical products\newline
\hspace*{1em}-- Non-metallic mineral products
& Eurostat \\
\\[-0.35em]

\multicolumn{2}{@{}l}{\textbf{Real energy prices}}\\[-0.15em]
Brent crude oil & London Stock Exchange Group (LSEG) \\
TTF natural gas, front-month and front-year & London Stock Exchange Group (LSEG) \\
ARA API-2 coal, front-year & London Stock Exchange Group (LSEG) \\
German power price, front-year & London Stock Exchange Group (LSEG) \\
Clean dark spread and clean spark spread, front-month & London Stock Exchange Group (LSEG) \\
\\[-0.35em]

\multicolumn{2}{@{}l}{\textbf{Technical and financial indicators}}\\[-0.15em]
Auction coverage ratio & European Energy Exchange \\
Auction clearing price & European Energy Exchange \\
Monthly auction price range & European Energy Exchange \\
Euro Stoxx 50 & London Stock Exchange Group (LSEG) \\
\\[-0.35em]

\multicolumn{2}{@{}l}{\textbf{Weather conditions}}\\[-0.15em]
Temperature and precipitation anomalies, EU-19 & International Energy Agency, Weather for Energy Tracker \\[.5cm] \hline

\bottomrule
\end{tabularx}

\vspace{0.75em}
{\footnotesize
\textit{Notes:} Nominal EUA futures prices, real energy prices, the auction clearing price, and the Euro Stoxx 50 are deflated using the Euro-area HICP where applicable. Annual verified emissions are temporally disaggregated to monthly frequency using the Chow--Lin method. \\

\textit{Disclaimer}: London Stock Exchange Group (LSEG) data are proprietary and subject to licensing restrictions; the underlying series therefore cannot be disclosed in raw form, redistributed, or made publicly available by the authors.
}
\end{minipage}
\endgroup

\end{document}